\documentclass{aa}

\usepackage{graphicx}
\usepackage{longtable}
\usepackage[varg]{txfonts}
\usepackage{xcolor}

\usepackage{hyperref}

\begin{document}

\title{MUSE imaging spectroscopy of the fullerene planetary nebula Tc~1}  
\titlerunning{MUSE study of Tc~1}
\author{
   J.~R.~Walsh\inst{1} \and
   M.~J.~Barlow\inst{2} \and
   A.~Monreal-Ibero\inst{3} \and
   J.~Cami\inst{4,5} \and
   E.~Peeters\inst{4,5} \and
   R.~Wesson\inst{6} \and
   J.~Bernard-Salas\inst{7} \and
   N.~L.~J.~Cox\inst{7} \and
   G.~F.~J.~Watt\inst{4}
}
     
\institute{
European Southern Observatory, Karl-Schwarzschild Strasse 2, D-85748 Garching, Germany \\
\email{jwalsh@eso.org}
\and
Department of Physics and Astronomy, University College London, Gower Street, London WC1E 6BT, UK
\and
Leiden Observatory, Leiden University, P.O. Box 9513, 2300 RA Leiden, the Netherlands
\and
Department of Physics and Astronomy, The University of Western Ontario, London, 
ON N6A 3K7, Canada
\and
Institute for Earth and Space Exploration, The University of Western Ontario, London, ON N6A 3K7, Canada
\and
Cardiff Hub for Astrophysics Research and Technology (CHART), School of Physics and Astronomy, Cardiff University, The Parade, Cardiff CF24 3AA, UK
\and
ACRI-ST, Centre d’Etudes et de Recherche de Grasse (CERGA), 10 Av. Nicolas Copernic, 06130 Grasse, France
}

\date{Received: 1 March 2026; accepted: 18 June 2026}
 
\abstract
{}
{The planetary nebula Tc~1 (PN G345.2 -08.8), one of the rare group of Galactic PNe showing 
fullerene emission in the infrared, is spatially extended with high surface brightness. 
Optical imaging spectroscopy enables the line and continuum structure and
diagnostics to be examined, and to search for any correlations to the presence of 
fullerene dust.
} 
{Tc~1 was observed with MUSE wide 
field mode with adaptive optics, wavelength range 4750-9300 \AA, at three exposure levels to 
ensure unsaturated emission lines. Extinction, electron temperature ($T_{\rm e}$) and density 
($N_{\rm e}$) diagnostic images are presented from collisionally excited and recombination line
ratios.
} 
{The nebula has a high surface brightness core, diameter 12$''$, with an elliptical ring 
of major axis 2.8$''$ around the central star and some low ionization knots, and an extended 
halo 55$''$ in diameter; between the core and halo is an annulus with intermediate properties, 
including higher $T_{\rm e}$ and lower $N_{\rm e}$ than in the core. The spectrum of the 
central star was extracted and fitted by a 31000\,K model atmosphere and is of 
type O7.5I(f). The image of optical extinction from H Balmer line ratios is highly structured, 
and shows an annulus, adjacent to the core, of low extinction, lower than the 
line-of-sight interstellar extinction. Instrumental effects to account for this 
anomalously low extinction area are investigated and intrinsic effects from the scattering 
properties of the nebular dust; neither can entirely explain the low-extinction region and
the most likely cause is a local non-standard dust reddening law. 
This low extinction region additionally shows an anomalously high \ion{He}{I} 7281/6678\AA\ line 
ratio, possibly caused by a contaminating line, but none were conclusively identified; 
its origin remains unresolved. 
Spectra over extended regions were also analysed and distinct enhancement of the continuum 
above the expected nebular continuum (as also seen in some other PNe observed with MUSE) 
was found.
}
{The annulus of low extinction occurs outside the region of strongest fullerene 
emission, in the zone where $N_{\rm e}$ declines and $T_{\rm e}$ rises. A change in 
dust properties linked to conditions in this transition region between the higher 
density core nebula and lower density halo is deduced.
}


\maketitle

\nolinenumbers

\section{Introduction}
\label{Sect:Intro}

\object{Tc 1} \citep{Thackeray1950}, alternative name IC~1266 (PN G345.2 -08.8), is an 
apparently normal low ionization planetary nebula (PN), but was the first PN to have fullerene 
emission (C$_{\rm 60}$) detected in the mid-infrared \citep{Camietal2010}. The number of such 
planetary nebulae (PNe) with fullerene emission is still low (25 - \cite{GarciaHernandez2012}, 
\cite{Otsuka2019} in the Galaxy, Large Magellanic Cloud (LMC) and Small Magellanic Cloud (SMC)). 
Fullerenes are now detected in many other interstellar environments, 
such as post-asymptotic giant branch (AGB) circumbinary disks \citep{Gielenetal2011}, 
the Orion Nebula \citep{Rubinetal2011}, reflection nebulae \citep{Sellgrenetal2010} and 
even the diffuse interstellar medium \citep{Berne2017}. 

Tc1 has a high surface brightness core $\sim$12$''$ in diameter surrounded by
a much fainter halo extending to radii $\sim$25$''$ \citep{Schwarzetal1992,  Corradietal2003}.
It is a low ionization PN: the highest ionization species observed is [\ion{O}{III}]
(ionization energy 35.1 -- 54.9 $eV$), and is classified as C-rich (log(C/O) of $+$0.14) on 
the basis of optical \citep{Williamsetal2008} and ultra-violet (IUE) spectra 
\citep{Pottaschetal2011}. The central star (CD -46$^{\circ}$11816) is bright 
(Gaia G mag. = 11.269) with a parallax of 0.2672 $\pm$ 0.0344 mas, a Galactic Plane object 
in the direction to the Galactic Bulge; \cite{Chornay2021} derive a distance of 
3530 $^{+380}_{-329}$ pc. 

\cite{Alemanetal2019} presented deep spectroscopy of Tc~1 with VLT X-shooter 
\citep{Vernetetal2011} for a slit offset from the nebula central star (CSt) 
over the bright core. 
The X-shooter slit (length $11''$) sampled the bright core, but not the halo. UV-blue
and red spectra were presented (wavelength coverage 3250--5550 and 5660--10050\,\AA)
and the emission lines analysed for extinction, physical diagnostics -- electron temperature
($T_{\rm e}$) and volume electron density ($N_{\rm e}$, cm$^{-3}$) -- and ionic and 
total abundances. The total abundances derived from a photoionization model were 
log(He/H) = $-$0.098 and for O, log(O/H) = $-$3.60 from the collisionally 
excited lines (CEL) and $-$3.06 from the optical recombination lines (ORL); with CEL 
log(N/O) = $-$1.00 (\cite{Alemanetal2019}, their Table 4). From the photoionization 
model, the CSt was matched by a 30\,000 or 32\,000\,K black 
body; using the Gaia distance, the luminosity is then 8200 or 9000 L$_{\odot}$.

Given the deep spectroscopic study conducted by \cite{Alemanetal2019}, although with only 
a single short slit, the next step in optical observations of Tc~1 is integral field 
spectroscopy (IFS). Apart from the central star's low effective temperature and high 
nebular C/O ratio,
the X-shooter spectroscopy did not reveal any particular aspect of the nebula marking it out
as exceptional relative to the more common fullerene non-detected PNe. Using the large field 
MUSE IFS on the VLT \citep{Bacon2010}, with a field which covers the core and the complete 
halo in a single pointing, we have investigated in depth possible spectro-spatial features that 
might betray the role of fullerenes in the optical structure and properties of Tc~1.

The outline of the paper is as follows: Section 2 presents the MUSE observations 
and a brief description of the reduction of instrument signature. This is followed by
presentation of the emission line imaging results in Section 3, including correction 
for the effect of field stars on the images over the area of the MUSE field and 
combination of images from the three exposure times. Section 4 presents the 
derived diagnostics:- extinction from Balmer line ratios and physical conditions 
$T_{\rm e}$ and $N_{\rm e}$ from the CEL and ORL ratios. Section 5 presents a wide 
ranging discussion on the emission line morphology, the complex 
extinction structure and a possible model to explain the area of anomalously 
low extinction, the parameters of the CSt as extracted from the MUSE cube, 
summed spectra for selected spatial areas of Tc~1 and finally on the role of critical 
further observations, particularly in the infrared. The papers ends with Conclusions. 
Four appendices consider: correction procedure of field star spectra; 
application of image 
restoration to investigate the reality of the sharp changes in areal extinction;
a dust scattering model to try to explain these features; comparison of a residual 
continuum signal in the summed spectra of several PNe observed with MUSE, not 
explained by nebular and stellar contributions. A fifth appendix tabulates the summed
fluxes for the emission lines in the three spatial areas and the total imaged area for 
Tc~1.  
    
\section[]{Observations and data reduction}
\label{Sec:observations}
Tc~1 was observed in service mode in 2022 in dark time with the Very Large Telescope (VLT) Multi-Unit 
Spectroscopic Explorer (MUSE) integral field instrument \citep{Baconetal2014} in wide field 
(59.9$\times$60$''$) mode (WFM) with adaptive optics (AO) correction fed by the 
Adaptive Optics Facility (AOF) instrument GALACSI in ground layer AO (GLAO) mode 
\citep{Stuiketal2006, Stroebele2012}. The MUSE nominal wavelength coverage 
(4700-9300\,\AA) was chosen (however, on account of the contamination in 
the region of the \ion{Na}{I} D lines at 5890.0, 5895.9\,\AA\ from the AOF 
laser guide stars, a blocking filter masks the spectral range 5806 -- 5965\,\AA). 
The mean (2 pixel) spectral resolving power of MUSE is $\simeq$2800. 

Three exposure times were specified: the shortest 40\,s to ensure all, including the strongest
emission lines (H$\alpha$, [\ion{O}{III}] 5007\,\AA) were detected unsaturated; 101\,s for lines 
of intermediate strength to be unsaturated; and 595\,s for detection of faint lines (e.g.,
$\ll$1\% of H$\beta$). The range of exposure times was chosen so that the fluxes of the faintest 
and strongest lines over a range of $>$100 could be compared. The total exposure time was 4.29 
hours divided between 48 40\,s exposures (0.53 h), 16 101\,s exposures 
(0.45 h) and 20 of 595\,s (3.31 h), taken over four 
nights in VLT service mode. 90$^{\circ}$ rotations (PAs of 0 and 90 $^{\circ}$) and dithers of 
0.4$''$ offsets were applied in order to help mitigate the pattern of the 24 individual integral 
field units of MUSE. Since the full extent of Tc~1 from previous studies did not fill the MUSE field of view, the 
background sky could be estimated from the edges of the field. Table \ref{tab:Obs} provides a 
log of these observations. Column 8 lists the mean airmass per exposure and column 9 the 
differential image motion monitor (DIMM) image quality. On 2022-06-16 four 595\,s exposures 
were obtained, but the seeing was poor ($>$1.20$''$) and these data were excluded from the 
final combination of the longest exposure set.  

\begin{table*}
\caption{Log of MUSE observations of Tc~1}
\centering
\begin{tabular}{lrrrrrrrr}
\hline\hline
Target   & RA~~~~          & Dec~~~                  & T~exp & PA           & Date  & Dataset   & Airmass & DIMM   \\
         & ($h$ $m$ $s$)~~ & ($^{\circ}$ $'$ $''$)~~ & (s)   & ($^{\circ}$) &       & UT(h:m:s) &         & ($''$) \\
\hline
 Tc~1 & 17 45 35.44 & -46 05 26.3 &  40.0 &  0 & 2022-06-29 & 02:29:41.267 & 1.134 & 0.56 \\ 
 Tc~1 & 17 45 35.44 & -46 05 26.3 &  40.0 &  0 & 2022-06-29 & 02:31:35.649 & 1.131 & 0.51 \\ 
 Tc~1 & 17 45 35.44 & -46 05 26.3 &  40.0 &  0 & 2022-06-29 & 02:33:29.903 & 1.129 & 0.44 \\ 
 Tc~1 & 17 45 35.50 & -46 05 26.3 &  40.0 &  0 & 2022-06-29 & 02:36:02.866 & 1.125 & 0.53 \\ 
 Tc~1 & 17 45 35.50 & -46 05 26.3 &  40.0 &  0 & 2022-06-29 & 02:37:57.262 & 1.123 & 0.57 \\ 
 Tc~1 & 17 45 35.50 & -46 05 26.3 &  40.0 &  0 & 2022-06-29 & 02:39:51.465 & 1.121 & 0.57 \\ 
 Tc~1 & 17 45 35.50 & -46 05 25.7 &  40.0 &  0 & 2022-06-29 & 02:42:23.818 & 1.118 & 0.44 \\
 Tc~1 & 17 45 35.50 & -46:05:25.7 &  40.0 &  0 & 2022-06-29 & 02:44:18.371 & 1.116 & 0.50 \\
 Tc~1 & 17 45 35.50 & -46 05 25.7 &  40.0 &  0 & 2022-06-29 & 02:46:12.929 & 1.113 & 0.55 \\
 Tc~1 & 17 45 35.44 & -46 05 25.7 &  40.0 &  0 & 2022-06-29 & 02:48:45.497 & 1.111 & 0.45 \\ 
\hline
\end{tabular}
\tablefoot{The column labelled DIMM lists the value of the VLT Differential Image Motion 
Monitor image quality (`seeing') for the observation. \\
Only the first 10 entries of this table are shown; the full table is 
contained in the on-line material.
}
\label{tab:Obs}
\end{table*}

The MUSE WFM-AO-N observations of Tc~1 were reduced with $EsoRex$ (version 3.13)
scripts and the MUSE instrument pipeline, version 2.8.7 \citep{Weilbacheretal2014,
Weilbacheretal2020}. Compatible master bias, master flat, master dark, wavelength 
calibration table (WAVECAL), trace table, twilight cube (for sky flat correction), and 
spectrophotometric response (STD\_RESPONSE), all local in time to the observations, were 
downloaded from the ESO MUSE calibration database \footnote{When applying the calibration 
database reference files, bright and dark rows were found 
on extracted images from the final data cubes from several nights of observation, 
irrespective of exposure time. This problem was tracked down to the database slit 
illumination files used (tagged ILLUM in the MUSE pipeline) which produced slightly 
shifted tracing of the IFU slits from those recorded each night. When applying the correct 
daily illumination files to the basic calibration steps, the brighter/darker rows were 
found to disappear.}. The spectral trace table from the 
calibration database dated from 2021 and the Paranal atmospheric extinction table from 
2014. The spectrophotometric standard star used to establish the flux calibration was 
GD\,153 \citep{Moehleretal2014}.

The correction for the presence of Raman scattered lines, induced by the AOF Na laser 
guide stars \citep{Vogtetal2017}, was made as included in the MUSE $sci\_basic$ 
task. Using exposures in a $V'$-band filter (square profile, 5350--5650\,\AA) and 
centroiding (IRAF\footnote{IRAF is distributed by the National Optical Astronomy 
Observatories, which are operated by the Association of Universities for Research
in Astronomy, Inc., under cooperative agreement with the National
Science Foundation.} $imexam$) the bright CSt, the offsets between exposures were determined
and used in precisely aligning the individual exposures to produce a single combined
data cube for each exposure level.

The resultant sky-subtracted cubes have dimensions 318 ($\alpha$) by 319 ($\delta$)
[63.6 $\times$ 63.8$''$ with 0.2$''$ spaxels] by 3681 ($\lambda$) voxels (4700–9300\,\AA) 
at the default binning of 1.25\,\AA. 
Extracting the same $V'$ filter images from the final cubes and measuring the same set 
of (17) well separated and apparently single stars, the mean and standard deviation
of the full width at half maximum (FWHM) Gaussian image quality was measured as: 
0.579 $\pm$ 0.023 $''$ for the 40\,s exposure; for the 101\,s set, 0.602 $\pm$ 0.027$''$; 
and 0.645 $\pm$ 0.021$''$ for 595\,s exposure.
The absolute astrometry was adjusted using the coordinates 
of the CSt from the \textit{Gaia} DR3 catalogue \citep{Lindegrenetal2021}:
(Gaia DR3 5954912374289120896, $\alpha$, $\delta$ 17$^{h}$ 45$^{m}$ 35.28$^{s}$, 
-46$^{\circ}$ 05$'$ 23.87$''$) for the 40\,s exposure where the $V'$ stellar image was 
unsaturated.

\section{Results: imaging}
The emission lines in each combined cube at the three exposure levels were analysed with a
semi-automatic Gaussian line fitting task, described in \cite{Walshetal2018}. This
program fits lines using a pre-defined list of emission lines expected in the spectrum and 
the list was produced from the Tc~1 line list in \cite{Alemanetal2019}. If a line from
this list was not detected to a given $\sigma$ above the background in the MUSE spectrum,
no Gaussian fit was performed. The Gaussian fitted lines in each spaxel were converted into
emission line flux and error images (and line wavelength and width images were also recorded);
the Gaussian fit errors were propagated from the MUSE variance cube.

\begin{figure*}[t]
\centering
\resizebox{\hsize}{!}{
\includegraphics[trim={0cm 1cm 2cm 1cm}, clip]{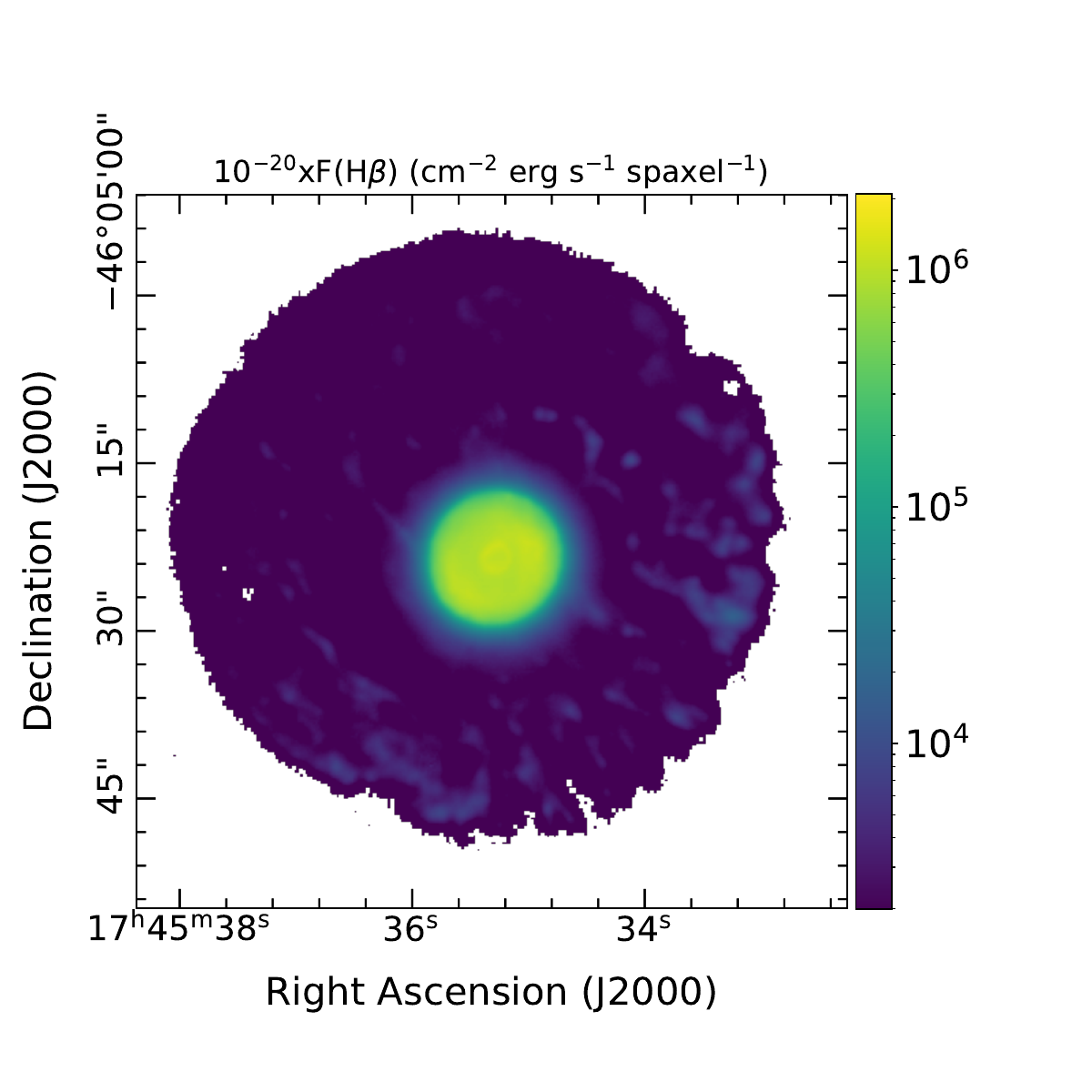}
\includegraphics[trim={0cm 1cm 2cm 1cm}, clip]{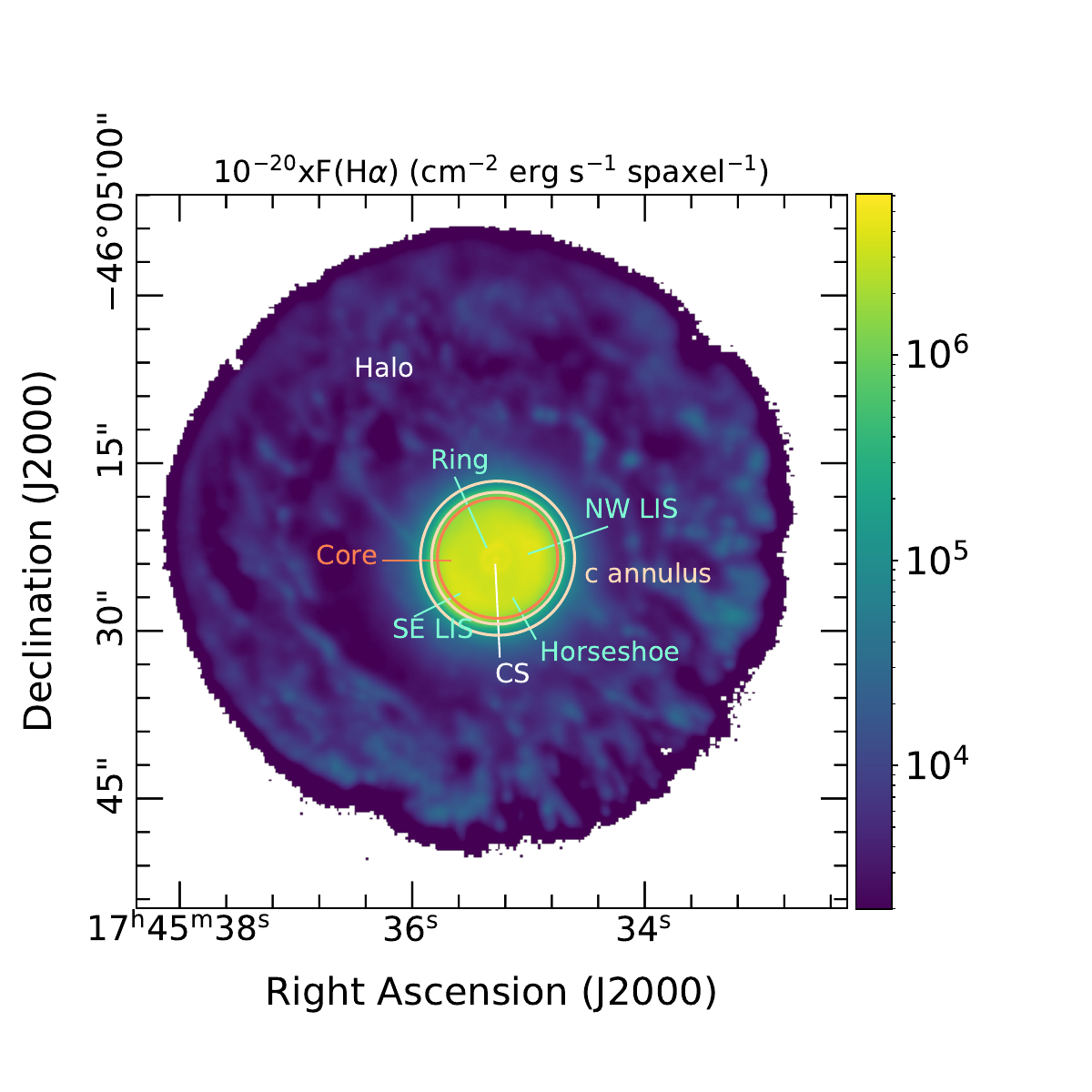}}
\caption{The morphology of Tc~1 in H Balmer emission. \\
Left: Log flux (erg cm$^{-2}$ s$^{-1}$) of H$\beta$ extracted from 
the 40\,s cube, displaying the bright core and low surface brightness halo
to 3$\sigma$ line flux detection per 0.2$\times$0.2$''$ spaxel; \\
Right: Composite image of Tc~1, composed of the bright core in H$\alpha$ 
(linear stretch) and the fainter surroundings and halo in [\ion{N}{II}] 6583\AA\ 
(log stretch), with the morphological features of the nebula indicated. \\
All images are oriented: E left, N up.
}
\label{fig:EmLines1}
\end{figure*}

\begin{figure*}[t]
\centering
\resizebox{\hsize}{!}{
\includegraphics[trim={0cm 1cm 2cm 1cm}, clip]{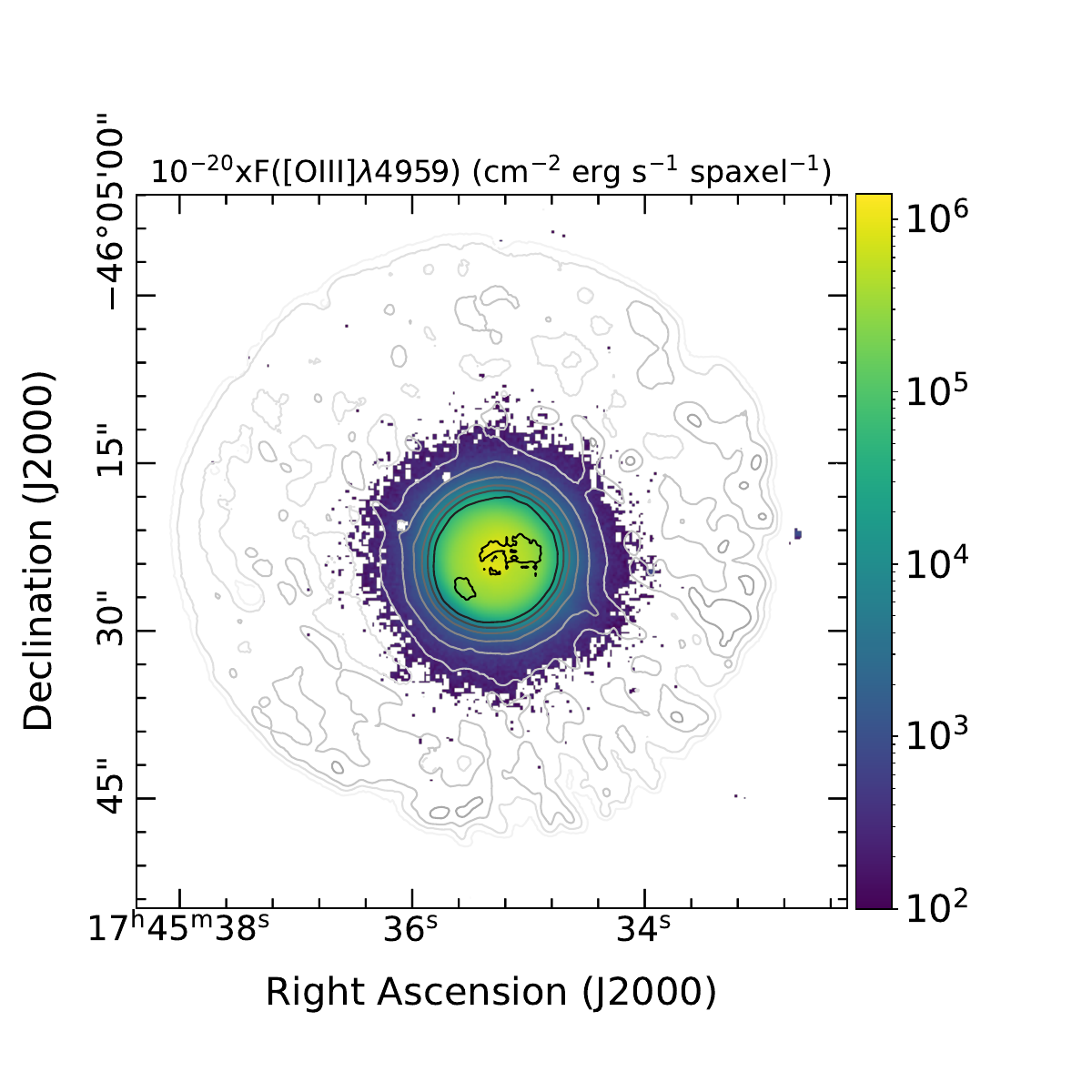}
\includegraphics[trim={0cm 1cm 2cm 1cm}, clip]{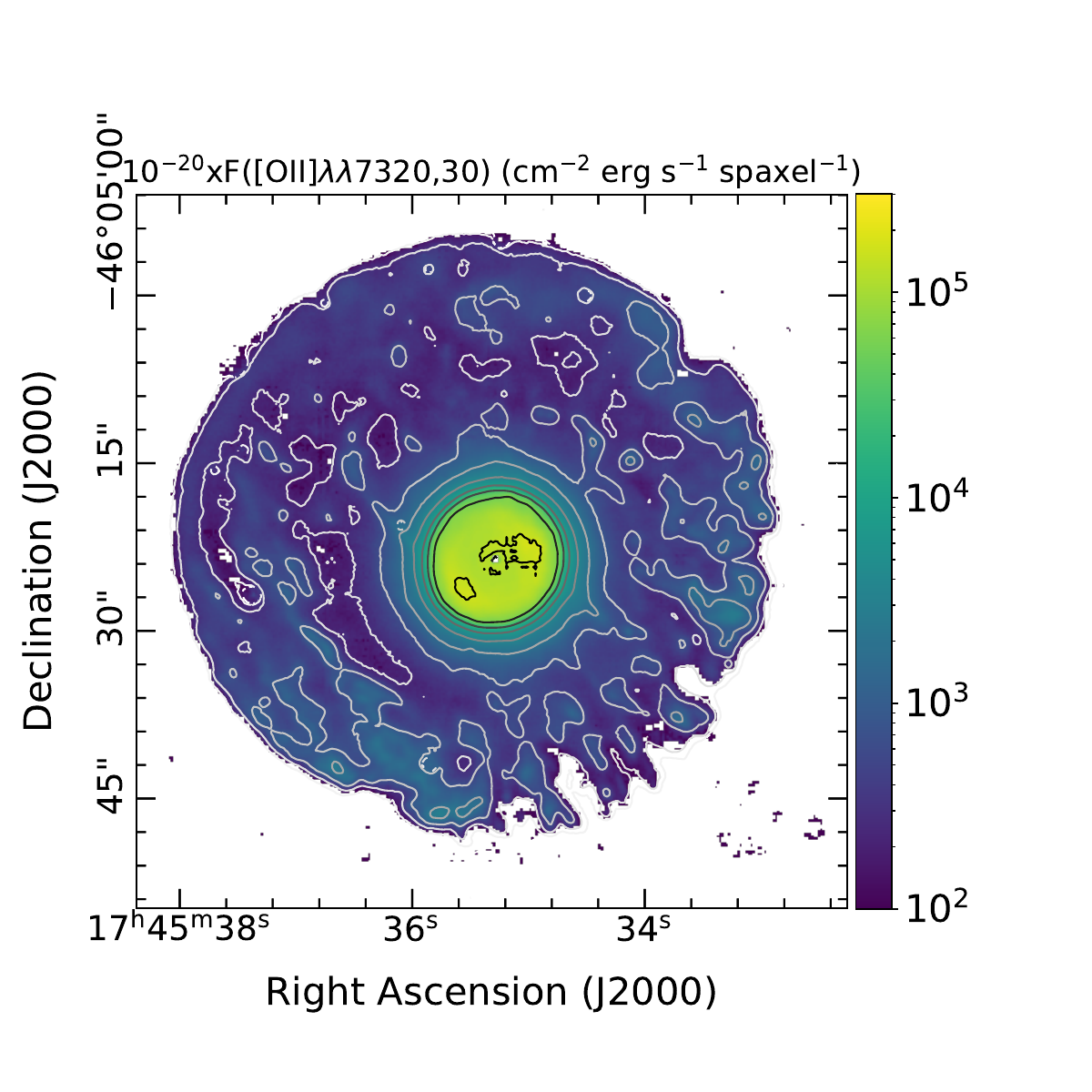}
}
\resizebox{\hsize}{!}{
\includegraphics[trim={0cm 1cm 2cm 1cm}, clip]{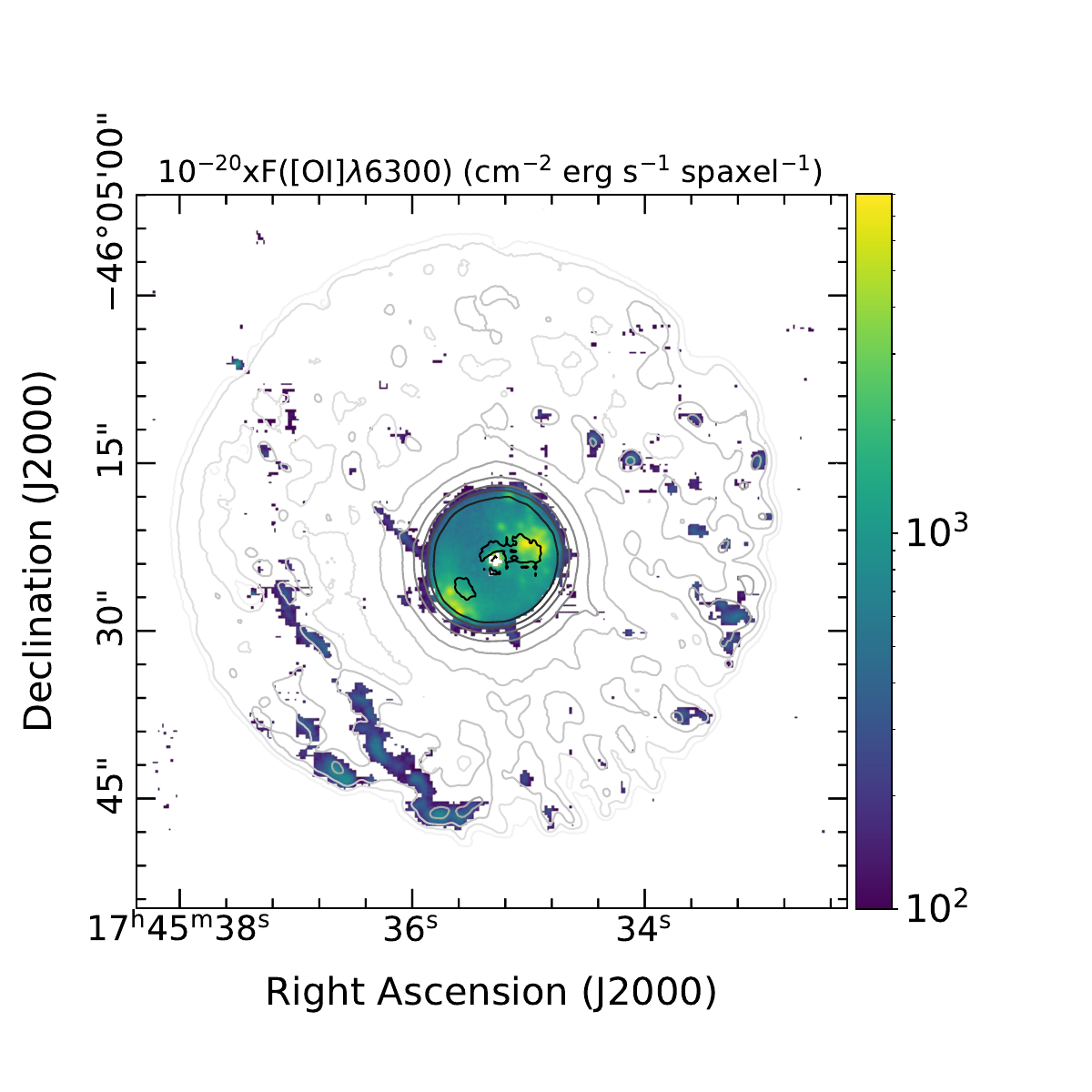}
\includegraphics[trim={0cm 1cm 2cm 1cm}, clip]{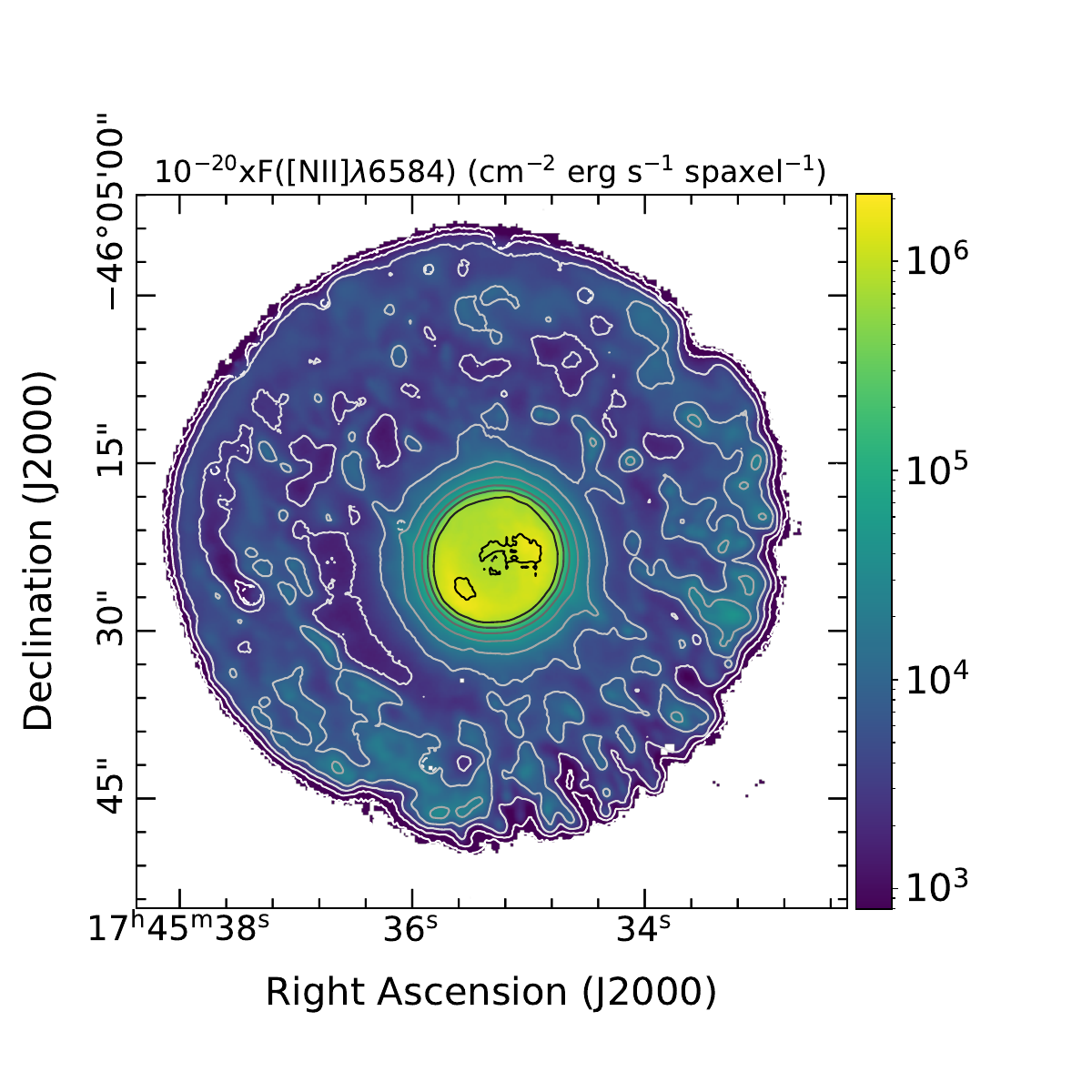}
\vspace{-0.2truecm}
}

\caption{The morphology of Tc~1 in a some CELs. \\
Upper left: Log flux (erg cm$^{-2}$ s$^{-1}$) of [\ion{O}{III}] 5007\,\AA, extracted for 
the 40\,s cube, displaying the bright, rather featureless core and low surface brightness 
diffuse halo; \\
Upper right: Log flux (erg cm$^{-2}$ s$^{-1}$) of [\ion{O}{II}] 7320+7330\,\AA, from the
40\,s cube, exhibiting the contrasting low ionization compact features in the
core and the halo; \\
Lower left: Linear flux (erg cm$^{-2}$ s$^{-1}$) of [\ion{O}{I}] 6300\,\AA\
line emission from the 595\,s cube emphasising the compact knots over the core and 
filaments in the halo; \\
Lower right: Linear flux (erg cm$^{-2}$ s$^{-1}$) of [\ion{N}{II}] 6583\,\AA, again from
the 40\,s cube, showing the contrasting appearance of the core and halo with respect to 
lines of O. \\
The H$\beta$ line contour map is superposed on the collisionally excited line 
images and all images are displayed to a cutoff of 3$\sigma$ line flux per spaxel.
}
\label{fig:EmLines2}
\end{figure*}

Figure \ref{fig:EmLines1} shows the appearance of Tc~1 in \ion{H}{} recombination  lines
and Figure \ref{fig:EmLines2} for CELs of varying ionization. 
H$\beta$ illustrates the general features of the nebula morphology (Figure \ref{fig:EmLines1}, 
left). The outer cut-off in H$\beta$ flux is at 3$\sigma$ detection per spaxel 
(0.04$''^{2}$).
A composite image of Tc~1, composed of the bright core in H$\alpha$ (linear stretch) 
and the fainter surroundings and halo in [\ion{N}{II}] 6583\AA\ (log stretch), both 
extracted from the 40s cube, is shown in Figure \ref{fig:EmLines1}, right. The main 
features of the nebula, which are subsequently 
discussed (see Sect. \ref{Morphology}), are indicated. The two regions 
of low ionization systems (LIS), analysed by \citet{Bouvisetal2025}, are also labelled. 
The half mean surface brightness extent of the core (radius = 5.4$''$) is 
indicated in orange and the annulus of lower extinction values (`c annulus', see Sects. 
\ref{SubSec: Av} and \ref{c_struct}) between radii 5.9$''$ and 6.9$''$ is indicated in 
pink. 
 
The highest ionization energy emission detected, that of [\ion{O}{III}] 5007\,\AA, is shown 
in Figure \ref{fig:EmLines2}, upper left; intermediate O ionization [\ion{O}{II}] (sum of 
7320+7330\,\AA\ line images) is shown upper right and the neutral oxygen 
emission [\ion{O}{I}] (lower left). The contrasting morphology in the strongest low 
ionization line [\ion{N}{II}] 6583\,\AA\ is shown lower right. The striking difference is 
the strength and structure in the outer extent of the nebula, beyond the bright core, 
subsequently referred to as the halo. The presence of a halo can be deduced from the image 
in \citet{Corradietal2003}, their Fig. 16,  but its detailed structure is now revealed for 
the first time. While there is diffuse emission in [\ion{O}{III}] outside the bright core, 
it is very faint and lacks the structure seen in lower ionization ([\ion{O}{II}] and 
[\ion{N}{II}]). The [\ion{O}{III}] emission surface brightness in the halo is on average 19 
times fainter than for [\ion{N}{II}] for the same (14$''$) radial offset. In [\ion{O}{I}], 
the morphology emphasises compact knots both within the core (including the LIS) and in 
the halo.

\subsection{Effect of field stars}
\label{SubSec: FieldStarRemoval}

\begin{figure}
\centering
\resizebox{\hsize}{!}{
\includegraphics[trim={0cm 1cm 2cm 1cm}, clip]{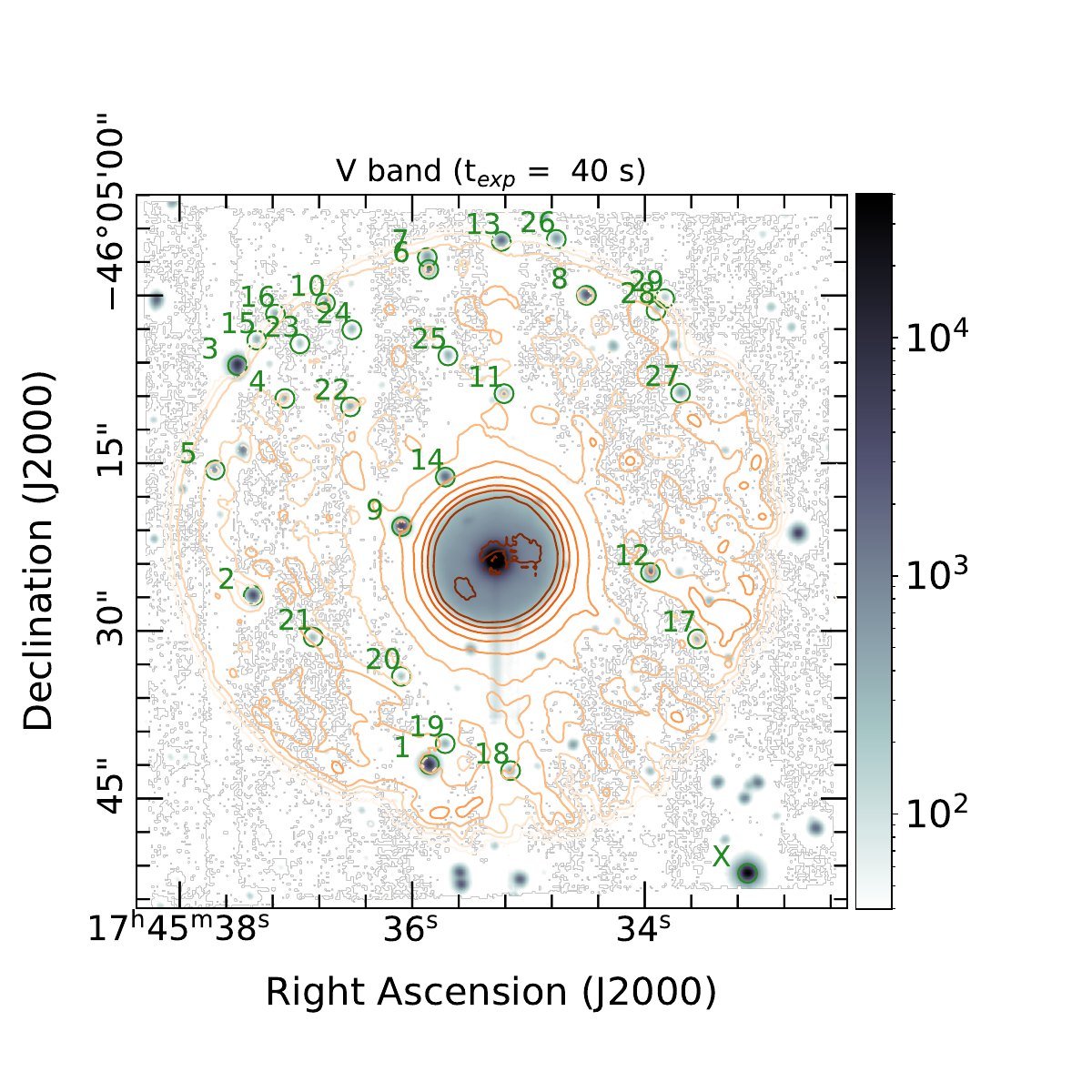}
}
\caption{The $v'$ band (5400--5600\,\AA) image in a linear flux scaling (from the 40\,s 
exposure cube) showing the stars in and around Tc~1. The stars whose spectra were matched to 
correct for the presence of underlying H$\beta$ and H$\alpha$ over the halo are numbered 
(see text for details). Star X is the next brightest star after the CSt and occurs outside 
the halo; it was used as a PSF star (see Sect. \ref{App:PSF restoration}). 
The contour map is derived from the H$\beta$ image (Fig. \ref{fig:EmLines1} upper left). 
}
\label{fig:v_40s}
\end{figure}

Before accurate emission line maps can be analysed, the problems posed by the presence 
of the many field stars requires consideration. Over the halo where the emission surface 
brightness is much lower than in the core, an emission line fitted on a late type stellar 
continuum spectrum can lead to mis-estimation of line fluxes, particularly for Balmer lines,
since the Gaussian line task only fits a low order spline to the continuum adjacent to 
emission lines. The narrow $v'$ (5400--5600\,\AA) continuum image shown in Figure \ref{fig:v_40s} 
demonstrates the scope of the problem: a poorly fitted emission line in the halo can be 
mistaken for an emission line knot, of which a few are coincident or close to field stars; 
an erroneous extinction estimate from the H$\alpha$/H$\beta$ ratio at the position of a field 
star can lead to deduction of a spurious higher or lower extinction knot relative to the 
surroundings. This problem was not manifest on the \ion{He} or CEL images, since in general
strong stellar absorption lines do not coincide with these lines.

Using a spectral library of observed Galactic stars, the spectra of the field stars over
the nebula face were fitted and removed before the emission lines, affected by the presence
of field star absorption lines, were analysed. Appendix \ref{App:Fitting of field stars} 
details this correction procedure.

\subsection{Combination of line images}
\label{SubSec: Combinationimages}
The signal-to-noise (S/N) of the emission line images is high over the bright core for the 40\,s cube 
but low for all but the brightest lines over the halo; whilst the strongest emission lines are
saturated over the bright core but well-exposed over the halo for the 101\,s and 595\,s cubes. 
With the aim of producing the deepest possible map of the extinction across the nebula, 
combination of the Balmer line images from each exposure level was made. The H emission lines images are well-exposed over the core in the 40\,s images; the 101\,s 
images, particularly for H$\alpha$, are saturated
in regions over the core but well-exposed in the immediate surrounding; the images from 
the 595\,s cube are saturated to radii beyond the core but well-exposed over the halo.
The c(H$\beta$), log extinction at H$\beta$, image was computed separately for each 
exposure level using masked images and the resulting c(H$\beta$) images combined. 
(detailed in Sect. \ref{SubSec: Av}).

\begin{figure*}
\centering
\resizebox{\hsize}{!}{
\includegraphics[trim={0cm 1cm 2cm 1cm}, clip]{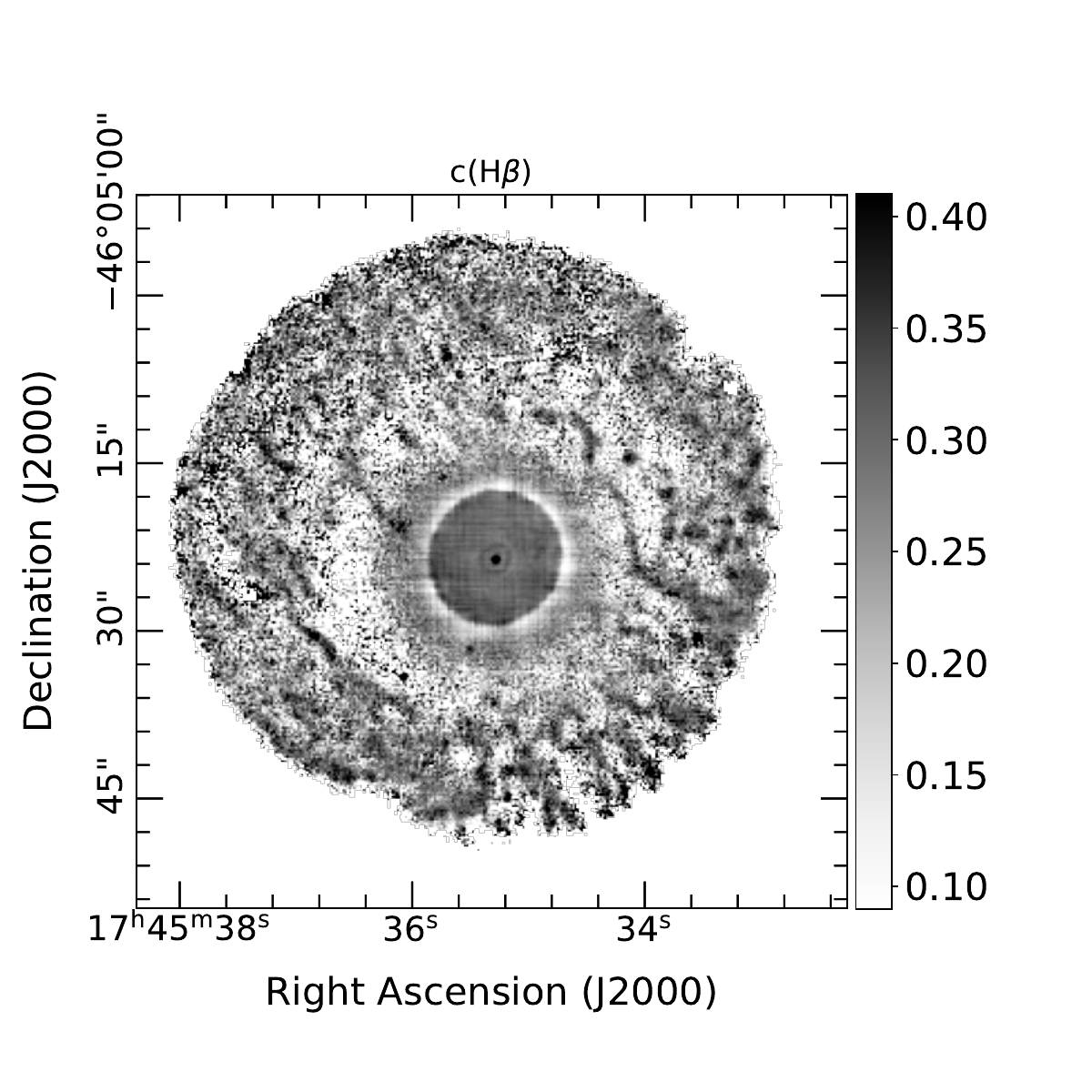}
}
\caption{Image of the Tc1 extinction, log c(H$\beta$), from H$\alpha$/H$\beta$ 
compared to the Case B value combining the ratios from the 40\,s, 101\,s and 595\,s 
cubes. See text for detail.  
}
\label{fig:c_map}
\end{figure*}

\section{Imaging of physical conditions -- extinction, density and temperature}

\subsection{Extinction mapping}
\label{SubSec: Av}
From images of the ratios of Balmer and Paschen line fluxes 
compared to the \citet{MenzelBaker} Case B values, a map of the
extinction across the nebula can be constructed. The value of the
electron temperature, $T_{\rm e}$, and electron density, $N_{\rm e}$, 
must be provided, such as from the CEL indicators;
for example, [\ion{S}{III}] 6312/9069\,\AA\ and [\ion{Cl}{III}] 5517/5538\,\AA\
respectively. The combined c(H$\beta$) map used the three exposure levels 
to derive optimally exposed H$\alpha$ and H$\beta$ images which were converted 
to c(H$\beta$) by comparison to the Case B H$\alpha$/H$\beta$ ratio: 40\,s cube 
for core region, using radius $r \leq 5.8''$; 101\,s for the 
annulus between core and halo, $6.0 \geq r \leq 10.2''$; 595\,s cube for the halo 
$r \geq 10.4''$. 

Based on first guess values of $T_{\rm e}$ and $N_{\rm e}$ of 
10\,000\,K and 2000 cm$^{-3}$ 
respectively for the core and 9\,000\,K for the halo, the Case B value of the 
\ion{H}{} line ratio was employed to calculate an initial c(H$\beta$) image.
Then from the variation of $T_{\rm e}$ and $N_{\rm e}$ over the reddening corrected 
emission line maps (Sect. \ref{SubsubSec: CELTeNe}), simple mean values in two zones 
(inner $r \leq 6.0''$) based on [\ion{S}{III}] for $T_{\rm e}$ and [\ion{Cl}{III}] 
for $N_{\rm e}$ and outer ($r > 6.0''$) based on [\ion{N}{II}] $T_{\rm e}$ and [\ion{S}{II}]
$N_{\rm e}$ were adopted for calculating the final Case B H$\alpha$/H$\beta$ ratio.
Fig. \ref{fig:c_map} shows the resulting c(H$\beta$) image. Only small differences
in appearance of this image are found if single constant values of $T_{\rm e}$ and 
$N_{\rm e}$ are used (10\,000\,K and 2000 cm$^{-2}$ respectively) for the Case B 
computation; c(H$\beta$) means differ by $<$ 0.01 for a spatial mean value of 0.233. 
An extinction map was also constructed 
including Paschen 9, 10, 11 and 12 lines in addition to the H$\alpha$ and H$\beta$ 
lines, and the results are again very similar, though systematically lower, within 
$\pm$ 0.02 for the halo and $\pm$ 0.04 for the halo. Given the lower S/N of the
Paschen lines in comparison to the Balmer lines, this small difference is 
considered marginally significant.  

All the emission line flux (and error) images were then dereddened using 
the c(H$\beta$) image (Fig. \ref{fig:c_map}) and its corresponding error (the 
latter produced by propagating the flux ratio errors, but not $T_{\rm e}$ and 
$N_{\rm e}$) errors.

\subsection{Mapping of $T_{\rm e}$ and $N_{\rm e}$} 
\label{SubSec: NeTe}

\subsubsection{CEL diagnostics}
\label{SubsubSec: CELTeNe}

The available CEL diagnostic ratios in the MUSE spectrum of Tc~1 are [\ion{N}{II}] and 
[\ion{S}{III}] for $T_{\rm e}$, [\ion{S}{II}] and [\ion{Cl}{III}] for $N_{\rm e}$. Whilst 
[\ion{Ar}{III}] is also detected, the 5192\,\AA\ line is weak and has very low 
S/N, resulting in $T_{\rm e}$ values at the original spaxel resolution (0.2$''$) with 
large errors or values of the 5192/7136\,\AA\ ratio outside the theoretical range.
All H, He and metal CEL ratio calculations were made with PyNeb 
\citep{Luridiana2015} and relevant sources of the atomic data for
the CEL computations are listed in Appendix B of \cite{Walshetal2024}. The c(H$\beta$) image 
(Fig. \ref{fig:c_map}) was used to deredden all the emission line maps required for
$T_{\rm e}$ and $N_{\rm e}$ measurement and the final maps were made by combining the
40\,s images for the bright core, with the 595\,s images for the surroundings and the halo.

\begin{figure*}[t]
\centering
\resizebox{\hsize}{!}{
\includegraphics[trim={0cm 1cm 1cm 1cm}, clip]{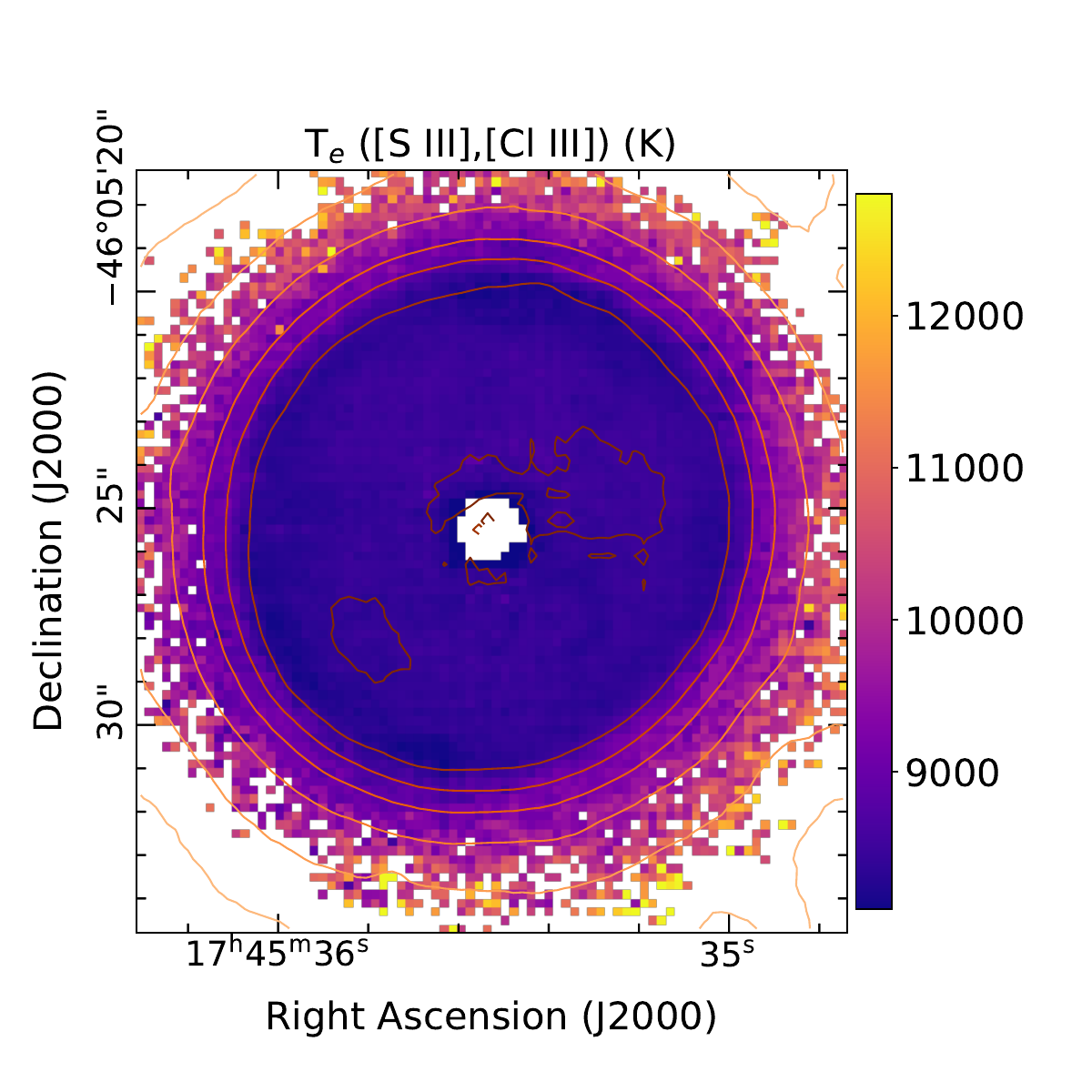}
\hrulefill
\includegraphics[trim={0cm 1cm 1cm 1cm}, clip]{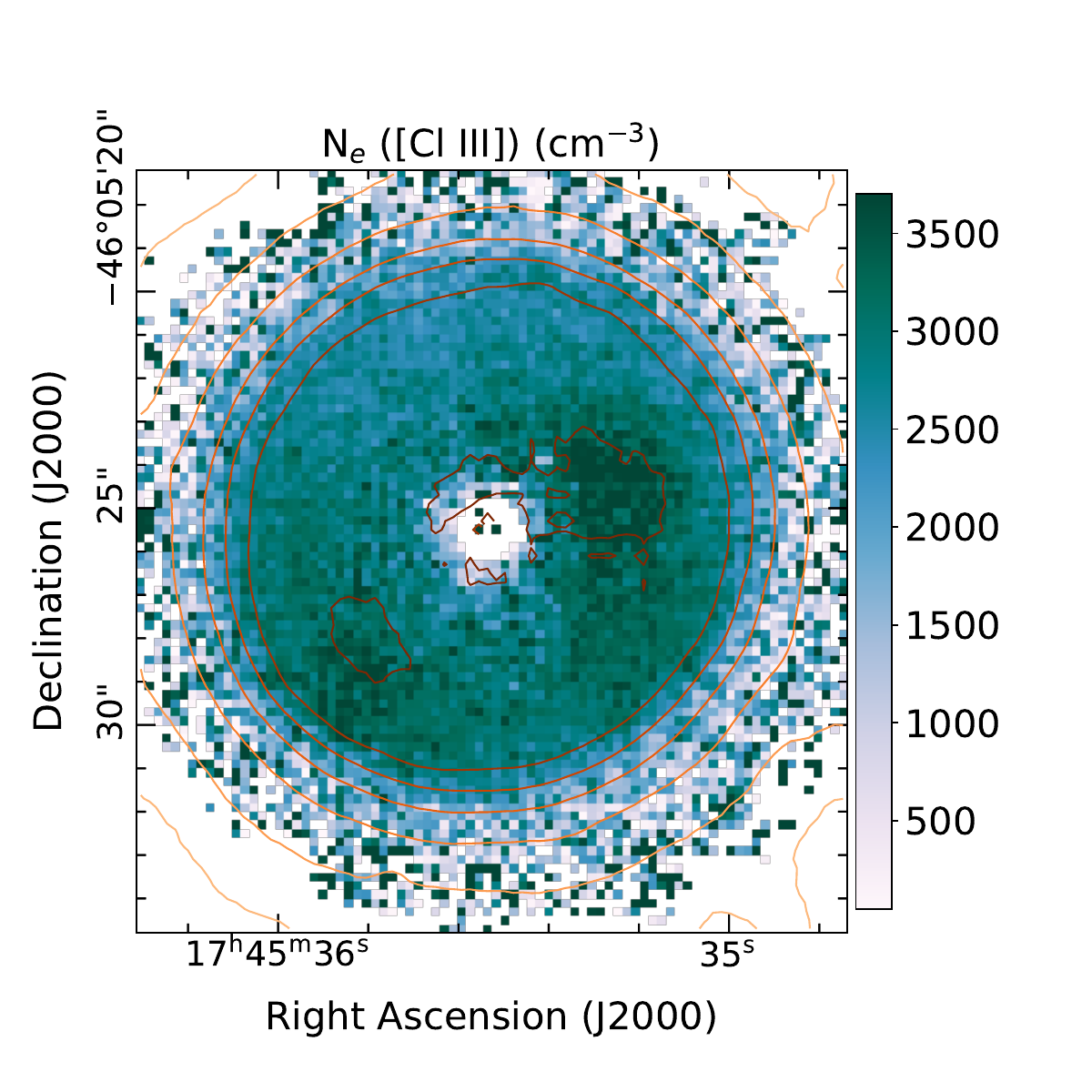}
}
\caption{$T_{\rm e}$, $N_{\rm e}$ images for the higher ionization core area
only, from [\ion{S}{III}] 6312/9069\,\AA\ (left) and [\ion{Cl}{III}] 5517/5537\,\AA\
(right). The colour bars indicate the $T_{\rm e}$ and $N_{\rm e}$ ranges  
and the H$\beta$ line contour map is superposed and the cut-off in S/N is 3
per 0.2$''$ spaxel. 
}
\label{fig:TeNe3maps}
\end{figure*}

\begin{figure*}[t]
\centering
\resizebox{\hsize}{!}{
\includegraphics[trim={0cm 1cm 1cm 1cm}, clip]{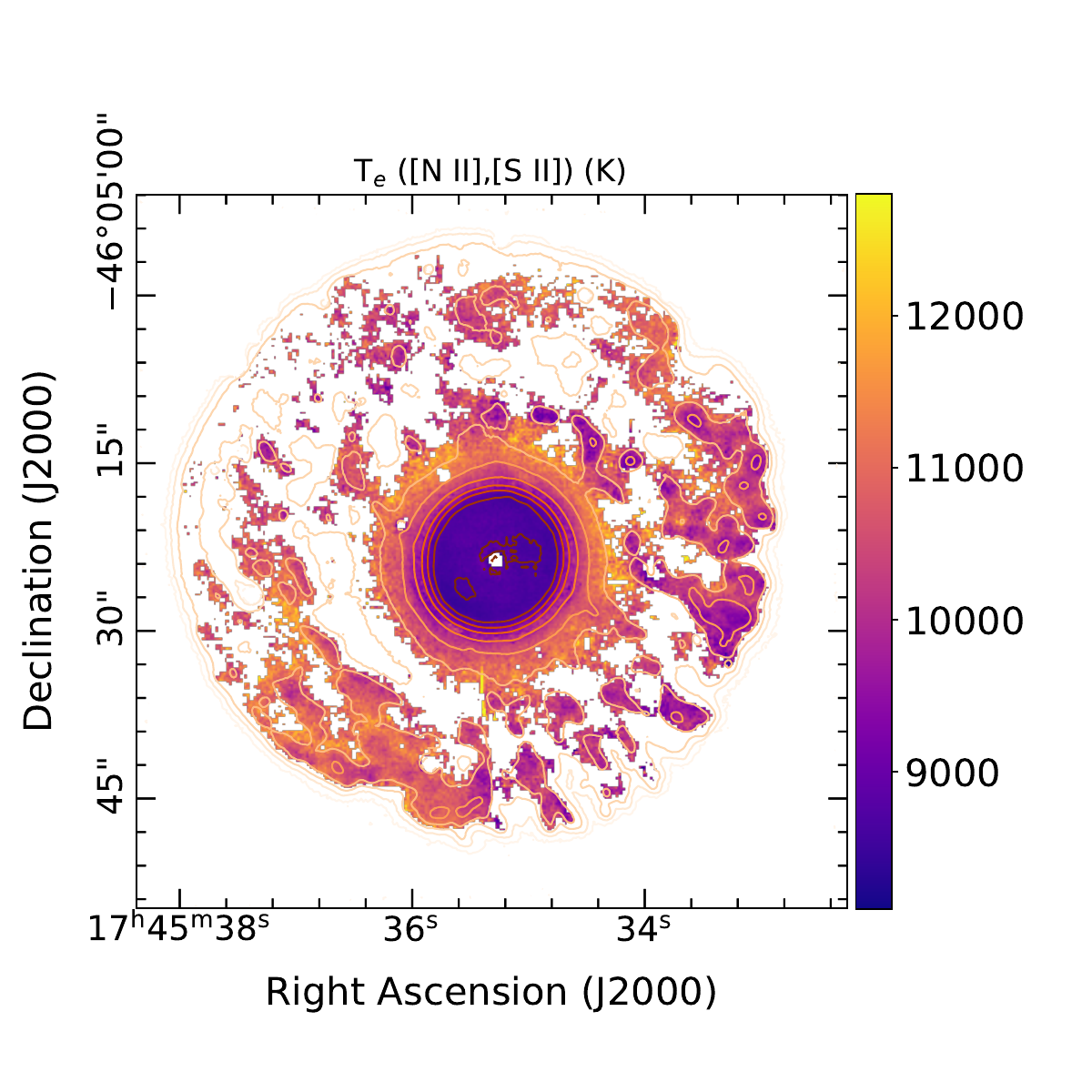}
\hrulefill
\includegraphics[trim={0cm 1cm 1cm 1cm}, clip]{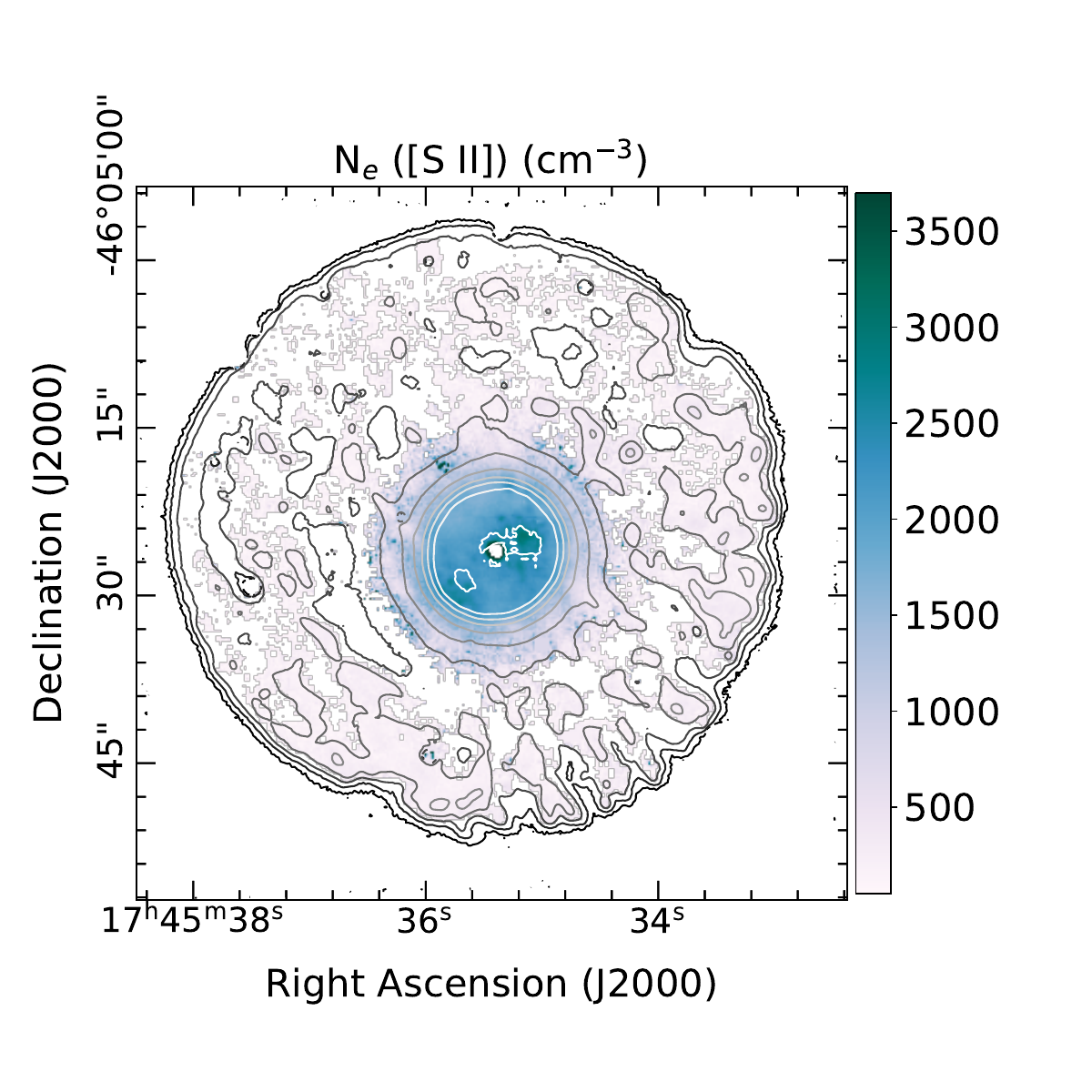}}
\caption{$T_{\rm e}$, $N_{\rm e}$ images of the entire nebula for the 
lower ionization
medium from [\ion{N}{II}]5755/6583\,\AA\ (left) and [\ion{S}{II}]6716/6731\,\AA\ 
(right). The colour bars indicate the $T_{\rm e}$ and $N_{\rm e}$ ranges 
and the H$\beta$ line contour map is superposed; the cut-off in S/N is 3 per
0.2$''$ spaxel.  
}
\label{fig:TeNe2maps}
\end{figure*}

Figure \ref{fig:TeNe3maps} shows the $T_{\rm e}$ and $N_{\rm e}$ images from 
[\ion{S}{III}] 6312/9069\,\AA\ (left) and [\ion{Cl}{III}] 5517/5537\,\AA\ (right) for the 
higher ionization gas. Figure \ref{fig:TeNe2maps} shows $T_{\rm e}$ and $N_{\rm e}$ 
images for the lower ionization gas from [\ion{N}{II}]5755/6583\,\AA\ (left) and
[\ion{S}{II}]6716/6731\,\AA\ (right). All the maps were computed using the PyNeb {
\it getCrossTemDen} code using the simultaneously computed $N_{\rm e}$ maps from
[\ion{Cl}{III}] and [\ion{S}{II}] ratio images. Spaxels in these images without an 
$N_{\rm e}$ and/or a $T_{\rm e}$ value correspond to S/N values for the weak auroral 
lines in the diagnostic ratio below a cut-off value of 3. In the halo particularly, 
there are many spaxels with [\ion{Cl}{III}] and [\ion{S}{III}] lines below this S/N cut, 
but fewer for the lower ionization [\ion{S}{II}] and [\ion{N}{II}] line ratios.

\begin{figure}[t]
\centering
\resizebox{\hsize}{!}{
\includegraphics[trim={0.5cm 0.5cm 1.6cm .7cm}, clip=]{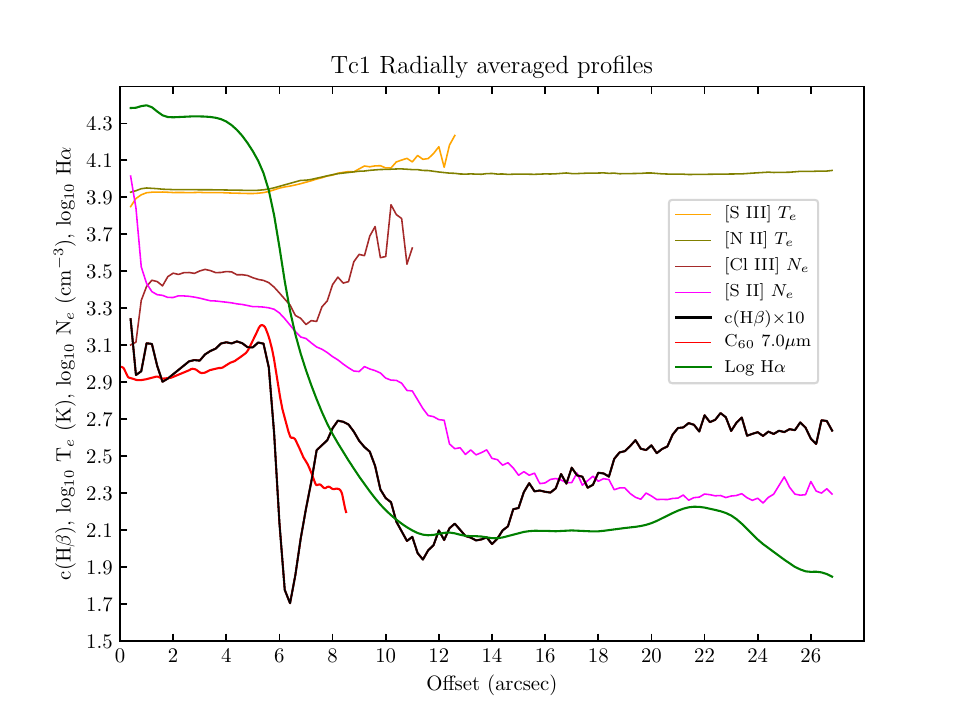}
}
\caption{The radial variation of the CEL diagnostics with offset from the
position of the CSt for $T_{\rm e}$  and $N_{\rm e}$, are plotted as log$_{10}$. For
comparison the radial variation for c(H$\beta$), scaled by a factor 10
to be similar in values to log$_{10}$ $T_{\rm e}$ and $N_{\rm e}$, and the 
scaled log$_{10}$ H$\alpha$ surface brightness, are also 
shown. The low c annulus occurs between radii 6--7$''$.
Also plotted (in red) is the log$_{10}$ surface brightness profile of the 
C$_{60}$ 7.0 $\mu$m band (erg cm$^{-2}$ s$^{-1}$ sr$^{-1}$ $\times$ an arbitrary 
constant) for comparison (c.f., Giese et al. in prep.).
The C$_{60}$ profile was however determined over a limited angular extent of 
40$^{\circ}$. 
}
\label{fig:RadialProfiles}
\end{figure}

Figure \ref{fig:RadialProfiles} summarizes the CEL diagnostics with a plot
of the radially averaged profile with offset from the position of the CSt for 
$T_{\rm e}$ from [\ion{S}{III}] 6312/9069\,\AA\ and [\ion{N}{II}]5755/6583\,\AA,
and $N_{\rm e}$ from [\ion{Cl}{III}] 5517/5537\,\AA\ and [\ion{S}{II}]6716/6731\,\AA. 
These are compared to the radial variation of extinction (c(H$\beta$) from Fig. 
\ref{fig:c_map}) and the H$\alpha$ surface brightness. For the diagnostics 
from [\ion{Cl}{III}] and [\ion{S}{III}], the line fluxes, and hence the S/N, 
were too low beyond the last plotted radial points for a meaningful comparison. 
The increase in
$T_{\rm e}$ begins where the density drops (radius $\sim$5.7$''$), although 
$N_{\rm e}$([\ion{Cl}{III}]) appears to increase again beyond radii 7.5$''$, 
where however there were few spaxels with sufficient S/N to calculate $N_{\rm e}$.
This turning point in radial $N_{\rm e}$ and $T_{\rm e}$ values occurs as
the extinction first decreases in an annulus around the core (see Figure 
\ref{fig:c_map} and Sect. \ref{c_struct} for a discussion). The extinction is
also lower in the radial range 10--15$''$ but here the H lines were faint
and there are many spaxels (particularly to the E and SE) without sufficient 
S/N to calculate c(H$\beta$). 

\subsubsection{ORL diagnostics}
\label{SubsubSec: ORLTeNe}

Diagnostic ratios of line and continuum from the recombination lines of H$^{+}$ 
and He$^{+}$ within the MUSE wavelength range can be used for electron temperature
and electron density determination. These include the ratio of the high Paschen 
lines as an electron density estimator \citep[c.f.,][]{Zhang2004}, the \ion{H}{I} 
Paschen continuum jump at 8204\,\AA\ and the ratios of He$^{+}$ singlet lines, 
which are primarily sensitive to $T_{\rm e}$. 

\paragraph{\large \ion{H}{I} Paschen Jump $T_{\rm e}$:} 
\label{SubsubSec: TePasJum}

~~~The Balmer and Paschen series continuum jumps for bound-free ($b-f$) transitions of 
\ion{H}{I} are sensitive to electron temperature \citep{Peimbert1971} and 
\citet{FangLiu2011} determined $T_{\rm e}$ from the Paschen Jump, normalizing the 
jump (defined as $F$(8194\,\AA) - $F$(8269\,\AA)) by the Paschen 11 (P11) flux 
(8863\,\AA). \cite{Walshetal2018} (in Appendix) presented a measurement method 
suitable for multiple spectra of an IFU by taking the mean continuum in several 
line-free windows on both sides of the Paschen Jump and a custom conversion of this 
difference with respect to the flux of the P11 line as a function of $T_{\rm e}$,  
$N_{\rm e}$ and He$^{+}$/H$^{+}$.

\begin{figure}[t]
\resizebox{\hsize}{!}{
\includegraphics[trim={0cm 1cm 1cm 1.5cm},clip]{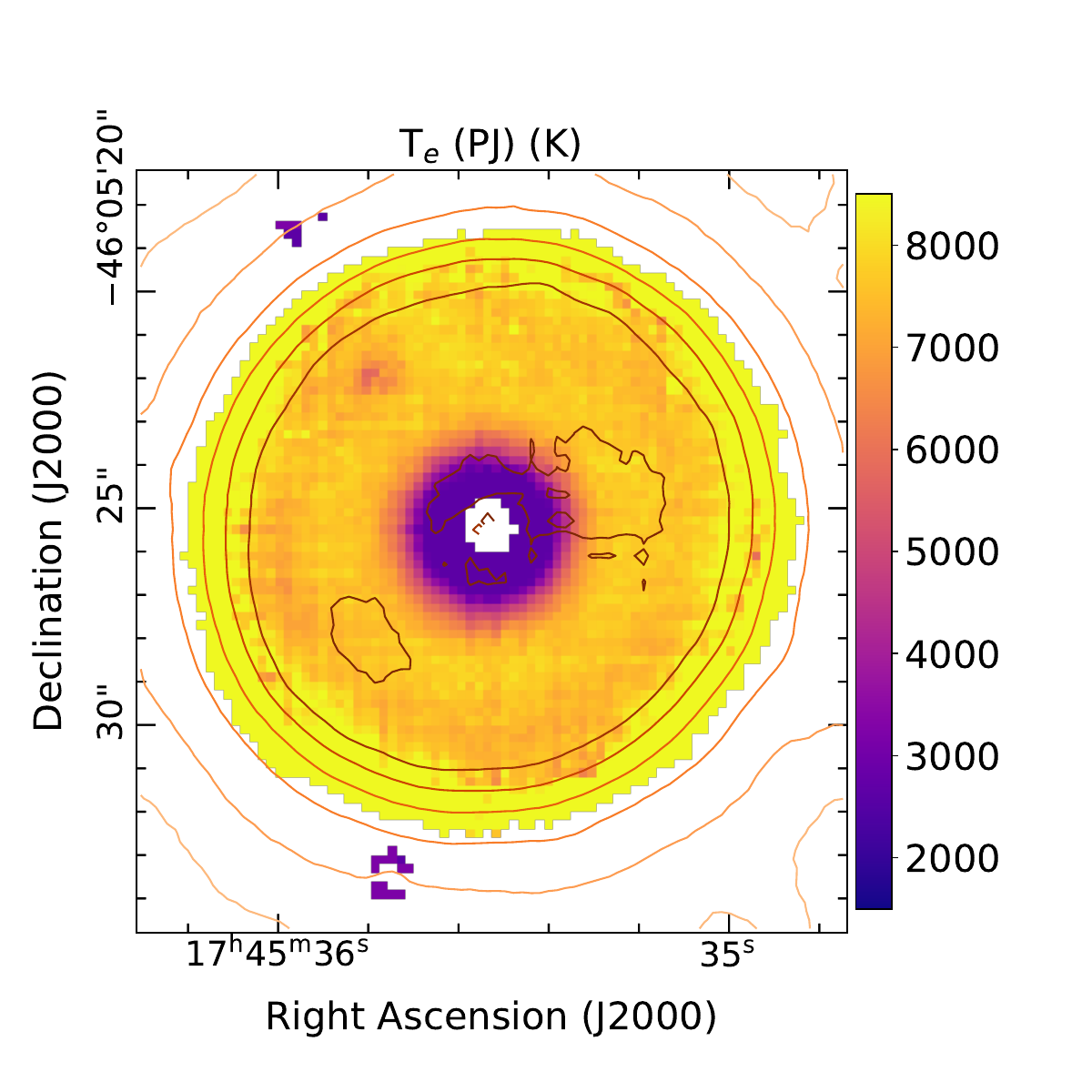}
}
\caption{Map of $T_{\rm e}$ for the core region only, derived from the 
Paschen continuum jump at 8250\,\AA\ ratioed by the dereddened 
\ion{H}{I} Paschen 11 (8862.8\,\AA) emission line strength. Initial 
estimates of $T_{\rm e}$ and $N_{\rm e}$ from the CEL diagnostic ratios
(Sect. \ref{SubsubSec: CELTeNe}) for calculation of the dependence of PJ/P11 on 
$T_{\rm e}$ were employed. The contours are from the H$\beta$ image
(Figure \ref{fig:EmLines1}) and the cut-off is at S/N 3 per 0.2$''$ 
spaxel.
}
\label{fig:PJTemap}
\end{figure}

Figure \ref{fig:PJTemap} shows the resulting map of the Paschen Jump (PJ) 
$T_{\rm e}$. Since He$^{+}$ also contributes to the nebular continuum, the He 
ionic fraction was calculated from the maps of \ion{He}{I} 6678 and 7065\,\AA\ 
line flux, but not 7281\,\AA\ on account of the anomalous flux of this line 
(see following subsection and below for details). Initial 
estimates of $T_{\rm e}$ and $N_{\rm e}$ for the dependence of PJ/P11 were taken 
as single values of $T_{\rm e}$ for the core (8900\,K) and halo (10750\,K), and 
$N_{\rm e}$ of 2300 and 500 cm$^{-3}$ respectively,  based on the 
CEL $T_{\rm e}$ and $N_{\rm e}$ diagnostics (Sect. \ref{SubsubSec: CELTeNe}). 
There are two distinct zones: the bright nebular core with 
mean value is 7550\,K with a root mean square (RMS) of 280\,K (3 $\times$ 3 $\sigma$ 
clipped mean); the ring of higher $T_{\rm e}$ (broadly coincident with the
annulus of low extinction) with a mean value of 11500\,K, RMS 1900\,K.
(The lower $T_{\rm e}$ values close to the CSt are caused by the
presence of continuum from the wings of the PSF increasing the measured 
PJ blue-red continuum ratio.)
The mean signal-to-error over the map, based on 100 Monte Carlo 
trials using the propagated errors on the measured PJ, is 12. Three small regions 
over the ring where there were background stars present had to be interpolated.
Only over the bright core is the PJ $T_{\rm e}$ lower than for the CEL 
$T_{\rm e}$ value; in the periphery the values are very similar.

\paragraph{\large \ion{He}{I} $T_{\rm e}$:} 
\label{SubsubSec: He1Te}

~~~\cite{Zhang2005} showed that the ratio of the two \ion{He}{I} singlet lines
7281.4 and 6678.2\,\AA\ provides a suitable diagnostic of the He$^{+}$ 
electron temperature and thus a useful ORL temperature. The determination 
of $T_{\rm e}$ from 7281/6678\,\AA\ was based on emissivities from Porter 
\citep{Porter2012, Porter2013} tabulated for $ 5000 < T_{\rm e} < 25\,000$K.
The observed dereddened \ion{He}{I} 7281/6678\,\AA\ ratio was converted to 
$T_{\rm e}$ and $N_{\rm e}$ minimizing the residual with the theoretical 
ratio. First guess values of $T_{\rm e}$ and $N_{\rm e}$ were taken as 
single values for the core and halo based on the CEL $T_{\rm e}$ and 
$N_{\rm e}$, as in the calculation of PJ $T_{\rm e}$ (see previous 
subsection). The \ion{He}{I} ratio is much more sensitive to 
$T_{\rm e}$ than $N_{\rm e}$ and the initial estimate of $T_{\rm e}$ was 
adopted to compute the emissivities; $N_{\rm e}$ was held fixed within 
narrow constraints during the minimization. 

\begin{figure}[t]
\centering
\resizebox{\hsize}{!}{
\includegraphics[trim={0.5cm 1.6cm 1.6cm 1.5cm},clip]{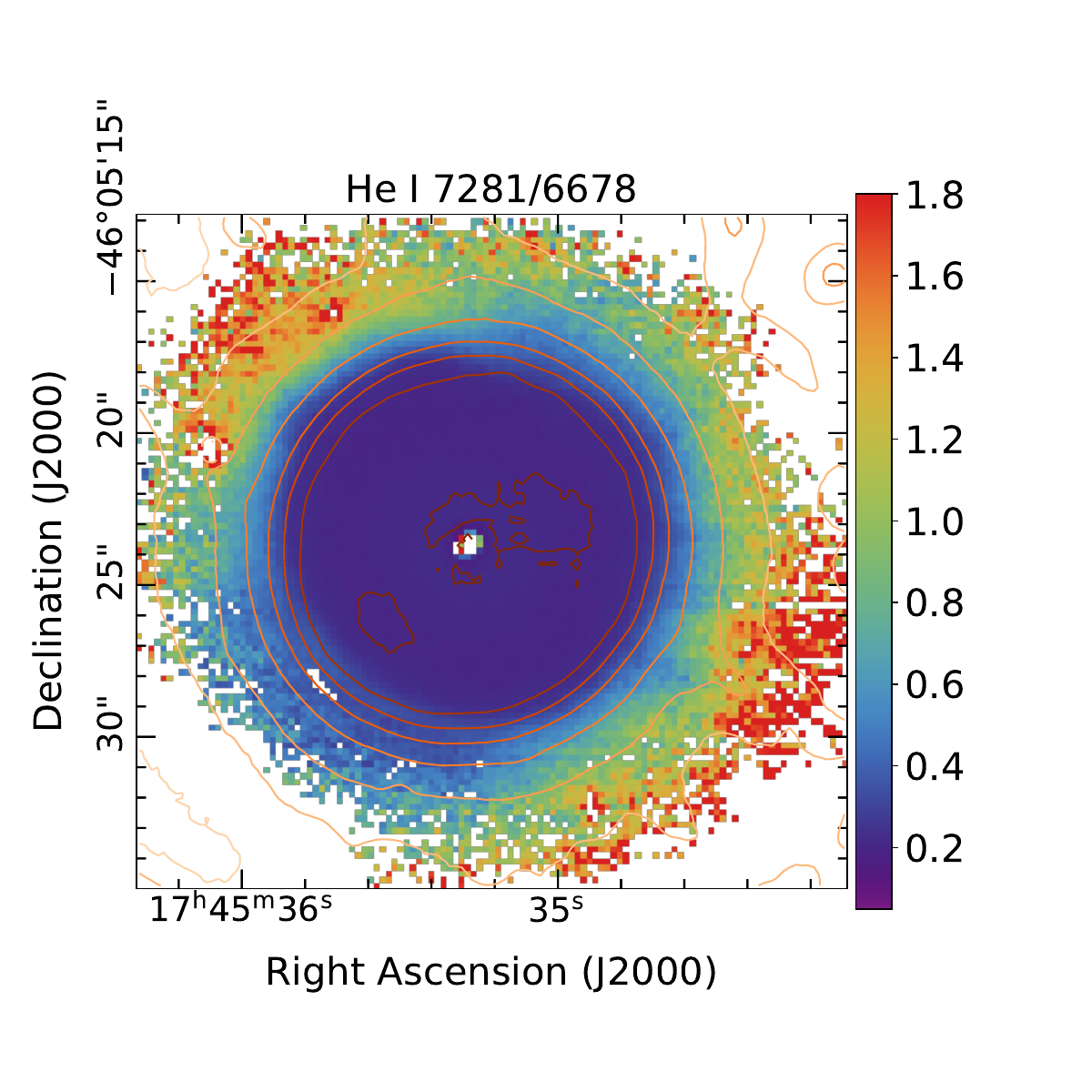}
}
\caption{Image of the Tc~1 core region for the ratio of the fluxes of the 
\ion{He}{I} 7281 and 6678\,\AA\ lines. Spaxel values above 0.31 are outside 
the feasible range of the \ion{He}{I}
ratio, indicating a contaminating line in the region outside of the bright core.
The contours are for the H$\beta$ flux.
}  
\label{fig:He17881_6678}
\end{figure}

Figure \ref{fig:He17881_6678} shows however that in an annulus beyond radius
5.5$''$ the observed \ion{He}{I} 7281/6678\,\AA\ ratio exceeds the allowable
range of this ratio of 0.13--0.31 across the range $T_{\rm e}$ 5000--25000\,K and 
log $N_{\rm e}$ 1--7. Over the core, within 5.5$''$, the computed mean value of 
\ion{He}{I} $T_{\rm e}$ is 10650 $\pm$ 450\,K (3 $\times$ 3$\sigma$ clipped mean), 
but no value can be computed for the region outside this radius. The most 
probable explanation is that another emission line, close to 7281\,\AA, but 
within the spectral resolution of MUSE at this wavelength ($\sim$2.8\,\AA), 
contaminates the \ion{He}{I} line flux outside the extent of the bright core. 

In order to investigate the possibility of a contaminating line close to \ion{He}{I} 
7281\,\AA, VLT X-Shooter \citep{Vernetetal2011} spectra (Proposal ID: 097.D-1033(A), 
PI J. Cami) were examined. There are two slits with differing PA's
including the CSt and extending beyond the outer edge of the core, so into
the region where \ion{He}{I} 7281\,\AA\ is enhanced. The much higher spectral
resolution of X-shooter ($\sim$ 10\,000) than MUSE could allow a 
contaminating line within $\sim$0.5\,\AA\ of 7281\,\AA\ to be revealed. These 
spectra were reduced and the spectrum around 7280\,\AA\ examined. No evidence
of a resolved nebula line close to 7281\AA\ was found, nor any suggestion
of a broadening of 7281\,\AA\ compared to the nearby \ion{He}{I} 6678\AA\ 
line. Although the X-Shooter slit only covered a region slightly beyond the core, 
an increase in 7281/6678 \,\AA\ ratio is visible towards the edge of the slit 
(although the S/N is low). The conclusions from the MUSE spectra that there
is indeed an enhancement at 7281\AA\ outside the core, and above the expected
\ion{He}{I} theoretical ratio to 6678\AA, is thus confirmed.

In the Atomic Line List \citep{vanHoof2018}, the closest feasible nebula line is 
[\ion{Fe}{II}] 7281.66\AA\ displaced only 0.35\AA\ from the \ion{He}{I} 
7281.31\AA\ line. In order to check 
if this line could represent a significant contamination to the \ion{He}{I} flux,
the deep spectrum for IC~418 \citep{Sharpeeetal2003} (a nebula with similarities 
to Tc~1 in terms of ionization and fullerene detection) was consulted for detected 
[\ion{Fe}{II}] lines. The strongest [\ion{Fe}{II}] line in IC~418 in a relatively 
uncrowded region of the spectrum is 5495.8\AA\ with a dereddened line strength 
0.015 (H$\beta$=100), although listed by \cite{Sharpeeetal2003} as blended with 
\ion{N}{II}\footnote{The [\ion{Fe}{II}] 5035.1\AA\ line is however
stronger than 5495.8\AA\ but has several nearby lines at MUSE resolution,
making it more difficult to fit.}. 
Extracting a spectrum of Tc~1 in the region of the possible contaminated 
\ion{He}{I} line, reveals a detectable line at 5496\AA\ with a dereddened strength 
of 0.015, remarkably similar to IC~418. However, given that the ratio of 
[\ion{Fe}{II}] 7281.7/5495.8 at the $T_{\rm e}$ and $N_{\rm e}$ measured for this 
region (see Tab. \ref{tab:regiondiags}) would be 0.003, 
any contamination of 7281\AA\ by [\ion{Fe}{II}] is negligible.
Since a viable contaminating line to \ion{He}{I} 7281\,\AA\ could not 
be identified, we infer that the origin of the 7281\,\AA\ enhancement in the outer
annulus is unresolved based on the analysed observations.

\section{Discussion}

\subsection{Emission line morphology}
\label{Morphology}

Already known from previous imaging (e.g. \cite{Corradietal2003}) was the strong contrast 
between the almost circular bright core, diameter $\sim$12$''$, 0.22 pc and the halo,
size 55$''$. The mean contrast in H$\alpha$ core/halo is 490 
(530 for the dereddened image). While the CSt is situated at the centre of the
core nebula, the circular halo is displaced relative to the CSt in 
PA $\sim$15$^{\circ}$ by 2$''$ (this is particularly noticeable in the
extinction map, Fig. \ref{fig:c_map}). The Gaia stellar proper motion direction is towards
PA 194.09$^{\circ}$; from the offset of the halo to the N, it is apparent that
the CSt and core nebula together have a motion relative to the halo. If the halo is an older
structure, such as the remnant of the AGB envelope, then a modest plane of the sky
differential velocity between the CSt and the halo of $<$ 1km s$^{-1}$ in 5$\times$10$^{4}$ 
yr (indicative difference in age between the AGB halo and current CSt) could account for 
the observed CSt-halo offset. 
Given that the proper motion is in almost the opposite direction to the halo extension, 
there may be an effect of compression of the halo in the direction of the pm, and/or
extension in the opposite direction depending on the ISM pressure.

\begin{figure}[t]
\centering
\resizebox{\hsize}{!}{
\includegraphics[trim={0.5cm, 1cm, 1.7cm, 1.7cm}, clip=]{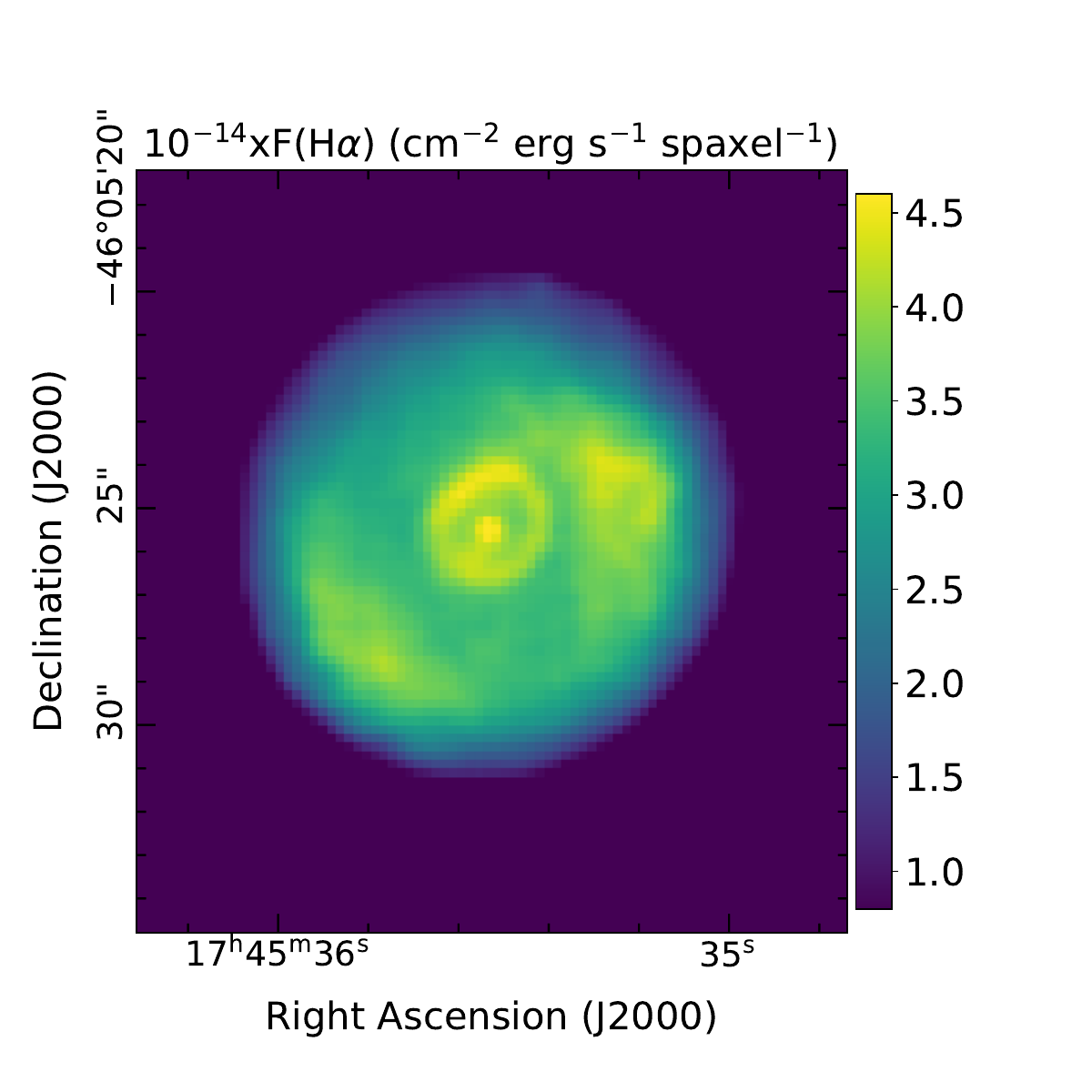}
}
\caption{Expanded image of the core region of Tc~1 in H$\alpha$ surface brightness 
(erg cm$^{-2}$ s$^{-1}$ spaxel$^{-1}$ $\times$ 10$^{14}$). 
}
\label{fig:HaCore}
\end{figure}

The core nebula has considerable structure (Figs. \ref{fig:EmLines1}, \ref{fig:EmLines2}), 
with the individual features marked in Figure \ref{fig:EmLines1}, right. 
The central 12$\times$12$''$ region in H$\alpha$ surface brightness is shown in 
Figure \ref{fig:HaCore}, emphasising the bright core region only.
The higher ionization morphology, typified by [\ion{O}{III}] is brightest 
near the CSt and steadily declines at larger radii of the core, and into the halo. 
The bright core is close to circular and the FWHM is about 10.8$''$ and there is 
a distinct inflection in the decrease of the H$\alpha$ surface brightness 
at radius 6.2$''$ from the CSt, giving an outer diameter of the core of 12.4$''$
(see also Fig. \ref{fig:RadialProfiles}). Closest to the star is an elliptical 
ring structure of axis lengths 2.8$\times$2.2$''$, position angle 
(PA, measured N through E) of the major axis 160$^{\circ}$, with 
centre offset by 0.3$''$ to the NW from the position of the CSt; the NE rim is 
notably brighter. This ring is visible in the permitted lines (H and He),
[\ion{O}{III}] and [\ion{O}{II}] but not [\ion{N}{II}] and its PA is displaced 
by $\sim$30$^{\circ}$ from the PA of the proper motion of the CSt. The ring is also 
enhanced in the extinction image (Fig. \ref{fig:c_map}) by about 0.03-0.04 dex 
relative to the immediate surroundings, implying some dust content. 

Outside of this ring beyond $>$2$''$ radii, the surface brightness is 
lower in the \ion{H}{} and low 
ionization emission but then rises again in a horseshoe shape extending from the NW 
around to SE, with the NW region brighter and composed of multiple knots, while that 
to the SE has a single bright knot. The horseshoe feature shows two regions of compact 
emission peaks (`knots') to the NW (brighter knots) and SE, strongest in lower ionization 
and in [\ion{O}{I}], but not seen in [\ion{O}{III}]. There are hints of arcs of emission 
extending from the central ring - the NW one extends inwards to the ring while that to 
the SE is more distinct. The contrast of these knots is greatest in [\ion{O}{I}] 
emission where the large-scale background emission is very low. Some of the
knots to the NW and the brightest one to the SE have been shown by 
\cite{Bouvisetal2025} to display enhanced metal line emission, characteristic of 
Low Ionization Systems (LIS) found in some PNe. A difference in radial velocity of 
these knots from the surrounding large scale emission is apparent, despite
the low spectral resolution of MUSE spectra: from the [\ion{O}{I}] image (595\,s cube),
the NW knots show velocities relative to systemic of about +20\,km s$^{-1}$
and the SE knots velocities -15 km s$^{-1}$. Given the low ionization of these features
and their velocity offsets with respect to systemic, it is suggested they compose
an outflow feature which lies outside the core and probably in the halo. Higher 
resolution spectra of these low ionization knots relative to their surroundings 
would be required to better establish their location.

Outside the core there is a shoulder in the H emission surface brightness,
between radii 6.4$''$ and 8.0$''$ leading to the halo where the surface 
brightness no longer declines ($ r \gtrsim 10''$); see Fig. \ref{fig:EmLines1}.
The halo can be broadly characterized as low ionization with only very faint 
and featureless emission of \ion{O}{}$^{++}$ (see Fig. \ref{fig:EmLines2}). 
Outside the core there is a ring of lower emission in H$\alpha$ and 
[\ion{O}{II}], [\ion{N}{II}] about 3-4$''$ in width (more extended to the SW) 
enclosed by many filamentary structures. Protruding from the core 
are two spikes at PA 70 and 240$^{\circ}$ but neither of the major axes of these
spikes pass through the position of the CSt; nor does the line joining the origins of 
these spikes at the edge of the core pass through the CSt. The spikes are visible in 
H lines and lower ionization but only the eastern spike (PA 70$^{\circ}$) on the 
extinction image and at [\ion{O}{I}]. Their nature and origin remain unexplained, 
but again higher spectral resolution may assist in detecting a velocity offset or 
gradient to these features. The outer filaments are often arcs but there are some 
radial ones; all are characterised by lower [\ion{N}{II}]6583\,\AA/H$\alpha$ 
than the large-scale extra-filament emission (ratio lower by $<$0.5 compared to
the mean of 1.00). The brightest feature in the halo is the wedge-shaped knot to
the WSW at $\Delta \alpha$, $\Delta \delta$ ~-21, -5$''$ relative to the position 
of the CSt, but this feature is only 10\% of the peak flux of the knots in the 
core already discussed. It is faint in 
[\ion{O}{I}] 6300\,\AA\ emission and the [\ion{N}{II}]6583\AA/H$\alpha$ 
ratio is low, implying different properties between the core and halo knots. The 
filaments seen in emission are visible in extinction with an increase in 
c(H$\beta$) with respect to the mean of the halo 
(c = 0.244) of up to 0.2 dex, indicating considerable dust content. None of the 
spikes or outer filaments show convincingly differing velocities from the systemic
velocity.  

The outer edge of the halo appears sharp to a limiting radius of 
$\sim$27.8$''$  (Fig. \ref{fig:EmLines1}), based on the brightest emission line, H$\alpha$. 
However, since the outer edge of the emission is actually defined by a S/N cutoff in the 
line flux detection, and in addition the outer edges of the field were employed for sky
background subtraction (Sect. \ref{Sec:observations}), the reality of an intrinsically sharp 
boundary to the nebula cannot be confirmed from these observations. 
At lower surface brightness the halo extent may be more than 28$''$ -- deep, wide-field
imaging is required to check this possibility.  If real, the boundary could occur 
between the halo (mean density 
$N_{\rm e}$ $\sim$ 600 cm$^{-3}$, Tab. \ref{tab:nebcons}) and the surrounding 
ISM (density-bounded), or the limit of photoionization for an ionization-bounded 
halo.

Based on the current expansion velocity of the low ionization species of $\simeq$ 20 
km s$^{-1}$ from \cite{Alemanetal2019}, the core of radius 6.2$''$ has a dynamical
age of 5200yr. To estimate the dynamical age of the halo requires an expansion
velocity that has not so far been measured (the emission lines over the halo
are spectrally unresolved by the MUSE data). Assuming the halo is the remnant of 
the AGB wind, then the compilation of \cite{Knappetal1998} provides a mean wind 
velocity of 13 km s$^{-1}$ based on their observations of 13 C star AGB envelopes
(mean 12.8 $\pm$ 4.8 kms$^{-1}$). With a radius of 28$''$, the dynamical time
of the Tc~1 halo is then 36\,000 yr.

\subsection{Dust extinction structure}
\label{c_struct}
The wealth of detail in the extinction image in Fig. \ref{fig:c_map} reveals 
for the first time the complexity of the dust distribution, in the case
of Tc~1 equalling or even surpassing the morphological detail in the emission 
line imaging. The structure to the extinction image in Fig. \ref{fig:c_map} 
can be broadly grouped into three areas: the bright core; the annulus between 
the core and faint halo; and the halo itself. The mean c(H$\beta$) value over the core 
is 0.306 (std 0.023) but there is a compact region to the WNW and a ring from 
PA 80--250$^{\circ}$ around the outer periphery of the bright core (radius 5.9$''$) 
with values elevated up to c=0.34. The peak c(H$\beta$) value occurs at the position of 
the CSt (maximum 0.59). At the centre is a slightly elliptical ring of enhanced 
extinction (by c(H$\beta$) of 0.02) which is spatially coincident with the ring 
of enhanced emission mentioned in Sect. \ref{Morphology}. 

Over the halo (radii $>$ 9.2$''$) the mean c(H$\beta$) value is 0.236 $\pm$ 0.097, 
but there is also structure with two distinct ridges to the W and SW plus a radial feature to 
the E, together with more compact c(H$\beta$) enhancements (larger in size than the 
point spread function, PSF). However most unusual is the annulus of lower c(H$\beta$) values 
around the periphery of the bright core (radii 5.9 - 6.9$''$). This feature in 
itself is not unique and from the MUSE observations of NGC~7009 \citep{Walshetal2016} 
a narrow region of lowered c(H$\beta$) values ($\Delta$ c(H$\beta$) $\sim$ 0.03) was found 
around the bright inner shell (see their Fig. 1). Here however, when the interstellar medium 
(ISM) extinction in the direction of Tc~1 is taken into account (E$_{B-V}$ = 0.158, 
\cite{Schlegel1998}, c(H$\beta$) $\sim$ 0.23 for a \cite{Howarth1983} reddening law 
with $R_{V}$=3.1, and very close to the mean for the halo), the annulus displays 'negative' 
extinction, with values to -0.07 below the ISM value (corresponding to 
H$\alpha$/H$\beta$ ratio = 2.99, well above the expected Case B value, so not
an extinction deficit in an absolute sense). This unusual structure, with seemingly 
negative values of extinction over an area $\sim$36$''^{2}$ surrounding the bright 
halo is unprecedented in a PN.

Given the unusual nature of this localised extinction behaviour it is valid to 
question whether the ISM value of extinction is perhaps too high and indeed
corresponds to the low value observed in the annulus. The values from
\cite{Schlegel1998} and G-TOMO \citep{Vergelyetal2022} however agree closely and
are concordant with the extinction inferred from the strength of the DIBs
\citep{DiazLuisetal2015}. The agreeement with these ISM values and the mean 
extinction over the halo, where $N_{\rm e}$ is low and, apart from the noted
extinction features, the extinction is rather uniform, tends to argue against 
a lower ISM extinction. An investigation into the extinction towards the 
stars in the halo of Tc~1 used to correct for the effect of their absorption 
lines on the emission line maps (Sec. \ref{SubSec: FieldStarRemoval}) might be productive.

In Fig. \ref{fig:RadialProfiles} the radial profile of the C$_{60}$ 7.0$mu$m band
from Giese et al. (2026 in prep.) is also included on the plot for direct
comparison with the visual extinction c(H$\beta$) and log$_{10}$ $T_{\rm e}$ and 
$N_{\rm e}$ radial profiles. However the C$_{60}$ profile is only formed over the 
limited range of PA's of 240--280$^{\circ}$ available from the James Webb Space 
Telescope (JWST) Mid-Infrared Instrument (MIRI) observations (Giese et al. 2026); 
it is the mean of the four profiles over 10$^{\circ}$ sections shown in Giese et al. 
(2026). The extinction displays a slight peak at offset 5.4$''$, coincident
with the peak emission of the C60 shell, although too much should not be drawn from
this comparison, given that the extinction profile is from a complete
radial profile and showed a slight elevation at this radius for PA 80--250$^{\circ}$.  
However beyond the peak in extinction and C$_{60}$ 7.0 $\mu$m emission both
decline, supporting links between the extinction and C$_{60}$ emission structures. 

The low extinction annulus occurs where $N_{\rm e}$ starts to drop, $T_{\rm e}$ 
starts to rise into the halo and C$_{60}$ emission drops. This concurrence of
four indicators of gas and dust properties changing within a short radial extent 
points to linked physical conditions at the edge of the inner bright core. The 
changes in dust properties interpreted from extinction as a higher proportion of small 
dust particles, or of larger particles, point to dust processing in this zone.
Two possibilities are suggested to cover these cases: destruction or ablation
of dust could occur at the shock front ahead of the outer rim of the higher density 
core nebula, leading to the low extinction annulus; ablation of larger dust particles, 
which are present in the halo, occurs as the ionization shock front moves outwards, 
resulting in a higher proportion of smaller grains inside the ionized core. The
increase in $T_{\rm e}$ can be linked to the decrease in $N_{\rm e}$ in this
annulus (see also Tab. \ref{tab:nebcons}), either through a change in ionizing
photon optical depth or radiation hardening (Sect. \ref{Subsec: Regions});
whether these changes in physical conditions are linked to the dust is not clear. 
If dust processing does occur, as strongly suggested, then this may provide increased 
heating; a detailed model would be required to determine if this mechanism could 
also contribute to the increase in $T_{\rm e}$.

\subsubsection{Instrumental effects on extinction determination}

Apparent negative extinction should not exist under standard nebular conditions
in a PN, so in order to verify an intrinsic effect within the nebula, possible 
observational or instrumental artefacts need to be scrutinised. The region of
the extinction deficit occurs in a narrow radial zone just outside the bright
core (radii 5.8 -- 7.2$''$ from the CSt) where the H$\alpha$ and H$\beta$ 
local gradients in surface brightness drop steeply 
(e.g., Fig. \ref{fig:RadialProfiles}). The observed spatial 
point spread function (PSF) is affected by both atmospheric effects and 
transfer through the instrument; the former has a well-known decrease to 
longer wavelengths \citep{Fried1966}, so even without instrument contribution, 
the radial profile in the blue is broader than in the red. Since the line 
surface brightness of the core is much higher than in the range of radii under 
scrutiny (by a factor $\sim$130 for the radially averaged profile), then 
light in the PSF wings can be scattered from the core into the surroundings, 
enhancing the blue flux and hence H$\beta$. Just such an enhanced H$\beta$ 
flux could then cause a lower extinction measured from H$\alpha$/H$\beta$. If the 
instrument contribution to the PSF was also broader in the blue, this 
would enhance the trend resulting from the atmospheric PSF; any significant 
contribution in the wings of the PSF, say $>$ 1.0$''$ would preferentially scatter
blue light from the core into the surroundings, 1--2$''$ offset from the
fairly sharp edge to the core (radius 5.5$''$). \cite{Walshetal2016} also
checked if this explanation could account for the (smaller) extinction 
decrease outside the bright shell in NGC~7009; it was discounted as the 
surface brightness contrast was lower. The chromatic character of the
PSF can be quantified by taking an observed PSF from one or more star 
images, or a model PSF, and deconvolving the observed H$\alpha$ and 
H$\beta$ images. Appendix \ref{App:PSF restoration} describes these 
experiments.

The conclusion of these tests was that, while there may be an influence
of the PSF chromaticity on the measured H$\alpha$/H$\beta$ in the range of 
radii adjacent to the bright core, it cannot explain the drop in extinction
to values below the ISM value. Therefore an astrophysical explanation 
of the extinction deficit is mandated and a process to locally alter the 
extinction law or optical depth effects on the H$\beta$ and/or H$\alpha$ line 
flux emitted towards the observer must be invoked. The latter can be strongly 
excluded since there are no strong changes in physical conditions over the 
radii in question (see Fig. \ref{fig:RadialProfiles}), other than a decrease 
in $N_{\rm e}$ by $\sim$ 2000 cm$^{-3}$ and an increase in $T_{\rm e}$ of 
$\sim$ 2000\,K (Fig. \ref{fig:TeNe2maps}). For these modest changes in physical 
conditions, no large change in H emissivities or optical depths is expected and 
no reason to suspect departure from Menzel \& Baker Case B. In addition the 
physical conditions are very similar to those over the halo and the extinction 
(above the ISM value) is low (and positive), as expected for the low density and 
hence low dust density, assuming constant gas/dust mass ratio.

\subsubsection{Dust radiative transfer effects}
Given that the dust extinction image (Fig. \ref{fig:c_map}) already shows 
gross effects of spatial variation in extinction, as between the core (c=0.31) 
and the halo (c=0.24), the role of dust transfer would appear to be a promising 
candidate to explain the low extinction annulus. Dust particles absorb radiation 
and scatter it out of
the incident direction and the combined effect produces extinction. If there 
was a local volume within a nebula where the particles were smaller, the 
resulting scattered light would be bluer than in the rest of the nebula;
then scattering of H$\beta$ would be elevated relative to the surroundings
leading to a decrease in observed H$\alpha$/H$\beta$ even though the mass of
the dust column may not spatially vary. 

A possible mechanism to produce this result could be sputtering of the bulk 
larger grains to produce a local enhancement in smaller grains (or even grains 
with a different composition). 
The threshold relative velocity for grain sputtering by atoms and ions is $>$ 
50 km s$^{-1}$, for a range of grain types \citep{Jonesetal1994}.  
However based on the velocity field, c.f., \cite{Alemanetal2019}, such gas 
velocities are unlikely to be present in Tc~1. Grain fragmentation as a result 
of grain-grain collisions has a much lower threshold velocity 
\citep[Table 2]{Kirchschlageretal2019} 
and such fragmentations produce a shower of smaller particles -- so this 
could provide a more likely mechanism for producing smaller grains than 
sputtering. A possible mechanism for producing grain-grain velocity 
differences could be provided by differential motions between grains in the 
bright core and those beyond. 
At the boundary of the bright core, where the ion density, and therefore the 
plasma drag, drops significantly, grains near the boundary may overshoot into 
the lower density nebular material and thereby collide with the existing 
grains in this region.

In order to explore the feasibility of smaller grains causing the localised 
lower extinction, a simple 3D dust scattering model was produced and details 
of the model and the approach are given in
Appendix \ref{App: ScatMod}. An annular reduction in extinction at the 
outer edge of the bright core could be produced if there was a shell 
of dust grains scattering light from the bright core; the line of sight 
extinction over the core itself would also be affected by this 
scattered component, but the effect is very small (since the scattering 
column is short). It is curious that the extinction from H$\alpha$/H$\beta$ 
in the vicinity of the CSt (Fig. \ref{fig:c_map}) is lower than measured to 
the CSt from stellar continuum fitting (see Sect. \ref{CenStar})
by 0.086 (c=0.299 from H$\alpha$/H$\beta$ in the 12$''^{2}$ around the
CSt compared to c=0.385 for the CSt SED fitting). It implies that there is small scale 
extinction along the sightline to the CSt, perhaps local to the CSt; 
this suggestion can be checked by comparison to infrared dust mapping.

The dust scattering models described in Appendix \ref{App: ScatMod} were
inconclusive in terms of explaining such a large apparent decrease in 
the extinction in the c(H$\beta$) annulus, although they are indicative that a
smaller effect could be produced ($\delta$ c $\sim$ 0.03). However the models
were confined to well-behaved dust (spherical particles of amorphous 
carbon or fullerene (assumed C60) molecules as very small grains). More 
complex dust species and distributions (ellipsoidal or cylindrical grains,
mixed C grain types, perhaps including fullerene admixture) should be 
considered. The James Webb Space Telescope near- and mid-infrared imaging 
and spectroscopy (Cami et al. 2026, in prep) can provide further data on the
spectral and spatial properties of the dust and fullerene emission. It
is already clear that the peak of the fullerene emission (offset 
$\sim$ 4.5--5.0$''$) occurs inside the ring of the apparent extinction deficit 
(5.9 -- 6.9$''$, Fig. \ref{fig:c_map}), so a direct association of a shell 
of fullerene emission with the region of lower extinction is not indicated. 

An alternative to small dust grains, giving enhanced scattering in the region outside the 
core of Tc~1, is large grains. Here the change is in the extinction properties: for larger 
grains the wavelength dependence of the extinction is flatter. The extinction 
properties can be described by the value $R_{V}$, the ratio of total ($A_{V}$) 
to the wavelength selective extinction (e.g., $E_{B-V}$). It has been found
that larger values of $R_{V}$ than the typical Milky Way value of 3.1 are characteristic of
larger grains (for a single size grain, $1/R_{V}$ $\propto$ slope of extinction curve between
$B$ and $V$, \cite{Draine2011}). Large values of $R_{V}$ have been measured in star-forming 
regions, such as H~II regions (e.g., \citet{Rogersetal2024}) and in the young 
starburst galaxy NGC~5253 \citep{Pruijtetal2026}. If the region outside the 
Tc~1 core was populated by a greater
fraction of large grains than within the core, then a higher value of R would be required
to determine the extinction from the observed to Case B H$\alpha$/H$\beta$ ratio. Then 
on account of the less steep extinction law compared to the core, the resulting c(H$\beta$) 
value for the region adjacent to the core would have a similar value to the region for
the halo, and to the interstellar extinction, obviating the need for 'negative extinction'. 

A simple experiment was conducted on the Region spectrum for the low c annulus 
(Sect. \ref{Subsec: Regions}), determining the extinction from the observed H$\alpha$/H$\beta$
of 3.29 with differing values of $R_{V}$ than used for constructing the extinction map
(Fig. \ref{fig:c_map}) with $R_{V}$=3.1. For these tests, the Clayton, Cardelli \& Mathis
\citep{CCM1989} extinction law, a parameterization that is dependent on the value of $R_{V}$,
was used. For example, the calculated value of c(H$\beta$) with $R_{V}$=5.0 was 0.25 for the 
low c annulus, actually larger than the interstellar extinction 
value of 0.23. The resulting extinction map with a confined annulus of $R_{V}$=5.0 grains 
would then show a flatter extinction from the outer rim of the core to the outer edge
of the halo. If the dust grain properties were so spatially stratified, then other regions
could show similar effects - such as part of the halo ($10 < r < 14 ''$) where c(H$\beta$) is 
low relative to the interstellar value (see the profile in Fig. \ref{fig:RadialProfiles}). 
Conversely if such a large value of $R_{V}$ was applicable in the core, then c(H$\beta$) would be 
higher than computed with $R_{V}$=3.1.

\subsection{Central star}
\label{CenStar}
The bright central star (CPD $-$46$^{\circ}$ 11816) was well detected in the shortest
exposure cube and allows an investigation of its spectral characteristics over the
MUSE wavelength range 4750--9300\,\AA. 

\subsubsection{The MUSE spectrum of the central star}
\begin{figure}[t]
\centering
\resizebox{\hsize}{!}{
\includegraphics[trim={0.5cm 0.2cm 1.5cm .8cm},clip]{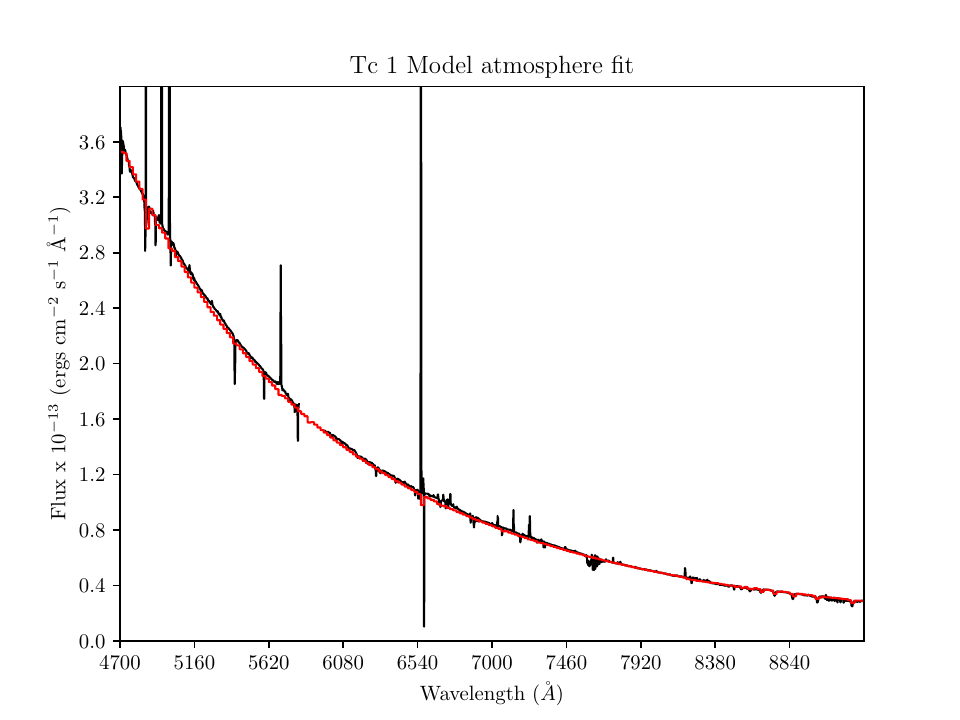}
}
\caption{MUSE spectrum of the CSt of Tc~1 extracted from the 40\,s cube. 
The black line shows the dereddened spectrum ($E_{B-V}$ = 0.265, c(H$\beta$) = 0.385) and 
the red line the matched Kurucz 9 LTE model atmosphere 
for $T_{eff}$ = 31000\,K, log $g$ = 3.5 \citep{CastelliKurucz2004}. See text for details. 
}
\label{fig:CSspec}
\end{figure}

The spectrum of the CSt was extracted from the 
40\,s datacube (the only exposure level where the star image was not saturated) by 
summing an almost circular area of 4.68$''^{2}$ centred on the star and subtracting 
the local emission as the mean in four cardinal regions around this area (annular 
radius 2.6$''$), encompassing an area of 25.44$''^{2}$. Figure \ref{fig:CSspec} 
shows the resulting observed spectrum (in black).
It is apparent that positive residuals of the strongest lines (H and 
[\ion{O}{III}]) remain after this subtraction; the H lines show a 15\% enhancement, 
the [\ion{O}{III}] lines a 130\% increase relative to the mean background annulus flux.
Thus excess emission close to the CSt, stronger for higher ionization species, is
spatially resolved. From the MUSE extinction map (Fig. \ref{fig:c_map}) there appears to 
be extra extinction to the CSt relative to the surroundings - the integrated c(H$\beta$) value
in an aperture of 0.4$''$ radius is 0.43 (0.46 from the integrated H$\alpha$/H$\beta$ 
flux ratio in the same aperture) and the peak value is 0.59, relative
to a mean value for the surroundings of 0.30. However given the problem of fitting
the narrow H$\beta$ stellar absorption line profile under the emission line (and to
a lesser extent the H$\alpha$ line), the enhancement in extinction is quite uncertain. 

Another approach is to fit the continuous spectrum of the CSt with a model 
atmosphere and search for the extinction and stellar ($T_{eff}$, log $g$) 
combination that best matches the observed stellar flux distribution. From their 
CLOUDY \citep{Ferland2017} photoionization modelling of Tc1, \cite{Alemanetal2019} 
found a blackbody best match for $T_{eff}$ between 30000 and 32000\,K. For 
the photoionization modelling of Tc~1, \cite{Otsukaetal2014} adopted 
$T_{eff}$ = 34060\,K and log $g$ = 3.4. From their NLTE stellar atmosphere 
modelling of the IUE high resolution UV (1150-2000\,\AA) spectrum of the CSt, 
\cite{Pauldrachetal2004} found a best fit for $T_{eff}$ = 35000\,K with log $g$ = 
3.62. 

Here fits to local thermodynamic equilibrium (LTE) spectral energy distributions 
(SEDs) from Kurucz 9 \cite{CastelliKurucz2004} model atmospheres for (a) $T_{eff}$ 
= 31000\,K, log $g$ = 3.5 and, (b) $T_{eff}$ = 35000\,K, log $g$ = 4.0 were tested.
The MUSE CSt spectrum was dereddened over a small grid of $E_{B-V}$ values using 
the \cite{Howarth1983} Galactic reddening law with $R_{V}$ = 3.1 and then 
normalising the two alternative Kurucz models to each dereddened CSt spectrum. The 
most suitable matches were selected by examining the blue-to-red slopes of the 
residual of the Kurucz model subtracted from the dereddened CSt spectrum. The 
residuals turn out to be very sensitive to the value of $E_{B-V}$ used for the 
dereddening. The 31000\,K, log $g$ 3.5 model, dereddened with $E_{B-V}$ 0.265 gave 
the most satisfactory fit; $E_{B-V}$ values differing by $\pm$0.005 showed 
residuals increasing / decreasing, respectively, towards the blue. For the case of 
the 35000\,K log $g$ 4.0 model SED, again $E_{B-V}$ of 0.265 produced the flattest 
residual slope, although the residuals are not as smooth as for the 31000\,K Kurucz 
model. This led us to adopt the 31000\,K, log $g$ 3.5 model as the most reliable. 
This value compares closely with the range 30\,000 -- 32\,000\,K found by 
\cite{Alemanetal2019} from photoionization models. Figure \ref{fig:CSspec} 
shows the match of this model atmosphere to the spectrum dereddened with 
$E_{B-V}$ = 0.265 mag. For the adopted Galactic reddening law, the corresponding 
value of c(H$\beta$) is 0.385, thus confirming the nebula analysis that 
there is an extra component of extinction towards the CSt above that for 
the ionized gas alone, although not as large as derived above from the nebular
extinction mapping. It 
is notable that this increased line-of-sight extinction to the CSt is larger than 
for the projected elliptical ring (discussed in Sec. \ref{Morphology}), so they may 
not necessarily be part of the same structure. 

For a distance of 3.53 kpc, the luminosity corresponding to the normalised Kurucz 9 
LTE model atmosphere is 8150~L$_\odot$. 
For this temperature and luminosity, the star is on the horizontal post-AGB track 
corresponding to a Solar metallicity initial mass of $\sim$2.0 M$_{\odot}$ 
\citep{MillerBertolami2016}. This mass estimate for Tc~1 is concordant with 
the mass range (1.5M$_{\odot}$ $<$ M $\lesssim$ 4--5M$_{\odot}$ at Solar 
metallicity) for production of a carbon-rich nebula \citep{Marigoetal2003}. 
However the dynamical expansion age of the nebula (Sect. \ref{Morphology}) 
is longer than predicted for a 2M$_{\odot}$ Solar abundance AGB star at $T_{eff}$ = 31000\,K 
by the \cite{MillerBertolami2016} tracks ($<$ 500 yr). Either the expansion velocity was higher 
in the past and has slowed, or the star has a lower mass with correspondingly slower evolution 
time, or both.
 
\subsubsection{The central star spectral type}
\label{CStype}

CPD $-$46$^{\circ}$ 11816 has been observed with the 8.2~m VLT UV Echelle 
Spectrograph (UVES) at a spectral resolving power of R = 37000 (Progam ID: 
095.D-0432(A)) and an extensive study of the diffuse interstellar bands (DIBs) was 
published \citep{DiazLuisetal2015}. The spectra were retrieved from the ESO Archive 
with the aim of determining the spectral type of the CSt. We also had access to a 
spectrum from July 1980 taken on the 3.9~m Anglo-Australian Telescope with a 
resolving power of 9000 over just the 4230-4730\,\AA\ region. This spectrum 
showed the stellar 4634,4640\,\AA\ \ion{N}{III} doublet in emission and 
no stellar \ion{He}{II} 4686\,\AA\ emission, implying an O(f) spectral type. From 
the UVES UV (3281 -- 4530\,\AA) spectrum the equivalent width (EW) of the 
\ion{He}{I} 4471\,\AA\ line\footnote{In the UVES spectrum,  the \ion{He}{I} 4471\,\AA\ 
line has a blue wing, extending to -300 km s$^{-1}$ from the absorption line centre. 
In the much lower resolution AAT spectrum the stellar absorption line is completely 
filled in by nebular emission.} is 690~m\,\AA, while the AAT spectrum 
gives an EW of 550~mA for the \ion{He}{II} 4542\,\AA\ line. The 4471/4542\,\AA\ EW 
ratio of 1.25 implies a spectral sub-class of O7.5, according to Table 3 of 
\cite{Conti1971} and Fig.1 of \cite{Martins2018} (assuming that the spectrum did 
not change between 1980 and 2015). A high mass Milky Way star with an O7.5I 
spectral type should have $T_{eff}$ = 34000\,K \citep{Masseyetal2005}{~Table 9}, 
similar to the estimates of 35000\,K \citep{Pauldrachetal2004} and 30000 -- 32000\,K 
\citep{Alemanetal2019}. We conclude that a spectral type of O7.5I(f) best matches 
the available stellar spectra.

\subsection{Region spectra}
\label{Subsec: Regions}

In order to check the controversial result of an extinction deficit in an
annulus around the Tc~1 core and explore if other effects were apparent in the 
line and continuum emission, and diagnostics (c.f. Fig. \ref{fig:RadialProfiles}), 
region spectra were constructed summing large numbers of spaxels over selected 
regions. Such spectra have the advantage of high S/N, bringing a check on areal 
analysis based on spaxel data only. Four regions were selected: \\
\begin{itemize}
\item an annulus over the bright core centred on the CSt (radius inner, outer, 
mean 4.4, 5.2, 4.8$''$, area 33.44$''^{2}$);
\item an annulus containing the region of low extinction (spaxels with 
c(H$\beta$) $<$ 0.22), inner, outer, mean radii 6.0, 6,9, 6.45$''$; this is an 
irregular annulus of 36.32$''^{2}$; 
\item a wider annulus over the faint halo (radii 8.6, 12.2, 10.4$''$, area 246.24$''^{2}$);
\item total nebula (area 2228.56$''^{2}$ based on detected H$\alpha$ flux $>$ 
1.0$\times$10$^{-17}$ erg cm$^{-2}$ s$^{-1}$ spaxel$^{-1}$ and excluding 5.08$''^{2}$
over the CSt).  
\end{itemize}
The shape of the first three regions was chosen as an annulus, so that any radial 
variation of parameters with respect to the position of the CSt could be examined.
The mean radii of these four regions and their areas are summarised in Table  
\ref{tab:regiondiags}.

\begin{table*}
\caption{Parameters and diagnostics for Region spectra}
\centering
\begin{tabular}{lrrrrrrrrrrrr}
\hline\hline
Region & Radius &       Area & c(H$\beta$) & Dered       & [\ion{S}{II}] $N_{\rm e}$ & [\ion{N}{II}] $T_{\rm e}$ & [\ion{Cl}{III}] $N_{\rm e}$ & [\ion{S}{III}] $T_{\rm e}$ \\
name   &        &            &             & F(H$\beta$) &  6716/6731\AA             &  6548+83/5755\AA           &  5517/5537\AA               &  9068/6312\AA  \\
       & ($''$) & ($''^{2}$) &             & (cgs)       &  (cm$^{-3}$)              & (K)                       & (cm$^{-3}$)                 & (K) \\
\hline
Core   &   4.8   &  33.44 & 0.32 & 1.19e-11 & 2100 &  8700  & 2900 &  8400 \\
Low c  &   6.5   &  36.32 & 0.18 & 8.30e-13 & 1800 &  9700  & 2000 &  9100 \\
Halo   &  10.3   &  246.2 & 0.24 & 2.63e-13 &  600 & 11100  &  680 & 10800 \\
Total  & $<$27.2 & 2228.6 & 0.31 & 4.39e-11 & 1900 &  9400  & 2700 &  8600 \\
\end{tabular}
\tablefoot{
All $N_{\rm e}$ and $T_{\rm e}$ values were calculated with PyNeb. See 
Appendix \ref{App:AnnSpects} for the line fluxes on which these diagnostics were based.
}
\label{tab:regiondiags}
\end{table*}

\subsubsection{Line emission}
The extinctions (from H$\alpha$/H$\beta$ compared to Case B at the relevant $T_{\rm e}$ 
and $N_{\rm e}$), dereddened H$\beta$ fluxes and CEL diagnostic physical parameters of 
the four regions are listed in Table \ref{tab:regiondiags}. Lists of the identified lines, 
observed and dereddened line fluxes for the four region spectra are provided in
Appendix \ref{App:AnnSpects}.

The diagnostics of the four regions listed in Tab. \ref{tab:regiondiags} confirm at higher S/N 
the trends shown in Figs. \ref{fig:TeNe3maps} and \ref{fig:TeNe2maps} for $T_{\rm e}$ and 
$N_{\rm e}$ at higher and lower ionization respectively, and the radial profiles in 
Fig. \ref{fig:RadialProfiles}. Most notable is the increase in $T_{\rm e}$ at the outer edge of 
the core from about 8500\,K (from both [\ion{S}{III}] and [\ion{N}{II}]) to 11000\,K and the 
decrease in $N_{\rm e}$ from $\sim$2500 to 650 cm$^{-3}$ (mean from [\ion{Cl}{III}] and 
[\ion{S}{II}]). The halo must also have a lower filling factor since the decrease in $N_{\rm e}^{2}$ 
between the core and halo is $\sim$20$\times$ less than the drop in surface brightness.

The difference in the $T_{\rm e}$ and $N_{\rm e}$ values between the values in Tab.
\ref{tab:regiondiags} and the images (and the radial profiles) can be understood both
in terms of S/N and weighting by the line fluxes contributing to the diagnostic ratios.
Taking [\ion{N}{II}] $T_{\rm e}$ as an example, for the area of the four regions, 
the mean unweighted spaxel values are (Core, Low c, Halo, Total) 8660, 9510, 11100, 10550\,K, 
to be compared with those in Tab. \ref{tab:regiondiags}, col. 9. The largest difference 
in diagnostics between integrated and spaxel-by-spaxel average values is for the Total nebula 
area, where the flux weighting by the spaxels of the bright core lower the integrated value 
in comparison to the areal mean of $T_{\rm e}$ values.

The increase in $T_{\rm e}$ from the rather uniform value in the core to the higher but also 
uniform value in the halo occurs over a rather narrow range of 6--9$''$ offset, precisely where 
$N_{\rm e}$ begins to decrease beyond 6$''$ radius into the halo. As the radial profiles 
demonstrate (Fig. \ref{fig:RadialProfiles}), it is this range where the extinction also 
decreases. The images of [\ion{O}{III}] emission (Fig. \ref{fig:EmLines2}, upper right)
show that some higher ionization emission escapes from the denser core into the surroundings,
so the core is not entirely optically thick to photons with energies $>$35 $eV$; the lower 
density of O$^{++}$ can lead to reduced cooling, hence a hotter halo. Also the effect of 
radiation hardening, whereby the frequency dependence of the H photoionization cross section 
above 13.6 $eV$ permits the larger fraction of higher frequency photons to produce 
a hotter halo. \cite{Sandinetal2008} observed a hotter halo than core (from [\ion{O}{III}] 
$T_{\rm e}$ measurements) in some PNe (NGC~6826, NGC~7662 and IC~3568) as did
\cite{MonrealIbero2005} for NGC~3242, finding
increases in $T_{\rm e}$ up to $\sim$4000\,K, larger than 2500\,K observed here for Tc~1. 
The MUSE observations provide sufficient 2D data for a detailed photoionization model of Tc~1,
so that the radiation transfer between core and halo could be studied.  

\subsubsection{Continuum emission}
\begin{figure*}[t]
\centering
\resizebox{\hsize}{!}{
\includegraphics[trim={0.8cm 0.2cm 1.5cm 0.8cm}, clip]{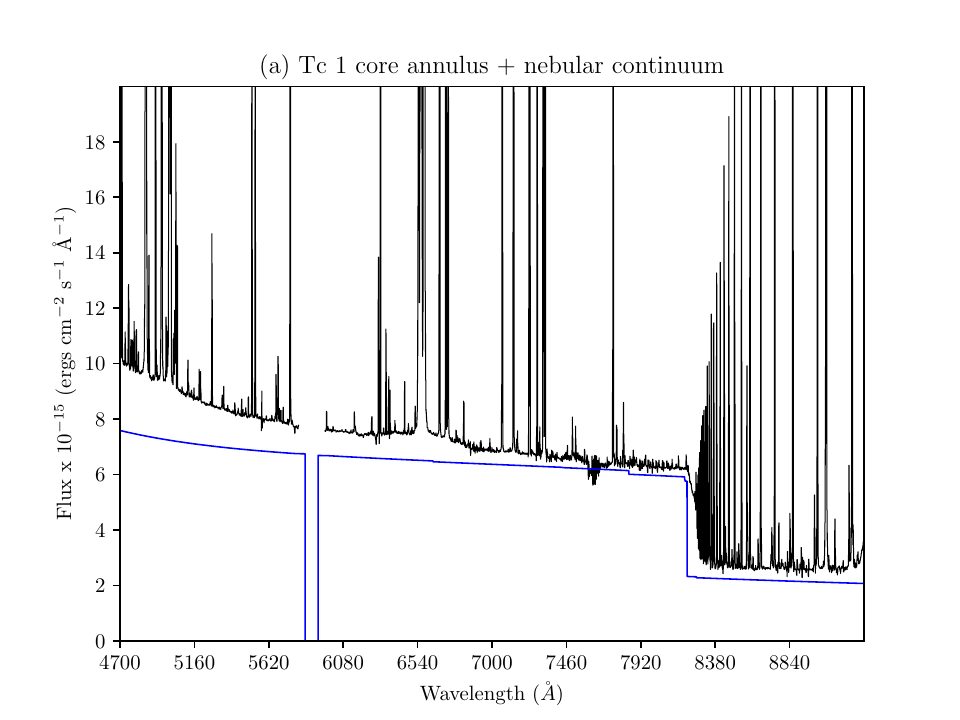}
\hrulefill
\includegraphics[trim={0.8cm 0.2cm 1.5cm 0.8cm}, clip]{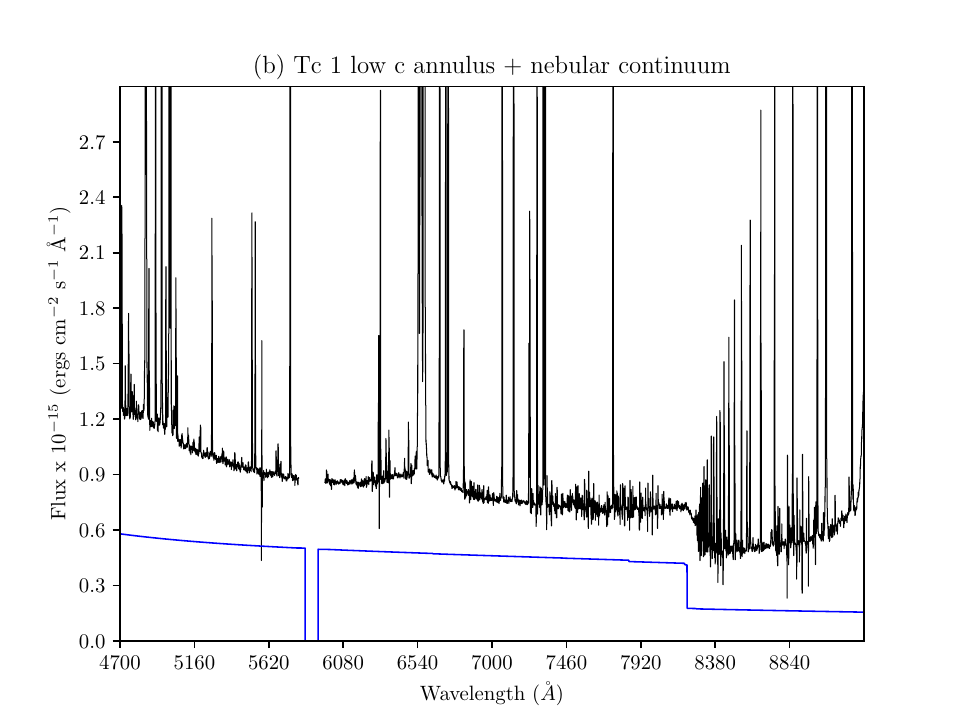}
}
\resizebox{\hsize}{!}{
\includegraphics[trim={0.8cm 0.2cm 1.5cm 0.8cm}, clip]{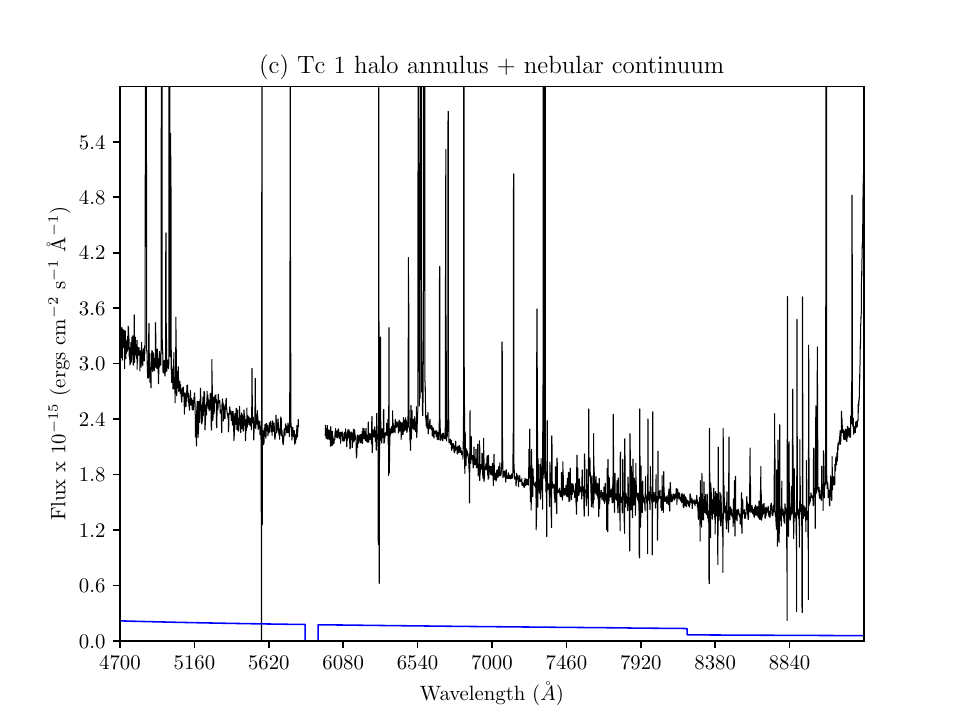}
\hrulefill
\includegraphics[trim={0.8cm 0.2cm 1.5cm 0.8cm}, clip]{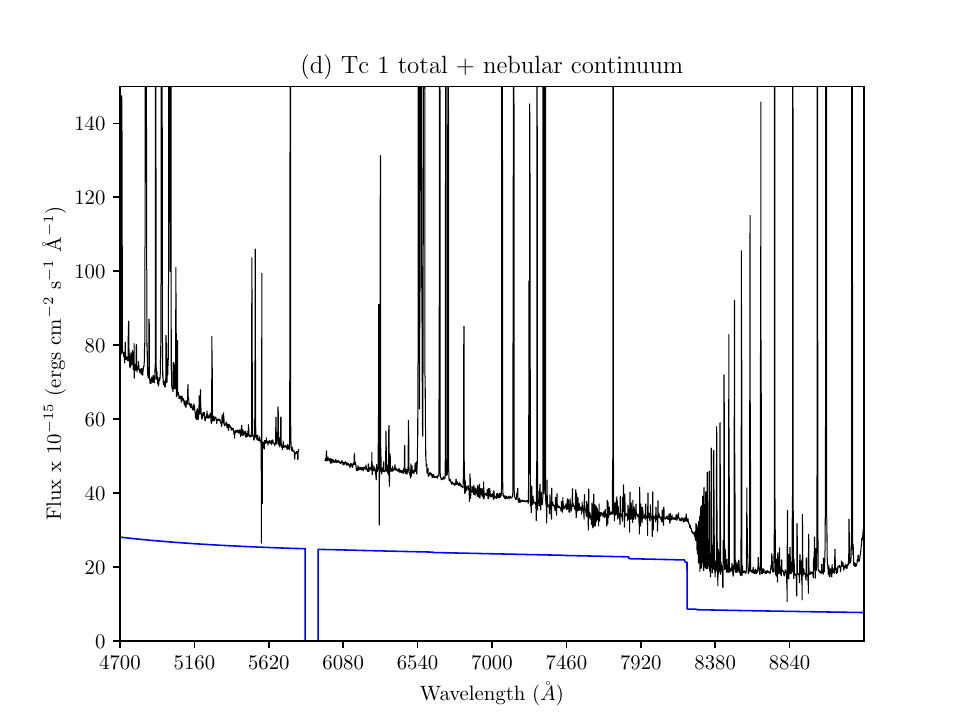}
}
\caption{Dereddened spectra of the four summed regions: \\
(a): bright core annulus; \\ 
(b): annulus of low c(H$\beta$) spaxels; \\
(c): halo annulus; \\
(d): total nebula; \\
in black and the fitted nebular continuum (sum of H$^{+}$ bound-free, 
free-free and 2-photon and He$^{+}$ bound-free) in blue.  The spectra 
highlight the continuum shape and level for the four spectra with the majority 
of the emission line peaks off scale. 
See text for further details for the four regions and Table \ref{tab:nebcons} for 
inputs to the nebular continuum calculations.
}  
\label{fig:annspecsnc}
\end{figure*}

Figure \ref{fig:annspecsnc} shows the four spectra, plotted in log$_{10}$ flux
and focussing on the continuum. For the inner nebula annulus, the continuum is 
strong and the Paschen jump at $\sim$ 8205\,\AA\ can be used as a $T_{\rm e}$
indicator; for the low extinction annulus, the continuum is weaker but shows
a Paschen jump, continuum generally increasing blueward and an abrupt increase
in flux redward of $\sim$8900\,\AA; the halo continuum is very weak with a discernible
Paschen jump but similar blue slope and rise in the far red. We attribute
the rise in flux at $\lambda \gtrsim$ 8500\,\AA\ to 2nd order contamination 
from the MUSE Volume Phase Holographic Grating of light at $\lambda \lesssim$ 
4700\,\AA\ (and an effect of defocus of the 2nd order for some slices), as 
documented in the MUSE instrument 
manual\footnote{\url{https://www.eso.org/sci/facilities/paranal/instruments/muse/doc.html$} }. 
Given that these aperture spectra are sums of many spaxels (IFUs and slicers),
then the fractional effect is noticeable, particularly where the red continuum is low 
(as in the two outer annuli). It is assumed that the bright CSt is the 
origin of this 2nd order continuum (since the MUSE PSF in the far red has significant 
'arms' in the cardinal directions), but there may also be a contribution from 
the blue nebular continuum ($\lambda <$ 4700\,\AA) from the bright core. As this 2nd 
order contamination of the spectra is explained as a purely instrumental effect, 
it will not be further mentioned in an astrophysical context.

\begin{figure*}[t]
\centering
\resizebox{\hsize}{!}{
\includegraphics[trim={0.7cm 0.2cm 1.0cm 0.9cm}, clip]{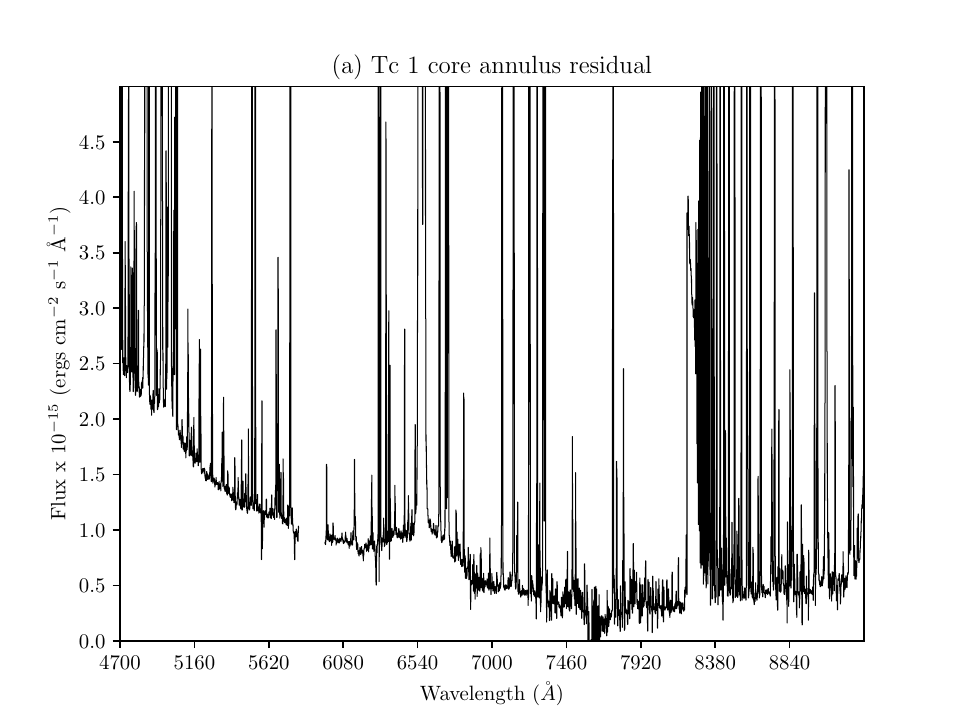}
\hspace{-0.6truecm}
\includegraphics[trim={0.7cm 0.2cm 1.0cm 0.9cm}, clip]{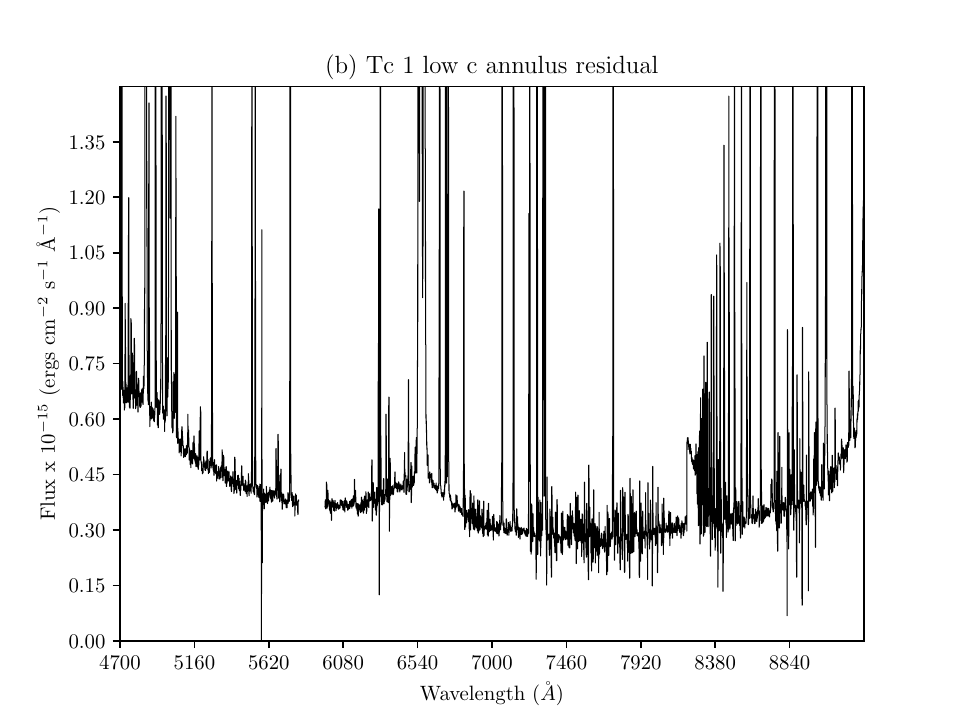}
}
\resizebox{\hsize}{!}{
\includegraphics[trim={0.7cm 0.2cm 1.0cm 0.9cm}, clip]{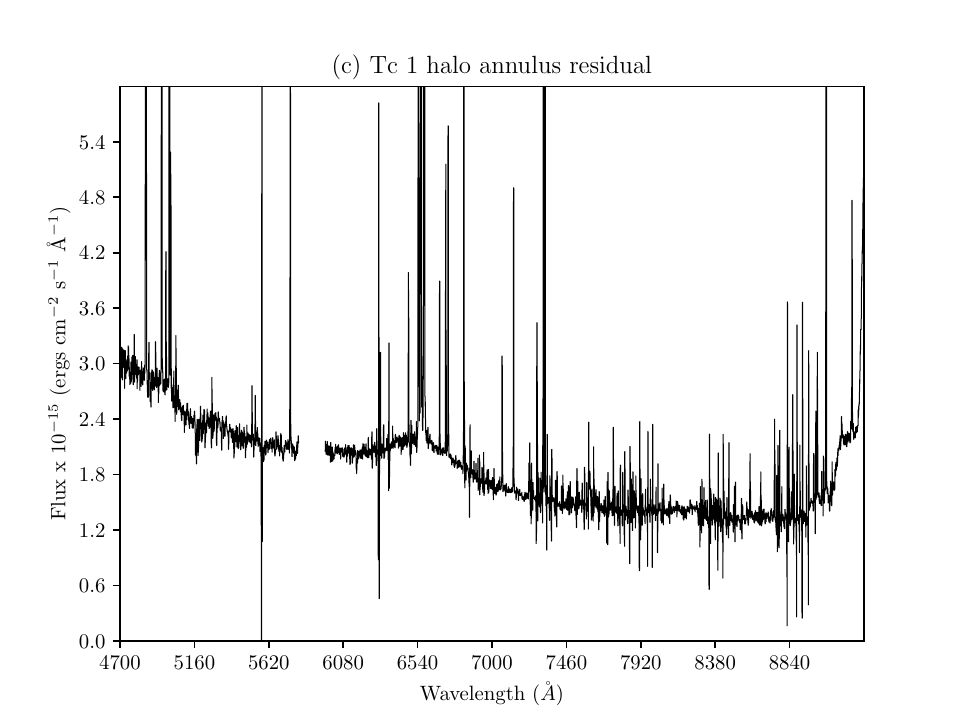}
\hspace{-0.6truecm}
\includegraphics[trim={0.7cm 0.2cm 1.0cm 0.9cm}, clip]{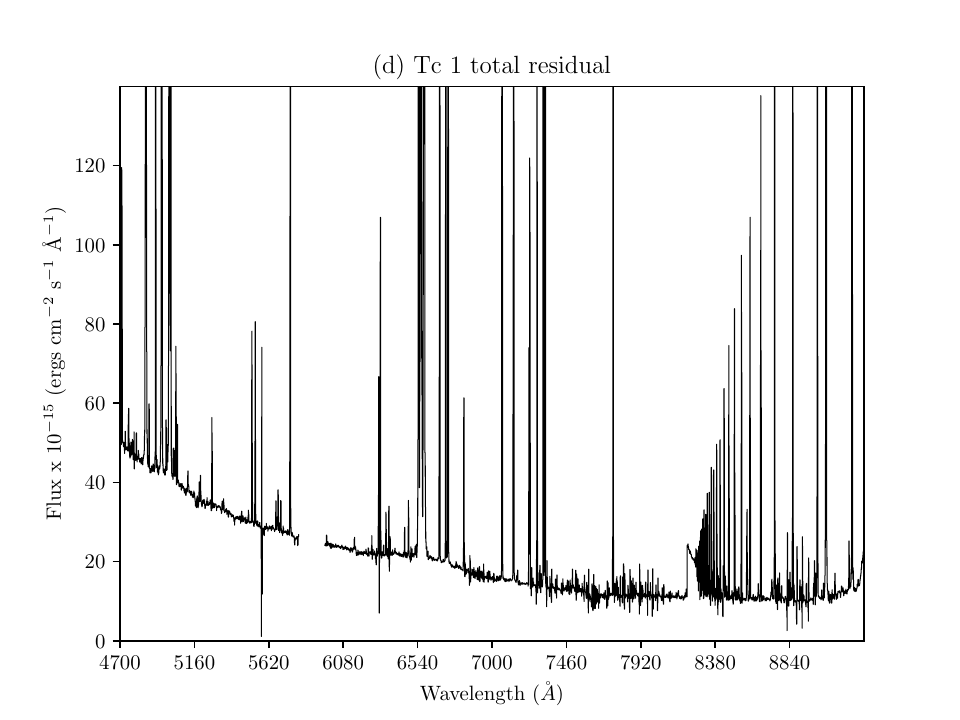}
}
\caption{Residual spectra of the dereddened spectrum of the four summed regions 
(rededdened spectra in Fig. \ref{fig:annspecsnc}) with the nebular continuum 
(see Fig. \ref{fig:annspecsnc}) subtracted: \\
(a): bright core annulus; \\ 
(b): annulus of low c(H$\beta$) spaxels; \\
(c): halo annulus; \\
(d): total nebula. \\
See Table \ref{tab:nebcons} for inputs to the nebular continuum calculations.
}  
\label{fig:ann_nebcons}
\end{figure*}

In order to determine the strength and spectral shape of any continuum in excess 
of the atomic nebular continuum, in each of the observed spectra the atomic 
nebular continuum (sum of bound-free, free-free and 2-photon for H$^{+}$ and 
He$^{+}$) was calculated and subtracted. This analysis requires the spectra to be 
dereddened (using the extinction derived from the observed H$\alpha$/H$\beta$ 
compared to Case B for the appropriate $T_{\rm e}$ and $N_{\rm e}$), the 
dereddened H$\beta$ flux and the He$^{+}$/H$^{+}$ number ratio; c.f. 
\cite[Appendix A]{Walshetal2018}. Here the results for 
the nebular continuum were calculated in the dipso spectral analysis package
\citep{Howarthetal2014}. Table \ref{tab:nebcons} lists the basic parameters of 
the nebular continuum fits and Fig. \ref{fig:ann_nebcons} shows the results of 
subtracting the nebular continuum from each annulus spectrum. The $T_{\rm e}$ 
values were not calculated from the size of the Paschen jump (e.g.
\cite{Zhang2004}, \cite{Walshetal2018}), but instead the values from the
CEL ratios were used (see footnote to Table \ref{tab:nebcons}).  

\begin{table*}
\caption{Adopted parameters of the nebular continuum fits to Tc~1 region spectra}
\centering
\begin{tabular}{lrrrrrr}
\hline\hline
Annulus & c(H$\beta$)   & Dered       & $T_{\rm e}$ & $T_{\rm e}$ & $N_{\rm e}$ & He$^{+}$/H$^{+}$ \\
        &               & F(H$\beta$) & \ion{H}{I}  & \ion{He}{I} &             &  \\
        &               & (erg cm$^{-2}$ s$^{-1}$) & (K) & (K)   & (cm$^{-3}$) &  \\
\hline
 Core  & 0.31 & 1.19$\times$10$^{-11}$ &  8570 & 9150  & 2200 & 0.092$^{1}$ \\
 Low c & 0.21 & 8.55$\times$10$^{-13}$ &  9250 & 10070 & 1800 & 0.074$^{2}$ \\
 Halo  & 0.26 & 2.74$\times$10$^{-13}$ & 11100 & 11100 &  670 & 0.049$^{3}$ \\
 Total & 0.31 & 4.77$\times$10$^{-11}$ &  8550 &  9150 & 2410 & 0.092$^{4}$ \\
\hline
\end{tabular}
\tablefoot{
$^{1}$ He$^{+}$/H$^{+}$ based on \ion{He}{I} 6678\,\AA/H$\beta$ and 4922\,\AA/H$\beta$ 
of 3.68 and 1.25, (I(H$\beta$)=100.0), with mean He$^{+}$/H$^{+}$ from [\ion{Cl}{III}] and 
[\ion{S}{III}] $N_{\rm e}$ and $T_{\rm e}$, weighted 3:1 respectively. \\
$^{2}$ He$^{+}$/H$^{+}$ based on \ion{He}{I} 6678\,\AA/H$\beta$ and 4922\,\AA/H$\beta$ of 
2.85 and 1.02. \\
$^{3}$ He$^{+}$/H$^{+}$ based on \ion{He}{I} 6678\,\AA/H$\beta$ of 1.93 only and 
[\ion{Cl}{III}] and [\ion{N}{II}] $N_{\rm e}$ and $T_{\rm e}$. \\
$^{4}$ He$^{+}$/H$^{+}$ based on \ion{He}{I} 6678\,\AA/H$\beta$ and 4922\,\AA/H$\beta$ 
of 3.530 and 1.175. \\
The $T_{\rm e}$ values for \ion{H}{I} are based on the temperature sensitive ratio 
[\ion{N}{II}]5755/6383\,\AA\ and $N_{\rm e}$ values are a mean from the density sensitive 
ratios [\ion{S}{II}]6716/6731\,\AA\ and [\ion{Cl}{III}] 5517/5537\,\AA. \\
The He$^{+}$/H$^{+}$ number ratios have been derived from recombination theory after 
correcting the \ion{H}{I} line flux for collisional excitation contributions, using the 
formulae of \cite{KingdonFerland1995}.
}
\label{tab:nebcons}
\end{table*}

All four residual spectra with the nebular continuum subtracted in Fig. 
\ref{fig:ann_nebcons} show a continuum over all wavelengths with an 
increase to the blue. We consider a number of explanations for this extra 
continuum:

\begin{enumerate}
\item Scattered continuum. The increase to the blue is steepest for
the annulus over the nebula which is closer to the CSt. Thus
the source could be instrument scattered light from the CSt with some addition
of intra-nebular scattered light. The level of the residual blue continuum (mean over 
4750--4850\,\AA), is 8.0e-17 erg cm$^{-2}$ s$^{-1}$ \AA$^{-1}$ arcsec$^{-2}$) 
for the mean offset of 4.4$''$ of the nebular annulus from the CSt; when 
compared to the dereddened stellar continuum (Sect. \ref{CenStar} and Fig. 
\ref{fig:CSspec}) over the same wavelength range, this value is 
2.4$\times$10$^{-4}$ $''^{-2}$.
However for the low c annulus, the shape of the continuum is less blue than 
that of the dereddened CSt, tending to rule out scattering within the nebula 
of the starlight (where typical ISM dust produces a blueing of the spectrum)
in this region. The reddened residual continuum (i.e., the residual spectrum 
of the dereddened continuum minus the nebular continuum, then reddened by the 
line of sight extinction) can also be compared to the observed CSt spectrum 
and is a probe of instrumental scattering. For the core annulus, the
reddened residual also shows a contribution increasing to the blue, 
(neglecting the 2nd order contamination above $\sim$8500\,\AA) relative to the CSt
spectrum, but not so for the low c or outer annuli with larger offsets. 

These comparisons of continuum slope in the spectra of the two inner annuli 
with that of the CSt, are consistent with a decreasing level of scattering 
with offset from the CSt. For the outer annulus located over the 
halo (mean offset 2.1$\times$ that of the core annulus) however, the residual is 
redder than the CSt spectrum. Here the
contamination in this aperture from field stars is considerable (estimated
at 55\% at 5500\,\AA\ by removing all the star images from the annulus region
in the $v$ image (Fig. \ref{fig:v_40s}) and these tend to be red (c.f.
the star colours in Table \ref{tab:FieldStars} with those of the CSt, 
$b'-v'$ -0.39, $v'-r'$ -0.45, $v'-i'$ -1.43).

\item Enhanced nebular continuum. Enhanced 2-photon continuum has been
observed in shock excited Herbig-Haro objects \citep{Dopitaetal1982}, so
a possible origin to the residual continuum could be an increase of 
2-photon continuum over that predicted. However the residual continuum 
is flatter than a 2-photon continuum for all four regions, refuting this
suggestion.

\item Instrumental scattering. For the total spectrum of Tc~1 with the
nebular continuum subtracted, the residual continuum is blue and this
has also been seen in the total spectra of other PNe observed with 
MUSE -- for NGC~4361 \citep{Walshetal2024}, their Fig. 9, and from
analysis of the total spectrum of NGC~7009 \citep{Walshetal2018}. 
Comparing the shapes of the residual continuum for the total flux in the
MUSE field of view (excluding the CSt) between the three nebulae 
shows good agreement in spectral shape between NGC~4361 and Tc~1, whilst 
for NGC~7009 there is a divergent broad feature under the H$\beta$ and [\ion{O}{III}] 
4959 \& 5007\,\AA\ lines extending to $\sim$ 5500\,\AA, probably arising from the
wings of the grating line spread function around the very strong emission lines.
This comparison is detailed in Appendix \ref{App:CompResids} and emphasises the
striking similarity in the shape of the residual continua of these three PNe
observed with MUSE.

\end{enumerate}

The most likely explanation for the excess continuum would appear to be in terms 
of instrument scattered light within the MUSE optics. 
Presumably the total continuum (sum
of starlight and nebular continuum, faint wings of the grating line profile
around the strong H$\beta$, H$\alpha$ and [\ion{O}{III}] lines, and perhaps also 
integrated line emission) supplies the scattered light. In Appendix 
\ref{App:CompResids}, a comparison of the residual continuum after nebular
continuum subtraction in the three MUSE datasets (NGC~7009, 
NGC~4361 and Tc~1) analysed for these effects, is presented.

One feature of the spectrum of the annuli that distinguishes them is the
broad wing around H$\alpha$, extending over $\sim$500\,\AA, which appears to
be present in all three annulus spectra, and the total spectrum, but
not detectable at H$\beta$. This broad profile is symmetric about the 
approximate observed wavelength of H$\alpha$ of FWHM $\sim$380\,\AA\
but 12 times stronger relative to the flux in
H$\alpha$ in the outer annulus than the low-$c$ annulus, and 50 times lower
for the core annulus (where the line emission is highest). This broad 
emission is present in all areas of
the nebula but strongest in surface brightness in the core, consistent with
a scattering origin from the centrally peaked (3.2$''$ diameter) 
high emission surface brightness region.  
\cite{Alemanetal2019} also remark on broad components to the 
bright lines, but only over FWHM extents of $\sim$1000\,km s$^{-1}$, and these 
may be the low intensity wings of the grating line profile. The origin of the
broad feature is unclear: the electron density is quite low so electron scattering
seems an unlikely origin.      

\subsection{Critical further observations}
Tc~1 is distinguished, in particular, from other PNe mapped in 2D by the two features revealed 
from these comprehensive optical IFU observations: the low c(H$\beta$) annulus around the core
and the halo of enhanced 7281\AA\ emission above that expected for \ion{He}{I}. Other
spectroscopic observations would be very helpful to conclusively determine if scattering effects
within the MUSE instrument were not responsible for these results. Scattering effects were 
explicitly considered in Appendix \ref{App:PSF restoration} in the case of the low c annulus, 
but for the 7281\AA\ excess it is difficult to understand why only this line should be more prone 
to scattering than the nearby \ion{He}{I} 6678\AA\ line. For example, longer slit observations at
several positions across the core-halo region, with and without inclusion of the CSt 
in the slit, could independently confirm these results. In addition higher spectral resolution
would be an advantage, to determine if any velocity shifts, or complex emission line profiles
occur in the core-to-halo transition region. Also imaging and spectroscopy in the infrared,
as performed by Cami et al. (2026, in prep.) with JWST, can study the dust emission 
distribution to determine if the low c annulus represents a deficiency of dust, or a spatial 
change in dust grain properties.
 
\section{Conclusions}
Deep MUSE wide field observations of the fullerene emitting PN, Tc~1, have been presented
for the optical range 4750--9300\AA. Tc~1 has been has been shown to be in most respects 
a typical PN in its early phase of transition from the AGB to white dwarf, with modest 
ionization, a rather symmetric structure and a central star with an effective temperature 
between 31\,000 K and 34\,000 K (see discussion in Sec. \ref{CStype}). 
Its core is 
larger (full diameter 0.21 pc) than would be expected for a younger PN and its 
dynamical time (Sect. \ref{Morphology}) is longer than for the 2M$_{\odot}$
mass evolutionary track suggested by the ($T_{eff}$, $L$) value (Sect. \ref{CenStar}).
The filamentary halo is extended (56$''$, 0.96 pc) with a long dynamical
time (Sect. \ref{Morphology}) and may be the remnant of the AGB wind.
MUSE data cubes at 
three exposure levels were analysed to provide combined unsaturated images in emission 
lines covering the range of surface brightness from H$\alpha$ to faint lines 
($<$ 10$^{-3}$ H$\beta$). These optical observations are highly complementary to recent 
JWST mid-infrared imaging spectroscopy with MIRI and NIR imaging with NIRSPEC (Cami 
et al, 2026 in prep.; Schuylenberg et al. 2026, in prep.).

Emission line images extracted from the datacubes are presented. The spectra of
33 field stars over the area of the faint halo affected the extinction determination from 
H$\alpha$/H$\beta$. Template spectra were fitted to these stars to subtract their spectra
and improve the fidelity of local extinction determination. Collisionally
excited line diagnostics for $T_{\rm e}$ and $N_{\rm e}$ were imaged; optical
recombination line diagnostics for Paschen Jump $T_{\rm e}$ and \ion{He}{I} $T_{\rm e}$
were also presented.

Three main morphological features of Tc~1 emerge from this study: the bright core, a 
periphery with lower surface brightness surrounded by the faint halo. The core shows a 
2$''$ elliptical ring around the CSt and groups of brighter knots of lower
ionization to the NW and SE. These knots have been shown by \cite{Bouvisetal2025} to
be similar to LIS found in other PNe and now revealed to have offset velocities from 
the systemic (Sect. \ref{Morphology}). 
The transition region between core and halo has lower $N_{\rm e}$ than in the core, 
and also higher $T_{\rm e}$, the latter probably 
on account of radiation hardening, and also shows two remarkable properties: a 
ring of lower extinction, which, when corrected by the larger scale interstellar 
extinction, implies negative extinction for a standard reddening law. 
In this region also the $T_{\rm e}$ sensitive 
\ion{He}{I} 7281/6678\AA\ ratio shows an increase above the maximum value allowed from 
radiative transfer: the presence of a contaminating line was investigated,
but no convincing candidate species could be established. 
The origin of this anomalous \ion{He}{I} line ratio therefore remains unanswered from 
these observations. 

In order to confirm the results from the spaxel images, four integrated regions (core, 
low c annulus, halo and total nebula), were defined and diagnostics from their high S/N 
spectra were analysed. These spectra allowed detailed examination of the continuum and 
in all cases an excess above the nebular continuum (bf, ff and 2$\nu$) was found. This 
effect has also been seen in previous MUSE spectra of PNe with bright central stars 
and instrumental scattering of starlight is suspected. However the shape of the excess
continuum does not match the CSt spectral shape. The spectrum of the bright CSt of Tc~1 
was extracted from the cubes 
and then dereddened over the 4750-9300~\AA\ range using different values of c(H$\beta$). 
Each of the dereddened continua was compared with LTE model atmospheres with differing 
effective temperatures and surface gravities. It was found that a CSt 
extinction of c(H$\beta$) = 0.385  and an LTE model with T$_{\rm eff}$ = 31,000~K and 
log~g = 3.5 minimised the differences between the continuum slopes of the dereddened 
central star and the model atmospheres. This effective temperature is within the range 
found by other investigators, while the deduced extinction towards the CSt is 
consistent with the increase in reddening towards the centre that was found in the 
extinction map (Fig. \ref{fig:c_map}).

Given that the annulus of low c values occurs just outside the bright core where the
nebular surface brightness decreases rapidly, an effect of the point spread function 
on extinction determination could be suspected and an investigation with image 
restoration at H$\beta$ and H$\alpha$ wavelengths was performed. While a small effect 
on the extinction could be induced, the depth of the low c annulus could not be 
reached, leaving the conclusion that this is a real effect. Two suggestions for this 
strong spatial change in c(H$\beta$) were discussed:
scattering by small dust grains in the nebula causing 'blueing' of the spectrum, hence
inducing increased H$\beta$/H$\alpha$ ratios; or a local change of the ratio of total 
to selective extinction, $R_{V}$, of the dust in this region. The former 
was investigated by a simple 
3D dust scattering model and at least a part of the drop in c(H$\beta$) could be 
thus explained. The latter effect, an increase in $R_{V}$ due to an increase in mean grain 
size, would appear to be the more promising in order to explain the apparent drop in 
c(H$\beta$) in the low c annulus.
The decline in c(H$\beta$) values occurs just beyond the peak of the fullerene 
C$_{60}$ emission rim studied by JWST MIRI spectra (Geise et al. 2026, in prep.)
suggesting a possible connection between the dust extinction and fullerene emission.
A change in the dust grain properties in the vicinity of the fullerene peak 
should be further investigated, such as through UV and near-infrared spectra to
determine the nature of the local reddening law or high spectral resolution 
observations to determine effects of an unusual velocity field or the presence of 
shocks.

MUSE again reveals the complexity of the extinction in a PN (c.f. NGC 7009,
\cite{Walshetal2018}), here complicated by the annulus of low c values. Given that 
c(H$\beta$) measures the line-of-sight extinction integrated along the column of 
emission, and is partly dependent on the emission distribution, the true dust structure 
is likely to be more complex than the image (Fig. \ref{fig:c_map}) suggests. Further 
optical -- infrared observations are required to better understand the unusual 
features of Tc~1 revealed in this study and if they 
are indicative of the presence of fullerenes.

\begin{acknowledgements}
Based on observations collected at the European Southern Observatory under ESO programme
105.20R7.001. We thank the VLT observing support team and telescope operators for enabling 
these fine observations.

We are grateful for discussions with Peter Weilbacher and Roland Bacon on the
possible origin of the residual continuum in the spectra of several PNe, as described 
in Appendix \ref{App:CompResids}. \\

We thank the referee, Joel Kastner, for his discerning comments that led 
to a sharpening of focus of the paper. \\

JC and EP acknowledge support from an NSERC Discovery Grant.

This research has made use of NASA’s Astrophysics Data System Bibliographic Services. 

This work has made use of data from the European Space Agency (ESA) mission
\textit{Gaia} (\url{https://www.cosmos.esa.int/gaia}), processed by the \textit{Gaia}
Data Processing and Analysis Consortium (DPAC,
\url{https://www.cosmos.esa.int/web/gaia/dpac/consortium}). Funding for the DPAC
has been provided by national institutions, in particular the institutions
participating in the Gaia Multilateral Agreement. \\

This research has greatly benefited from the use of PyNeb \citep{Luridiana2015}, 
for the nebular diagnostics. We are grateful to the communities who have developed the 
Python packages used in this research, such as MPDAF \citep{Piquerasetal2019}, 
Astropy \citep{AstropyCollaboration13, AstropyCollaboration18,AstropyCollaboration22}, 
numpy \citep{Walt11}, scipy \citep{Jones01} and matplotlib \citep{Hunter07}.
\end{acknowledgements}

\bibliographystyle{aa}
\bibliography{Tc1}

\clearpage

\newpage

\begin{appendix}

\nolinenumbers

\section{Procedure for fitting continuum of field stars}
\label{App:Fitting of field stars}
Given the complexity of the spectra of the generally late-type field stars, and the 
dangers in fitting high order functions to the continuum in order to 
measure the emission line fluxes, it was decided to use spectra of observed stars. 
The requirements are a library of spectra covering a wide range of spectral types well-
sampled by sub-type, and at a spectral resolution, higher or close to that of MUSE. The   
SDSS-IV MaStar spectral library \citep{Yanetal2019} was selected as the most appropriate, 
being large and based on observed spectra rather than a model library. The spectral coverage 
(3622--10354\,\AA) is wider than MUSE, the resolution at R$\sim$ 1800 close to that of MUSE; 
the library is extensive with a broad range of spectral types; the 1st release with 3321 star 
spectra was employed. The full MaStar release (24130 stars) is described in 
\citet{Abdurroufetal2022}, but only the first release was used in this work.

The magnitudes of the stars in the MaStar database were determined in four filters, chosen 
carefully to avoid stronger emission lines so that the comparison to magnitudes of stars 
on the MUSE cube would be minimally biased by the presence of emission. The 
intermediate width, square profile, filter bands were: $b'$ 4712.5 -- 4850.0\,\AA; 
$v'$ 5400 -- 5600\,\AA; $r'$ 6320 -- 6520\,\AA;  and $i'$ 7900 -- 8100\,\AA. 
Then the colours ($b' - v'$, $v' - r'$, $v' - i'$) of the field stars in the MUSE 
field over the area of the halo were compared to the same set
of synthetic magnitudes for all the stars in the MaSTar database in order to 
select a MaStar star, or stars, matching as closely as possible in spectral shape to
the observed stars. 29 stars were identified in the halo as influencing the determination of
c(H$\beta$) from comparison of H$\alpha$/H$\beta$ to Case B value: they are indicated on Fig. 
\ref{fig:v_40s}. The magnitudes of these stars were measured by simple aperture photometry 
using IRAF\footnote{IRAF is distributed by the National Optical Astronomy 
Observatories, which are operated by the Association of Universities for Research
in Astronomy, Inc., under cooperative agreement with the National
Science Foundation.} $imexam$ (photometric aperture 0.6$''$ radius with a background annulus
1.2--1.8$''$). Comparing the $v'$ magnitude of the brightest field stars with
the Gaia G magnitude and correcting to Johnson $V$ magnitude using the colour terms from 
\footnote{Carrasco (2021): \url{https://gea.esac.esa.int/archive/documentation/GDR2/Data_processing/chap_cu5pho/sec_cu5pho_calibr/sec_cu5pho_PhotTransf.html} }
enabled the indicative $V'$ mags of the stars to be estimated (neglecting colour terms 
and extinction). The positions of these stars, relative to the position of 
the CSt, their $V'$ mags and the ($b' - v'$), ($v' - r'$), ($v' - i'$) colours are listed 
in Table \ref{tab:FieldStars}; these stars range over $\Delta$ $V'$ of 4.2 mags. \footnote{A
further field star was found projected on the bright core (offsets 
$\Delta$$\alpha$$+$2.4, $\Delta$$\delta$$+$3.5 $''$) when estimating the contribution 
of all field stars to the continuum for the aperture over the halo (see Sect.   
\ref{Subsec: Regions}). In the $V'$ band this star had an estimated mag. of 19.7,
and contributed only 17\% to the local $V'$ surface brightness. Its effect on emission line 
measurements was not detectable and no feature is seen in the c(H$\beta$) extinction 
image (see Sect. \ref{SubSec: Av} and Figure \ref{fig:c_map}) at this position.}

\begin{table}
\caption{Position and photometry of interfering field stars}
\centering
\begin{tabular}{lrrrrrr}
\hline\hline
Star   & $\Delta$ $\alpha$ & $\Delta$ $\delta$ & ~V$'$   & ($b' - v'$) & ($v' - r'$) & ($v' - i'$) \\
       & ($''$)            & ($''$)            & mag. &  ~~mag.     & ~~mag.      & ~~mag.     \\
\hline
1  &   5.9 & -18.3 & 16.37 &  0.00 & -0.08 & -0.37 \\
2  &  21.7 &  -3.1 & 17.43 &  0.01 & -0.07 & -0.34 \\
3  &  23.1 &  17.4 & 16.28 &  0.05 & -0.04 & -0.28 \\
4  &  18.8 &  14.5 & 19.89 &  0.01 & -0.08 & -0.33 \\
5  &  25.1 &   8.1 & 17.65 &  1.52 & -1.59 & -1.82 \\
6  &   6.0 &  26.0 & 18.42 & -0.41 &  0.57 &  0.53 \\
7  &   6.1 &  27.1 & 17.85 &  0.19 & -0.03 & -0.10 \\
8  &  -8.1 &  23.7 & 17.66 &  0.10 &  0.00 & -0.16 \\
9  &   8.4 &   3.0 & 17.10 &  0.03 & -0.09 & -0.32 \\
10 &  15.2 &  23.0 & 19.10 &  0.09 & -0.05 & -0.24 \\
\hline
\end{tabular}
\tablefoot{
Column 4 lists the $V'$ mag., corrected from the measured 5400-5600\AA\ 
narrow band magnitude by the G mag. of several stars from Gaia; see text 
for details. \\ 
Only the first 10 entries of this table are shown; the full table is 
contained in the on-line material.
}
\label{tab:FieldStars}
\end{table}

The closest sets of matches to the ($b' - v'$), ($v' - r'$) and ($v' - i'$) colours of
the field stars were selected from the MaStar database and the spectra retrieved. For an
oversized area around each field star in Table \ref{tab:FieldStars} (typically 100 spaxels but 
depending on the brightness of the star), a MaStar spectrum was separately scaled to the 
continuum on both sides of the H$\alpha$ and H$\beta$ lines for each spaxel and subtracted 
from the observed spectrum. The resulting absorption line subtracted spectrum was then refitted 
by a Gaussian and the revised H$\alpha$ or H$\beta$ line flux inserted into the flux map. This
process required considerable oversight, as the indicative spectra from the colour matches
often did not clearly match the obvious H$\beta$ and H$\alpha$ absorption line shape and 
depth, and in some cases
only one of H$\alpha$ or H$\beta$ would match well for a given MaStar spectrum. In these 
cases fits by stars with similar ($b' - v'$), ($v' - r'$) and ($v' - i'$) colours were
trialed until a better match to the absorption line shape around the H$\alpha$ or H$\beta$ 
emission lines was found. This procedure
could be automated but it is clear that many spectra may be required to locate a good match and
a minimization algorithm would be needed to find `best' matches (of course within the limits of 
the available spectra in the database and differences in reddening for the database star 
and the observed field star). 

Applying this correction procedure for the effect of field star spectra, most of the apparent 
emission or absorption features in the H$\beta$ and H$\alpha$ images were removed so that no 
excursions in extinction from the local mean were observed at these positions. Since the field 
star absorption around H$\alpha$ could sometimes be broad, the [\ion{N}{II}] 6548, 6583\,\AA\ 
and [\ion{S}{II}] 6716, 6731\,\AA\ lines were also refitted in the corrected
spectra; however this step was only found necessary for the set of 595\,s exposures, as the effect 
of the star continuum fit for the shorter exposures was not significant for the flux images.

\section{Reality of the extinction structure -- image restoration}
\label{App:PSF restoration}

In order to obtain accurate quantitative images of the nebula at different 
wavelengths with the observed PSF deconvolved, a high fidelity PSF sampled at 
those wavelengths is required. This is feasible
with a grid of well-exposed and separated stars taken in the same field
(or possibly an adjacent field immediately pre- or post-ceding the nebula
target exposures) but the available set of images of Tc~1 cannot hope to achieve this. 
Models of the MUSE PSF (as parameters of a Moffat \citep{Moffat1969} profile) 
can be made using the $psfr$ task \citep{Feticketal2019} based on the seeing
and AO parameters of an exposure read from the image header. However given that the
MUSE cubes were formed from multiple exposures, all with differing atmospheric
and AO settings, a combined model of the PSF is required.

The closest fidelity to an image-based PSF is to use a selected star, or stars, 
in the image of interest, such as shown in Fig. \ref{fig:v_40s}. The brightest star 
is the CSt itself, but lies on a field of bright structured emission, so even 
selecting narrow wavelength ranges away from bright emission lines, nebular 
continuum and faint lines are always present. Choice of a suitable field star 
is limited by the nebular extent, and the brightest field star (coincidentally 
not situated on the nebula halo) is the one marked X in Fig. \ref{fig:v_40s} 
with V mag ~14.4 (see Table \ref{tab:FieldStars}) but this star is close to the 
field edge so the PSF wings cannot be sampled to large radii. Clearly any 
deconvolution will be subject to 
uncertainties, in addition to the typical ringing around point sources and noise
amplification induced by Fourier or iterative deconvolution methods. Thus we set 
a modest goal of trying to determine if the extinction deficit in an annulus around 
the bright core could be returned to a value close to that for the halo (and of 
the ISM) by restoring H$\alpha$ and H$\beta$ images and examining their ratio on the 
restored images. 

\begin{figure}[t]
\centering
\resizebox{\hsize}{!}{
\includegraphics[trim={0.8cm, 0.2cm, 1cm, .8cm},clip]{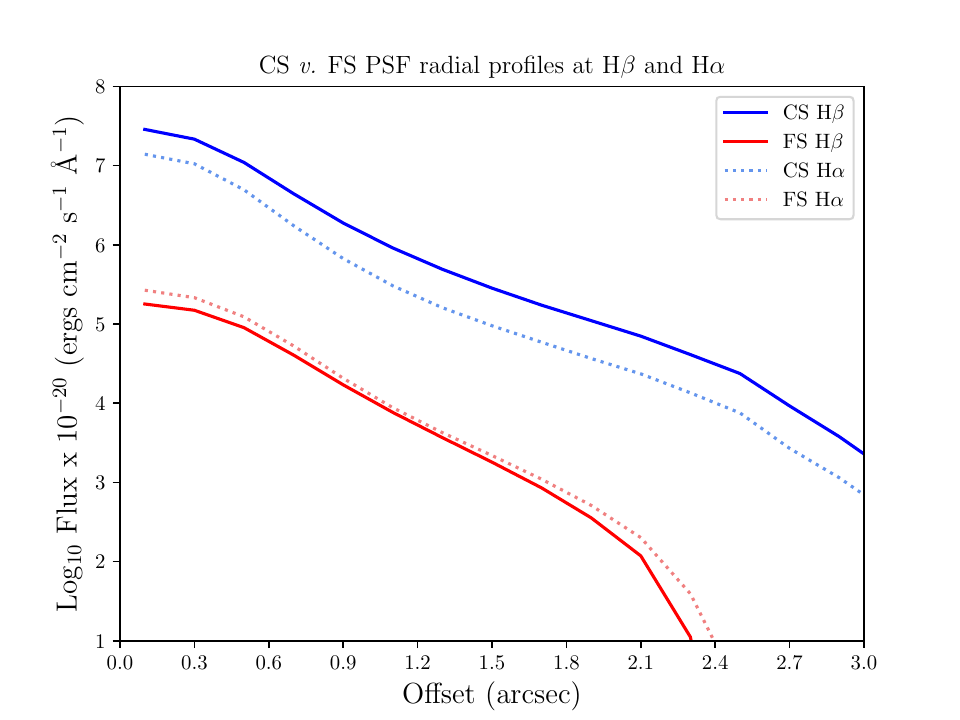}
}
\caption{Radially averaged profiles of the observed point spread functions 
for the central star (blue lines) and the field star, X (red lines), at
H$\beta$ (continuous lines) and H$\alpha$ (dotted lines).
Log$_{10}$ flux (spaxel$^{-1}$) is plotted against offset from the centre of
the peak spaxel.}  
\label{fig:PSFs}
\end{figure}

PSF's for the field star X and the CSt were made by extracting images from the 40\,s cube
(on the 101 and 595\,s cubes the CSt is saturated) at wavelengths blueward and redward 
of each line (for H$\beta$: 4820--4850 and 4870--4900\,\AA; and for H$\alpha$: 
6515--6545 and 6585--6615\,\AA\ avoiding the [\ion{N}{II}] 6548, 6583\,\AA\ lines) and 
averaging the pairs to form single H$\beta$ and H$\alpha$ PSFs. In the case of the 
PSF for the CSt an attempt 
was made to remove the nebula contribution by subtracting a scaled version of the
emission line map matched in mean level to the extended emission; however the
attempt did not result in far PSF wings with mean zero signal probably on account 
of the local structure of the ionized emission. Fig. \ref{fig:PSFs}
show the radially averaged PSFs of the CSt (blue lines) and the field star X 
(red lines) demonstrating the large wings to the CSt and the curtailed wings 
to the (fainter) field star (also on account of its proximity to the image edge). 
The CSt and field star PSFs show similar shape at the same wavelength to 2.0$''$
radius but for H$\alpha$ the CSt has a notably narrower PSF than at H$\beta$, 
dissimilar to the field star. 2D Gaussian and Moffat fits to these stars are 
listed in Table \ref{tab:PSFfits}.

\begin{table}
\caption{Fits to Point Spread Function stars}
\centering
\begin{tabular}{lrrrr}
\hline\hline
\ion{H}{I} line/      & Gaussian     & Moffat   & Moffat  & Moffat    \\ 
star                  & FWHM ($''$)  & $\alpha$~~~ & $\beta$~~~ & FWHM($''$) \\
\hline  
H$\beta$ & & & & \\
Tc~1 CSt      & 0.536 & 0.546 & 3.30 & 0.528 \\ 
Field Star X & 0.568 & 0.524 & 2.78 & 0.558 \\
\hline
H$\alpha$ & & & & \\
Tc~1 CSt      & 0.486 & 0.638 & 5.20 & 0.482 \\ 
Field Star X & 0.558 & 0.679 & 4.53 & 0.552 \\
\hline
\end{tabular}
\label{tab:PSFfits}
\end{table}

\begin{figure*}[t]
\centering
\resizebox{\hsize}{!}{
\includegraphics[trim={1cm 0.2cm 1.8cm 0.9cm},clip]{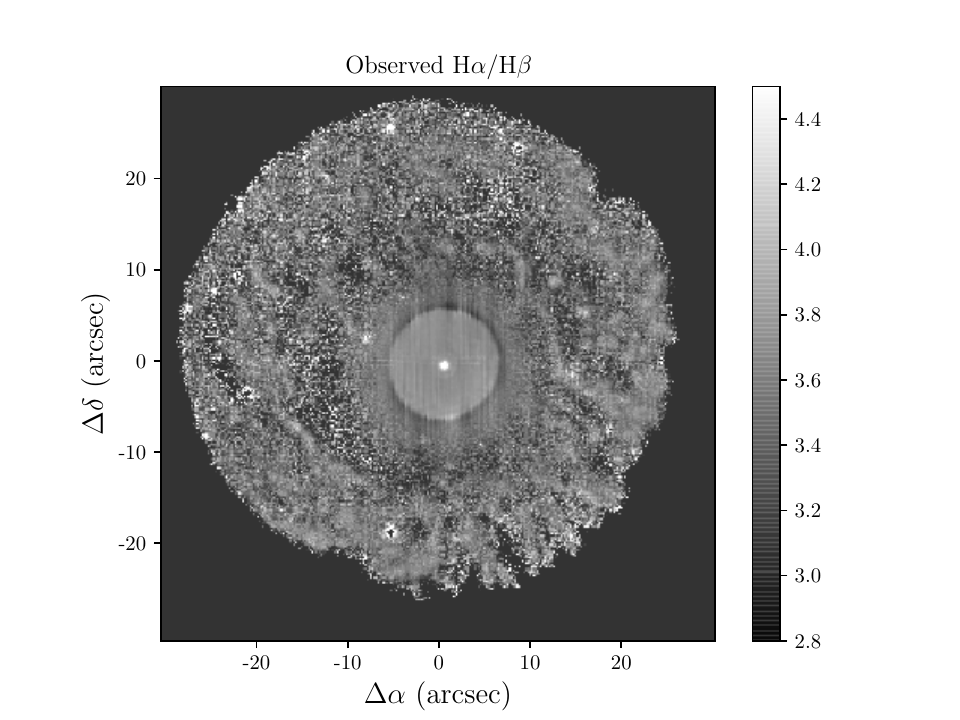}
\hrulefill
\includegraphics[trim={1cm 0.2cm 1.8cm 0.9cm},clip]{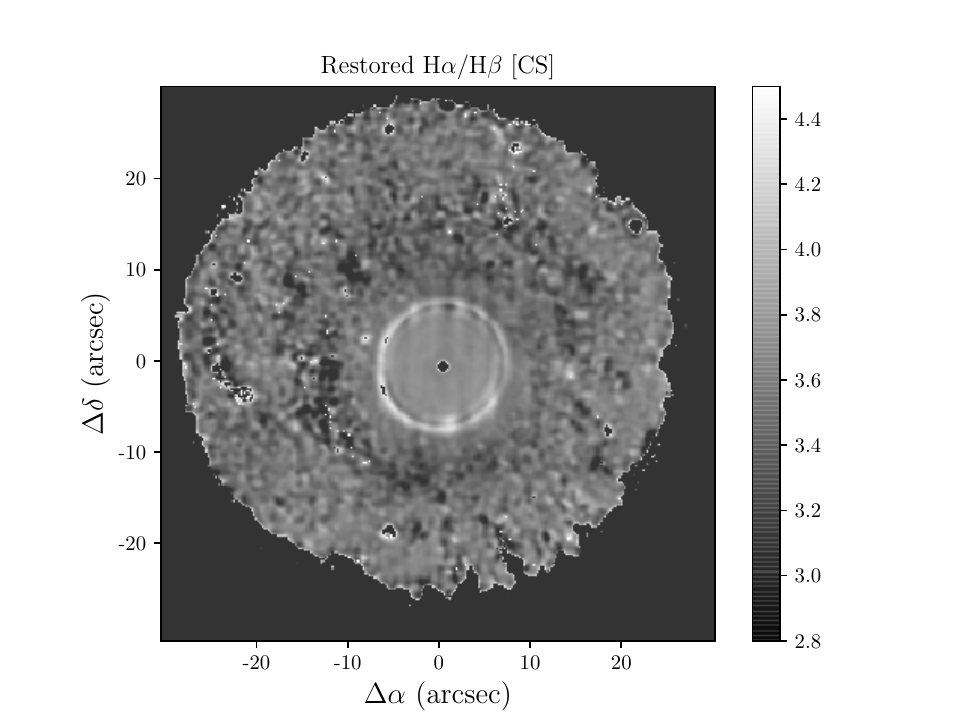}
\hrulefill
\includegraphics[trim={1cm 0.2cm 1.8cm 0.9cm},clip]{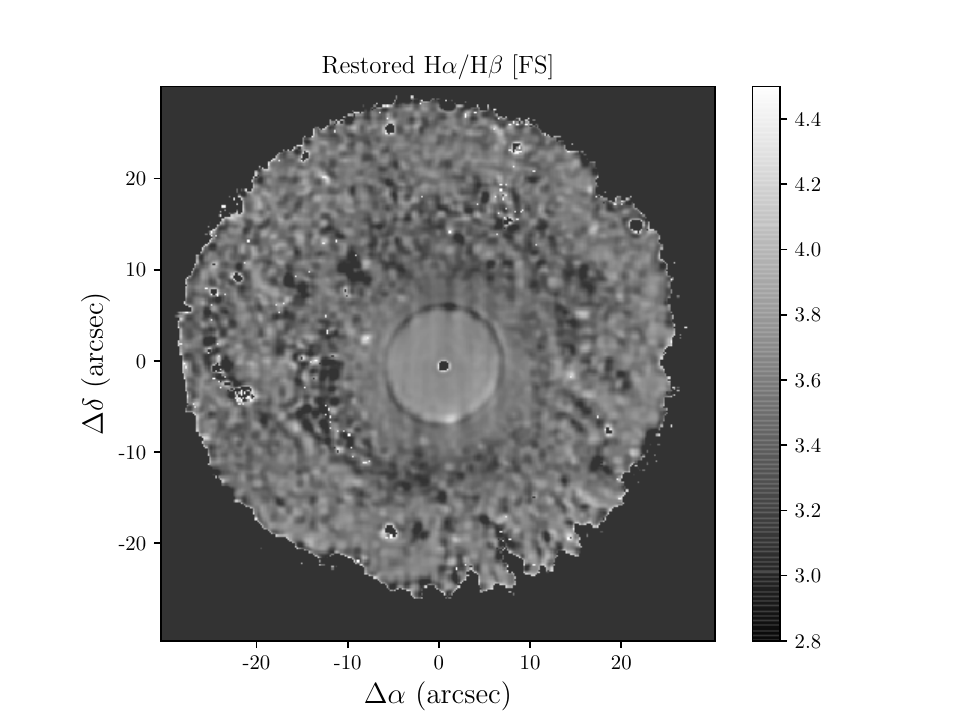}
}
\caption{Images of the H$\alpha$/H$\beta$ ratio of the Lucy restored H$\alpha$ and 
H$\beta$ images, using the central star [CSt] (centre) and the field star X [FSt] 
(right) as PSFs. The image of the observed H$\alpha$/H$\beta$ ratio is shown at left.
The axes display offsets with respect to the position of the CSt. For comparison, an 
H$\alpha$/H$\beta$ ratio of 3.40 corresponds to c(H$\beta$) = 0.237 for
$T_{\rm e}$ = 10\,000\,K and $N_{\rm e}$ = 2000 cm$^{-3}$.
}
\label{fig:restored}
\end{figure*}

These PSF images were used to restore the H$\beta$ and H$\alpha$ emission
line images using Richardson-Lucy (\cite{Richardson1972}, \cite{Lucy1974}) iterative
restoration\footnote{Employing Python SciKit: skimage.restoration.richardson\_lucy}. 
Since the deconvolved PSFs will not have the same width on the H$\beta$ and H$\alpha$ 
restored images, after the iterative restoration the images were convolved with a 
two-dimensional Gaussian of the same width (FWHM 0.6$''$, or 3 pixels) before forming 
the H$\alpha$/H$\beta$ ratio image. Fig. \ref{fig:restored} shows the comparison of
the H$\alpha$/H$\beta$ images restored from the original images employing 10
iterations with PSF's from the central star and the field star, X. A mask was applied 
to the noise-amplified spaxel values at the outer edge of the halo. 

The H$\alpha$/H$\beta$ ratio image restored with the CSt shows two extra rings which are 
not present on the original H$\alpha$/H$\beta$ ratio image (compare the extinction map 
Fig. \ref{fig:c_map}), and would appear to be restoration features presumably induced by 
the narrow CSt H$\alpha$ profile (see Fig. \ref{fig:PSFs}). An alternative explanation
could be that the restoration produces a thick ring of higher H$\alpha$/H$\beta$ which is 
not enough to eradicate the low ratio annulus entirely. The low extinction ring is 
partially filled in by these two bright rings but is still visible. In contrast, for the 
result restored with the FS, the low extinction ring in H$\alpha$/H$\beta$ remains but 
the ratio is slightly enhanced but not by enough to remove it. The mean c(H$\beta$) values 
in a circular annulus of 1.6$''$ width (inner radius 6.0$''$) centred on the CSt 
show that the FS restored image produces a value only 0.01 dex. higher than the observed 
({\it viz} unrestored) extinction map but the mean value for the image restored by the 
CSt is 0.14 dex. higher. Although these experiments are
not exhaustive they demonstrate that the annulus of low extinction cannot be completely
removed to match the extinction value to that of the halo, arguing that it is a real 
feature of the nebula dust distribution. An additional test was to 
restore the Paschen 9 (9229\,\AA) image and compute the ratio to the restored H$\beta$ 
image; for both restorations with CSt and FS, the results showed that the low P09/H$\beta$
annulus visible in the observed ratio image was not removed and indeed deepened. However       
the MUSE PSF is broader and has extensive structure in the wings at this long wavelength; 
instrument scattered light from the CSt is visible and produces bars in the
cardinal directions resulting from the combination of two instrument rotations.
 
\section{Modelling of gas$+$dust structure and dust extinction/scattering}
\label{App: ScatMod}

A simple spherically symmetric nebula model was constructed to test how
dust inside the nebula volume could affect the spatial variation of 
extinction and, in particular, if a shell of apparent extinction deficit
could feasibly be explained. The model consists of a gas volume (producing 
emission) and a dust volume (responsible for extinction and scattering); the 
latter can be independent of the former, so not necessarily confined to fixed 
dust/gas ratio of the gaseous emission volume. Models for H$\beta$ and H$\alpha$ 
were constructed so that extinction maps of the model nebula could be formed for 
comparison with the observed extinction map (Fig. \ref{fig:c_map}). Computations
were performed in Python on a 200$\times$200$\times$200 cartesian grid.

The core nebula was simulated by a sphere with a central empty zone (radius
0.35$''$) and two density regimes: inner to radius 2.75$''$, density 2200 cm$^{-3}$
and outer, radii 3.0 -- 6.0$''$, density 2300 cm$^{-3}$, The two regimes have
differing filling factors (FF), tuned to match the dereddened average H$\beta$ 
surface brightness of Tc~1 from the extinction corrected 40\,s cube flux map. The 
halo extended from radius 6.25 -- 24.75$''$ with a density 700 cm$^{-3}$ (from 
the nebula coadded region analysis, Sect. \ref{Subsec: Regions}). The 
justification for a very low filling factor for the halo is apparent from the 
very patchy emission (Fig. \ref{fig:EmLines1}). The quantitative details of the 
model components are 
listed in Table \ref{tab:ScatModfits}. For each region a gas/dust mass (G/D) ratio 
was adopted, based on \cite{Uetaetal2014}. The large scale dust was assumed to 
be amorphous carbon with optical constants from \cite{RouleauMartin1991} and an 
MRN \cite{Mathis1977} size distribution ($\propto a^{-3.5}, a_{min} 0.005, 
a_{max} 0.25$) with mass density 1.8 g cm$^{-3}$.    

Two dust shells were selectively added, an inner one (4.75 -- 5.75$''$) -- Dust 
Shell A -- and an outer one (5.0 -- 6.75$''$) -- Dust Shell B -- both with unity 
filling factors (see Table \ref{tab:ScatModfits}). Dust Shell B was selected only to
scatter radiation, whilst Dust Shell A could both scatter and extinguish
incident/transmitted radiation. In order to explore the possibility that C60 
could participate in the scattering of the Dust Shell B, optical constants from
\cite{DattaKumar2009} and \cite{Renetal1991} from analysis of C60 thin films 
were employed\footnote{Adopted ($n, -k$): H$\beta$ (1.999, 0.794); H$\alpha$ 
(2.205, 0.796).} assuming the C60 molecule acted as a 'grain' with a size of 
0.001$\mu$m. Amplitude functions for 
scattered light appropriate to these particles were generated at the wavelengths 
of H$\beta$ and H$\alpha$ using Mie theory (e.g. \cite{vandeHulst1981} as a 
function of scattering angle as tables, which were read by the Python script to 
model the 3D nebula. Also, the extinction cross section, $C_{ext}$, was computed from 
Mie theory at the wavelengths of H$\beta$ and H$\alpha$ for each dust species. For 
the amorphous carbon (AMC) MRN size distribution, all quantities of interest were 
integrated over the size range.

Each model was run for Case B H$\beta$ or H$\alpha$ emissivity to convert density 
voxels to emission flux with extinction by the line of sight dust from the emission
voxel to a distant observer. Scattered emission was summed for each dust voxel 
taking account of the local emission, appropriately extinction corrected for the 
summed dust path from the emission voxel to the dust voxel; the scattered flux 
in the direction to the observer was also extinction corrected for its dust 
column. The column of voxels in the
direction of the line of sight to the observer was then summed to produce maps
of e.g. H$\beta$ intrinsic and scattered emission for comparison to those observed.
From the known distance, and by adjusting the filling factor (col. 6 of Table
\ref{tab:ScatModfits}), the projected emission flux could be adjusted to match the
average values over the core and halo of Tc~1 to the dereddened images (c.f.,
Sec. \ref{SubSec: Av})). Extinction, c(H$\beta$), maps both with and without the 
scattering component included were 
produced from ratioing H$\alpha$/H$\beta$ and comparing the projected spaxel value 
to Case B. The difference of these two images showed the resultant effect of 
dust scattering on the measured extinction.

\begin{table}
\caption{Extinction/scattering model: Gas and dust volumes}
\centering
\begin{tabular}{lrrrrll}
\hline\hline
Component    & Inner  & Outer  & Density     & G/D   & FF   & Dust    \\ 
             & ($''$) & ($''$) & (cm$^{-3}$) & ratio &      & species  \\
\hline  
Core inner   &  0.35  & 2.75   & 2200        & 200   & 0.70  & AMC \\
Core outer   &  3.00  & 6.00   & 2300        & 200   & 0.80  & AMC \\
Halo         &  6.25  & 24.75  & 700         & 200   & 0.001 & AMC \\
             &        &        &             &       &       &      \\
Dust shell A & 4.75   & 5.75   & 5000        & 200   & 1.00  & AMC \\
Dust shell B & 5.00   & 7.00   & 5000        & 140   & 1.00  & AMC \\
             &        &        &             &       &       &      \\
Dust shell B & 5.00   & 6.75   & 1.0$\times$10$^{8}$ & 140   & 1.00  & C$_{60}$ \\

\hline
\end{tabular}
\tablefoot{Voxel size 0.25$''$, 4.28$\times$10$^{-3}$ pc (200$^{3}$ 
voxel grid); \\
H Balmer line emissivities calculated for $T_{\rm e}$ 9000\,K, 
$N_{\rm e}$ = 2500 cm$^{-3}$; \\
The fractional mass per H atom of 1.501 based on element abundances from 
Aleman et al. (2019); \\
The number of electrons per H atom of 1.11 was computed from the sum of 
the ionic abundances (Aleman et al. 2019); \\
AMC = amorphous carbon.
}
\label{tab:ScatModfits}
\end{table}

Many models were run slightly altering the dimensions of the components in Table 
\ref{tab:ScatModfits}, their densities, filling factors and G/D ratios. The
dust shells were explored by selectively turning off their scattering and or 
extinction of the emission. The power law of the AMC grains was also adjusted 
for lower or higher $a_{min}$ and $a_{max}$ to simulate a smaller or larger 
MRN mean size. With AMC grains and C60, the wavelength dependence of
the scattering properties (viz. grains smaller than the wavelength of light 
being scattered) naturally led to higher scattered flux at H$\beta$ relative 
to H$\alpha$). But none of the extensions of the dust properties proved capable 
of producing such a large decrease in apparent extinction in the rim around 
the Tc~1 bright core, as observed in Fig. \ref{fig:c_map}, maximum 0.07 dex. For 
example increasing the density of AMC scattering grains in DustShell B did not 
necessarily lead to larger scattered flux towards the observer, as the extinction 
along the line from the emission voxel to the scattering voxel reduces the 
incident radiation on the scattering voxels. 

\begin{figure}[t]
\centering
\resizebox{\hsize}{!}{
\includegraphics[trim={1.1cm 0.2cm 2.2cm 0.8cm}, clip]{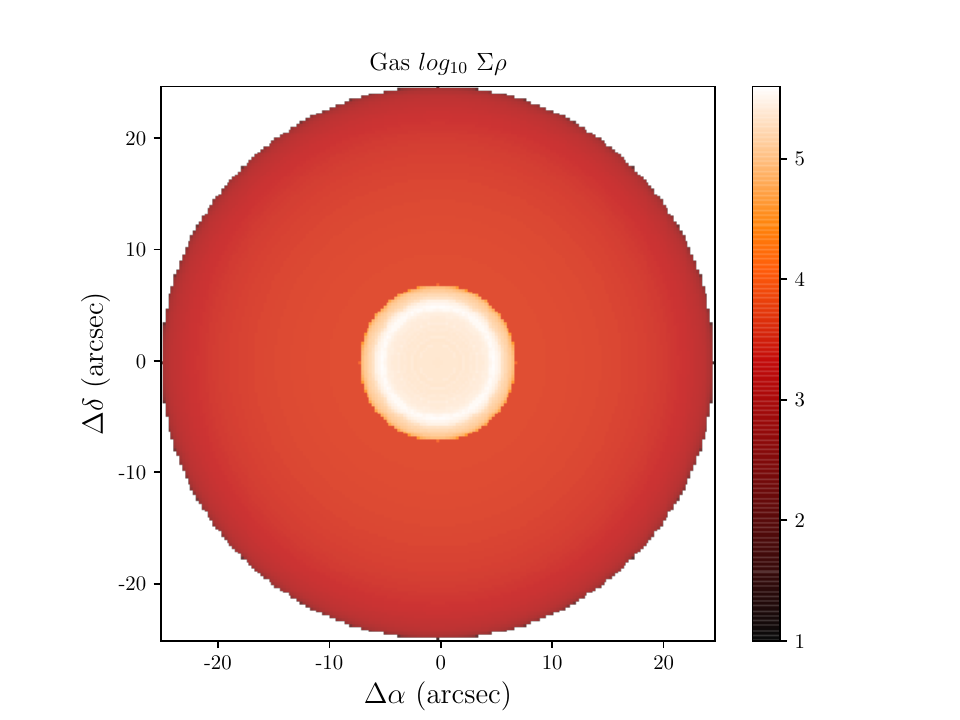}
\hrulefill
\includegraphics[trim={1.1cm 0.2cm 2.2cm 0.8cm}, clip]{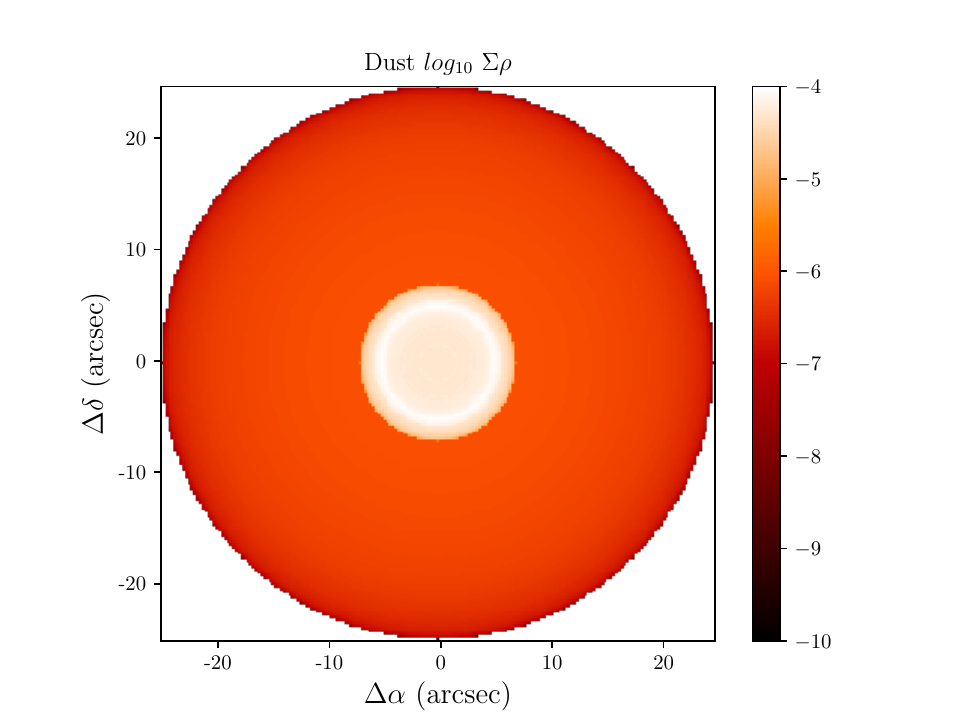}
}
\resizebox{\hsize}{!}{
\includegraphics[trim={1.1cm 0.2cm 2.2cm 0.8cm}, clip=]{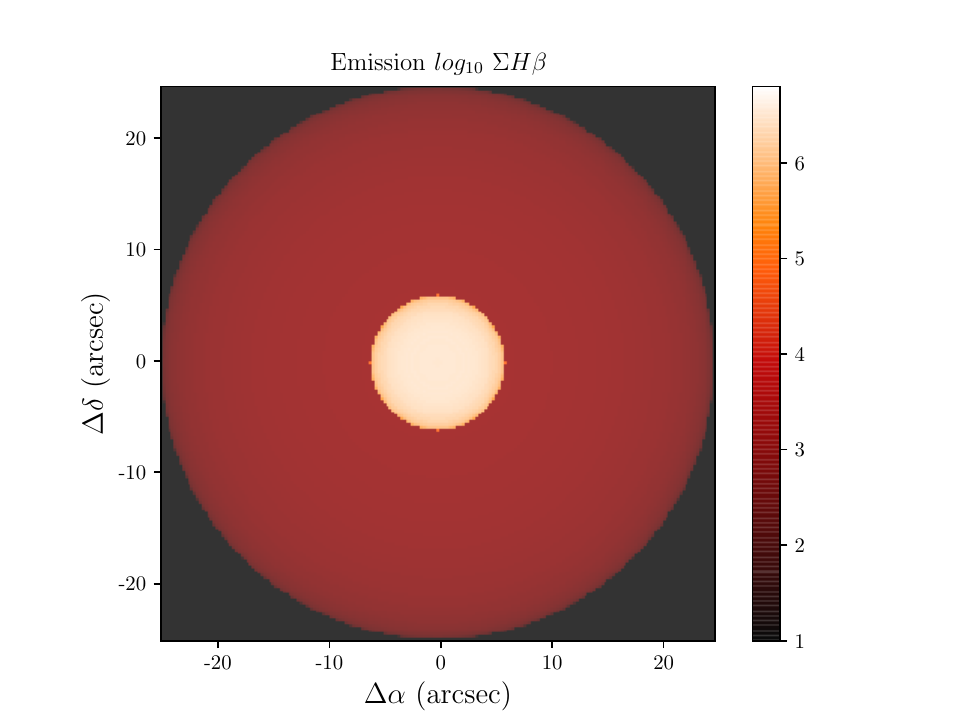}
\hrulefill
\includegraphics[trim={1.1cm 0.2cm 2.2cm 0.8cm}, clip=]{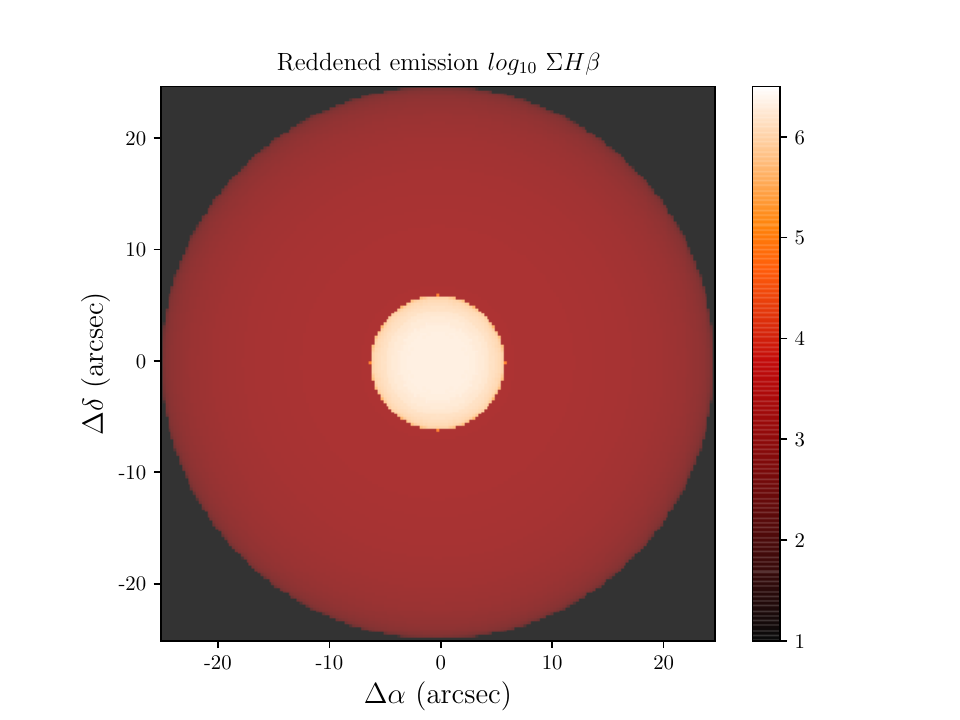}
}
\resizebox{\hsize}{!}{
\includegraphics[trim={1.1cm 0.2cm 2.2cm 0.8cm}, clip]{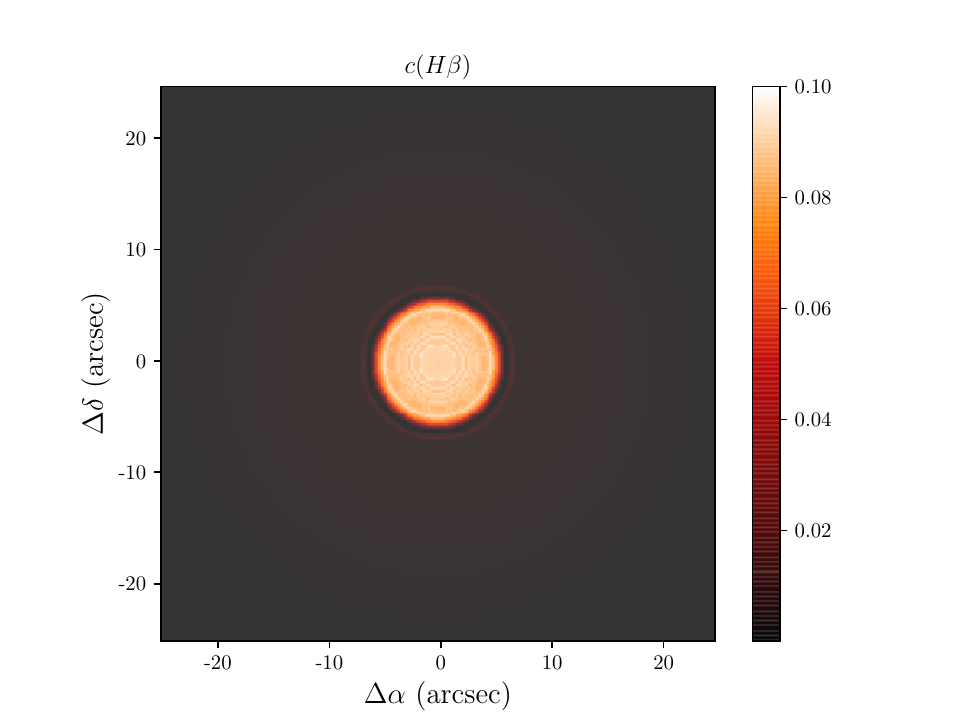}
\hrulefill
\includegraphics[trim={1.1cm 0.2cm 2.2cm 0.8cm},clip]{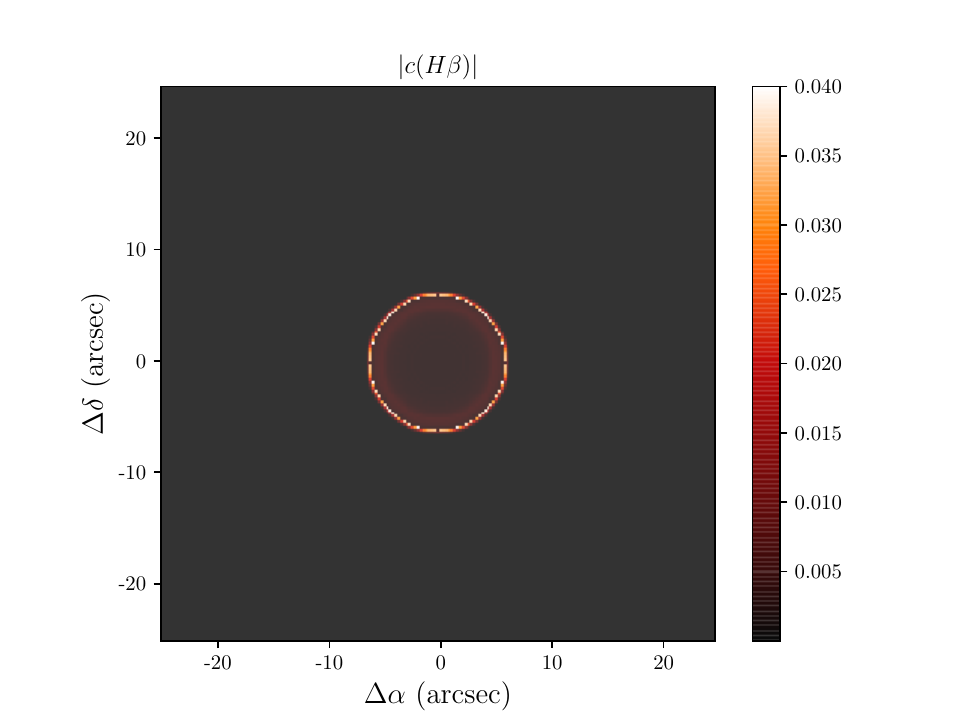}
}
\caption{An overview of the modelling results for a representative model 
employing AMC dust only and summarised in Table \ref{tab:ScatModfits}.
Upper row: the log$_{10}$ model surface gas density per cm$^{3}$ (left) and
log$_{10}$ surface dust density (right), designated by $\Sigma$; \\
Middle row: intrinsic (without line-of-sight extinction) log$_{10}$ surface 
density in H$\beta$ emission in erg cm$^{-2}$ s$^{-1}$ $\times$ 10$^{20}$ (left) and 
observed H$\beta$ emission with line-of-sight extinction (right); \\
Lower row: log extinction c(H$\beta$) from the model H$\alpha$/H$\beta$ ratio
images -- (left) for the extinction (i.e., sum of dust absorption and 
scattering), and (right) for the extinction decrement induced by the dust scattering 
only (expressed as $|$c(H$\beta$)$|$). \\
Axes show offsets with respect to the position of centre of the model (assumed
position of the CSt).  
}
\label{fig:ScatMod}
\end{figure}

Figure \ref{fig:ScatMod} shows a representative model employing AMC as 
described in Table \ref{tab:ScatModfits}. The projected gas and total dust 
density (sum of core, halo and Dust Shells A \& B contributions) of this model 
are shown in the upper panels; the middle panels show a log$_{10}$ image of 
intrinsic H$\beta$ emission (i.e. without extinction) and the observed H$\beta$ 
emission (with extinction by the intra-nebula dust); the lower panels show the 
image of the extinction, c(H$\beta$), derived from the ratio of the model 
H$\alpha$ and H$\beta$ images compared to Case B and the extinction 
difference (expressed as $|$c$|$) induced by the scattering from dust. 
While the results give a large-scale approximate match to the observed nebula, 
many details are lacking. For the extinction images, the mean value of extinction 
to the core nebula is reproduced ($c \cong$ 0.07, after subtracting the interstellar 
extinction), but the magnitude and morphology of the 
annulus of extinction deficit is poorly reproduced, with only a narrow rim of 
peak values, even though the 
Dust Shell 2 is much broader (Table \ref{tab:ScatModfits}). 

In the case of the scattering by C$_{60}$ molecules, the density was arbitrarily
increased to try to reach a reduction in extinction c(H$\beta$) comparable to the
observed value. Such decrements could be achieved but with a shell of very 
high density (1.0$\times$10$^{8}$ cm$^{-3}$ in the example in Table 
\ref{tab:ScatModfits}). For the set of parameters in the table, the
density of C60 molecules is then 5500 cm$^{-3}$. This would require a factor
$\sim$10$^{5}$ increase in C/H relative to the value for the core nebula
determined by \citet{Alemanetal2019} with a density $\sim$3000cm$^{-3}$.
Even allowing that one AMC grain in the MRN distribution representative of the
mean size could break down completely into about 9000 C60 molecules (based on
ratio of grain masses), such a density of C60 molecules is extreme.
This result would imply very exceptional conditions in the shell of dust around 
the Tc~1 core without a signature in the gaseous emission; it is concluded that 
scattering by fullerene molecules only cannot account for the apparent 
extinction deficit.

\section{Comparison of residual continuum (dereddened spectrum - nebular continuum)
for MUSE observations of NGC~7009, NGC~4361 and Tc~1}
\label{App:CompResids}

Already in \cite{Walshetal2018} and \cite{Walshetal2024} evidence was presented 
of an excess continuum above the nebular continuum for NGC~7009 (Fig. A.1) and
for NGC~4361 (Fig. 9). Here we compare these excess continua with that for Tc~1,
to investigate any similarities in the shape of the residual continuum for the
three PNe with MUSE observations and determinations of their nebular continuum.
  
\begin{figure}[t]
\centering
\resizebox{\hsize}{!}{
\includegraphics[trim={0.6cm 0.2cm 1.3cm 0.8cm},clip]{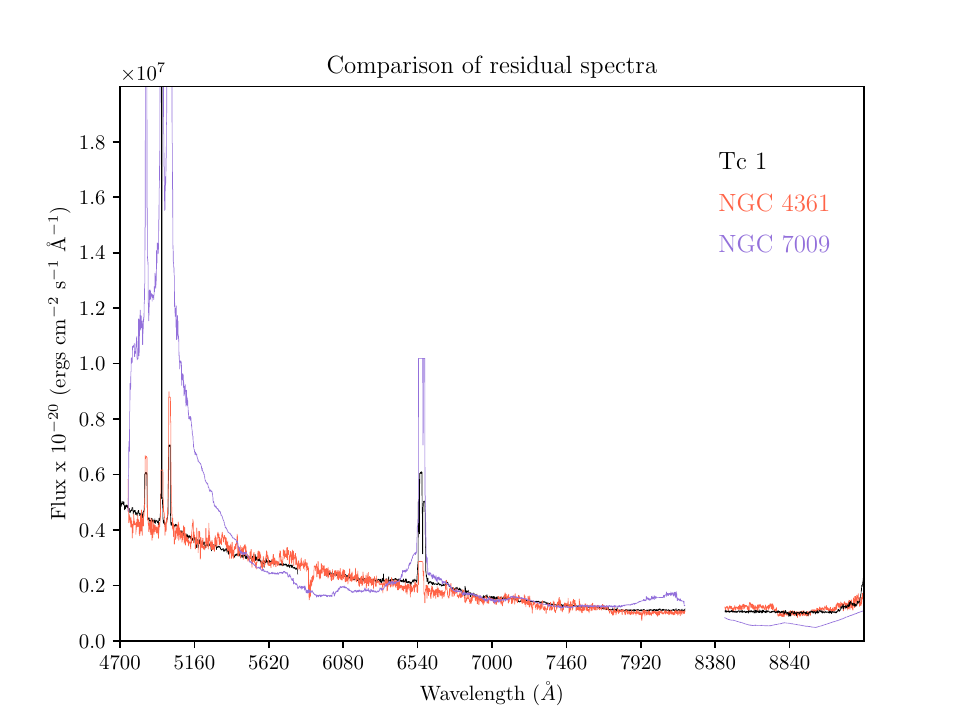}
}
\caption{Comparison of the residual spectra (i.e., dereddened spectrum of the
total nebular flux collected in the MUSE observations $-$ the fitted nebular 
continuum, as calculated from the parameters in Table \ref{tab:MUSE3nebcons}, 
is shown between Tc~1, NGC~4361 (Walsh et al. (2024) and NGC~7009 (Walsh et 
al. 2018). All spectra have been cleaned of narrow emission lines and sky 
subtraction residuals and normalised to the same flux at 7000\AA. See text 
for further details.
}
\label{fig:MUSE3nebcons}
\end{figure}

The residual spectra were calculated from the dereddened spectrum of the total 
nebular flux collected in the MUSE observations (with the area of the CSt 
point spread function excluded) $-$ the fitted nebular continuum 
(bound-free [bf] and free-free [ff] for \ion{H}{I}, bf for \ion{He}{I} and 
bf$+$ff for \ion{He}{II} and 2-photon for \ion{H}{I} and \ion{He}{II}). The 
adopted parameters and physical conditions for the 
nebular continuum computation are listed Table \ref{tab:MUSE3nebcons}.
The emission lines in the residual spectra were fitted by Gaussians and
removed but there were many fainter lines not in the list of fitted lines,
as well as residuals of sky subtraction (telluric emission and absorption). 
These features confused the comparison of the three spectra. The spectra were 
thus interactively cleaned to remove the traces of emission lines and
sky residuals with the aim to arrive at a representation of the continuum.    
Fig. \ref{fig:MUSE3nebcons} shows the cleaned residual continuum for Tc~1, 
NGC~4361 and NGC~7009. For the brightest lines which display broad line wings,
the residual profile was cut at some maximum for display purposes. To enable 
direct comparison, the three spectra have been normalised to the same mean flux 
at 7000\,\AA. The region between 8195 and 8440\,\AA\ is blanked out in this plot
since it was not possible to effectively fit all the closely-spaced high 
Paschen (and in the cases of NGC~4361 and NGC~7009, also the He$^{++}$, 
n=5-9 and n=6-36 ... 6-50) lines.  

\begin{table*}
\caption{Parameters of the nebular continuum fits to total spectra of three PNe}
\centering
\begin{tabular}{lrrrrrrr}
\hline\hline
Nebula & c(H$\beta$)   & Dered       & $T_{\rm e}$ & $T_{\rm e}$    & $N_{\rm e}$ & 
He$^{+}$/H$^{+}$ & He$^{++}$/H$^{+}$ \\
       &               & F(H$\beta$) & \ion{H}{I}  & \ion{He}{I/II} & (cm$^{-3}$) &                &                   \\
\hline
 Tc~1     & 0.31 & 4.77$\times$10$^{-11}$ & 8550 &  9150 & 2410 & 0.0920 &       \\
 NGC~4361 & 0.10 & 1.85$\times$10$^{-11}$ & 7500 & 17000 & 1500 & 0.0041 & 0.0951 \\
 NGC~7009 & 0.12 & 1.93$\times$10$^{-10}$ & 9600 &   9600 & 2500 & 0.0718 & 0.007 \\
\hline
\end{tabular}
\tablefoot{Adopted physical parameters used to calculate the nebular continuum for 
Tc~1, NGC~4361 and NGC~7009; compare to Table \ref{tab:nebcons} for Tc~1. See Walsh 
et al. (2024) for NGC~4361 and Walsh et al. (2018) for NGC~7009.
}
\label{tab:MUSE3nebcons}
\end{table*}

The shapes of the residual spectra show remarkably close similarity for Tc~1 
and NGC~4361, despite their very different extinction and ionization conditions. 
All three PNe residual spectra exhibit the rise at $\gtrsim$ 9000\,\AA\ caused by 
the MUSE 2nd order contamination. A power law in $\lambda^{-3}$ fits these two 
residual spectra quite well over the range 4700--8000\,\AA. 
For NGC~7009, there is some similarity in the overall shape, but 
the continuum S/N of this spectrum is lower than those for NGC~4361 and Tc~1 on 
account of the requisite of detecting unsaturated emission lines (exposure 
is a combination of 4 10\,s exposures). The broad wings under the very strong 
emission lines in NGC~7009 (\ion{O}{III}5007\,\AA\ is already saturated in this 
short exposure) complicate the comparison below 5600\,\AA. The broad pedestal to 
the strongest lines is interpreted as the grating line profile; for H$\alpha$ 
(as a single very strong line), its level is $\sim$7.7$\times$10$^{-5}$ at 100\,\AA\
offset from peak. A weaker pedestal is also visible to H$\alpha$ in 
Tc~1\footnote{This profile is much wider than the $\sim$1000 km s$^{-1}$ FWHM 
profiles reported on strong lines in Tc~1 by \cite{Alemanetal2019}.}.
Probably there is a nebular-specific component to these line wings in 
addition to instrumental effects, since their strengths relative to the line 
peak differ between Tc~1 and NGC~7009.  

Given that the three spectra all have the common presence of a hot 
star, while the nebular continuum spectra do not,
it seems most probable that some small fraction of the flux of the CSt
is distributed over the nebula. Although the $T_{eff}$ of the three 
PNe differ (31\,000 K for Tc~1, 85\,000 K for NGC~4361 and $\sim$95\,000 K
for NGC~7009), the optical spectra have very similar slopes 
($\propto \lambda^{-4}$). However the residual continuum follows 
$\lambda^{-3}$ so that some 'reddening' of the CSt continuum is required.
A possible mechanism could be internal extinction, but the amount required
to change a $\lambda^{-4}$ slope to $\lambda^{-3}$ is rather large
(c(H$\beta$) $\sim$ 0.5 dex.). An instrumental origin for this residual 
continuum still seems more likely, but has difficulties explaining how the 
extinction-corrected residual continua are so similar. The same residual 
spectra (dereddened $-$ nebular continuum) for other PNe with differing 
CSt properties (for example with a strong line-of-sight reddening to the CSt) 
could offer other possibilities to test the origin of this residual continuum.

\section{Spectra of the Tc~1 integrated regions -- core, low c, halo and total nebula}
\label{App:AnnSpects}

Lists of line identifications and Gaussian fluxes for the spectra of the three 
annular regions and the total spectrum of the full nebula (excluding an area around 
the CSt) are presented. See Sect. \ref{Subsec: Regions} for details on the 
extent of the integrated regions. Line identifications are taken from the line flux 
lists in \cite{Alemanetal2019}, Appendix A. The line flux errors are propagated 
from the flux error per pixel, provided by the MUSE data reduction pipeline, through 
the Gaussian fitting via the correlation matrix. The flux errors for the dereddened lines 
include the propagated contribution of the extinction error.     

\begin{table*}
\caption{Tc~1 MUSE Region Spectrum: Core}
\centering
\begin{tabular}{llrrrr}
\hline\hline
 Rest $\lambda$ & Line Species & Observed &         Error & Dered & Error  \\
 ~~~~~ (\AA)    &              & flux     &         flux  & flux  & flux  \\
\hline
  4713.1 & He~I             &     0.500 &     0.002 &     0.514 &     0.017 \\
  4754.8 & [Fe~III]         &     0.115 &     0.001 &     0.118 &     0.004 \\
  4861.2 & H4               &   100.000 &     0.000 &   100.000 &     0.000 \\
  4880.9 & [Fe~III]         &     0.116 &     0.001 &     0.116 &     0.004 \\
  4921.9 & He~I             &     1.253 &     0.002 &     1.239 &     0.040 \\
  4958.8 & [O~III]          &    15.417 &     0.006 &    15.147 &     0.487 \\
  4987.4 & [Fe~III]         &     0.063 &     0.001 &     0.062 &     0.002 \\
  5006.8 & [O~III]          &    46.535 &     0.012 &    45.327 &     1.450 \\
  5015.5 & He~I             &     2.733 &     0.002 &     2.658 &     0.085 \\

  5040.9 & Si~II            &     0.083 &     0.001 &     0.080 &     0.003 \\
\hline
\end{tabular}
\tablefoot{
Relative fluxes with respect to flux H$\beta$ = 100.0; \\
Observed \ion{H}{$\beta$} normalising flux: 5.67 $\times$10$^{-12}$ $\pm$ 
4.85$\times$10$^{-16}$ erg cm$^{-2}$ s$^{-1}$; \\
Dereddened \ion{H}{$\beta$} normalising flux: 1.19 $\times$10$^{-11}$ $\pm$ 
1.36$\times$10$^{-13}$ erg cm$^{-2}$ s$^{-1}$; \\
Extinction c = 0.320 $\pm$ 0.005 calculated with $T_{\rm e}$ = 9000 K; 
$N_{\rm e}$ = 2200 cm$^{-3}$. \\
Only the first ten entries of this table are shown; the full table is 
contained in the on-line material. \\
} 
\label{Tab:nebann}
\end{table*}

\begin{table*}
\caption{Tc~1 MUSE Region Spectrum: Low c}
\centering
\begin{tabular}{llrrrr}
\hline\hline
 Rest $\lambda$ & Line Species & Observed &         Error & Dered & Error  \\
 ~~~~~ (\AA)    &              & flux     &         flux  & flux  & flux  \\
\hline
  4713.1 & He~I             &     0.402 &     0.008 &     0.409 &     0.028 \\
  4754.7 & [Fe~III]         &     0.273 &     0.008 &     0.276 &     0.020 \\
  4861.2 & H4               &   100.000 &     0.000 &   100.000 &     0.000 \\
  4881.0 & [Fe~III]         &     0.306 &     0.006 &     0.306 &     0.021 \\
  4921.9 & He~I             &     1.021 &     0.007 &     1.015 &     0.066 \\
  4958.8 & [O~III]          &    10.574 &     0.015 &    10.470 &     0.674 \\
  4987.4 & [Fe~III]         &     0.366 &     0.006 &     0.362 &     0.024 \\
  5006.8 & [O~III]          &    31.808 &     0.026 &    31.341 &     2.005 \\
  5015.5 & He~I             &     2.557 &     0.009 &     2.518 &     0.161 \\
  5047.5 & He~I             &     0.283 &     0.004 &     0.278 &     0.018 \\
\hline
\end{tabular}
\tablefoot{
Relative fluxes with respect to H$\beta$ = 100.0; \\
Observed \ion{H}{$\beta$} normalising flux: 5.49 $\times$10$^{-13}$ $\pm$ 
1.16$\times$10$^{-16}$ erg cm$^{-2}$ s$^{-1}$ \\
Dereddened \ion{H}{$\beta$} normalising flux: 8.30 $\times$10$^{-13}$ $\pm$ 
1.91$\times$10$^{-14}$ erg cm$^{-2}$ s$^{-1}$ \\
Extinction c = 0.18 $\pm$ 0.01 calculated with $T_{\rm e}$ = 9000 K; 
$N_{\rm e}$ = 1500 cm$^{-3}$. \\
Only the first ten entries of this table are shown; the full table is 
contained in the on-line material. \\
} 
\label{Tab:cann}
\end{table*}

\begin{table*}
\caption{Tc~1 MUSE Region Spectrum: Halo}
\centering
\begin{tabular}{llrrrr}
\hline\hline
 Rest $\lambda$ & Line Species & Observed &         Error & Dered & Error  \\
 ~~~~~ (\AA)    &              & flux     &         flux  & flux  & flux  \\
\hline
  4713.6 & He~I             &     0.177 &     0.041 &     0.181 &     0.043 \\
  4754.1 & [Fe~III]         &     0.479 &     0.046 &     0.486 &     0.056 \\
  4780.4 & N~II             &     0.251 &     0.041 &     0.253 &     0.044 \\
  4789.9 & N~II             &     0.224 &     0.005 &     0.227 &     0.016 \\
  4861.2 & H4               &   100.000 &     0.000 &   100.000 &     0.000 \\
  4880.8 & [Fe~III]         &     0.457 &     0.051 &     0.456 &     0.059 \\
  4922.0 & He~I             &     0.400 &     0.023 &     0.397 &     0.034 \\
  4958.9 & [O~III]          &    12.843 &     0.055 &    12.674 &     0.817 \\
  4987.4 & [Fe~III]         &     1.716 &     0.046 &     1.688 &     0.117 \\
  5006.8 & [O~III]          &    38.773 &     0.076 &    38.015 &     2.433 \\
\hline
\end{tabular}
\tablefoot{
Relative fluxes with respect to H$\beta$ = 100.0; \\
Observed \ion{H}{$\beta$} normalising flux: 1.52 $\times$10$^{-13}$ $\pm$ 
7.27$\times$10$^{-17}$ erg cm$^{-2}$ s$^{-1}$ \\
Dereddened \ion{H}{$\beta$} normalising flux: 2.63 $\times$10$^{-13}$ $\pm$ 
6.07$\times$10$^{-15}$ erg cm$^{-2}$ s$^{-1}$ \\
Extinction c = 0.24 $\pm$ 0.01 calculated with $T_{\rm e}$ = 9000 K; 
$N_{\rm e}$ = 700 cm$^{-3}$. \\
Only the first ten entries of this table are shown; the full table is 
contained in the on-line material. \\
} 
\label{Tab:outerann}
\end{table*}

\begin{table*}
\caption{Tc~1 MUSE Region Spectrum: Total}
\centering
\begin{tabular}{llrrrr}
\hline\hline
 Rest $\lambda$ & Line Species & Observed &         Error & Dered & Error  \\
 ~~~~~ (\AA)    &              & flux     &         flux  & flux  & flux  \\
\hline
  4713.1 & He~I             &     0.483 &     0.002 &     0.496 &     0.033 \\
  4754.7 & [Fe~III]         &     0.114 &     0.001 &     0.116 &     0.008 \\
  4861.2 & H4               &   100.000 &     0.000 &   100.000 &     0.000 \\
  4880.9 & [Fe~III]         &     0.118 &     0.001 &     0.118 &     0.008 \\
  4921.9 & He~I             &     1.191 &     0.001 &     1.179 &     0.076 \\
  4958.8 & [O~III]          &    28.619 &     0.005 &    28.133 &     1.810 \\
  4987.4 & [Fe~III]         &     0.103 &     0.001 &     0.101 &     0.007 \\
  5006.8 & [O~III]          &    86.465 &     0.012 &    84.289 &     5.392 \\
  5015.5 & He~I             &     2.724 &     0.002 &     2.652 &     0.169 \\
  5035.5 & O~II             &     0.051 &     0.001 &     0.049 &     0.003 \\
\hline
\end{tabular}
\tablefoot{
Relative fluxes with respect to H$\beta$ = 100.0; \\
Observed \ion{H}{$\beta$} normalising flux: 2.15 $\times$10$^{-11}$ $\pm$ 
9.74$\times$10$^{-16}$ erg cm$^{-2}$ s$^{-1}$ \\
Dereddened \ion{H}{$\beta$} normalising flux: 4.39 $\times$10$^{-11}$ $\pm$ 
1.01$\times$10$^{-12}$ erg cm$^{-2}$ s$^{-1}$ \\
Extinction c = 0.31 $\pm$ 0.01 calculated with $T_{\rm e}$ = 9000 K; 
$N_{\rm e}$ = 1500 cm$^{-3}$. \\
Only the first ten entries of this table are shown; the full table is 
contained in the on-line material. \\
} 
\label{tab:Total-CS}
\end{table*}

\end{appendix}

\end{document}